\documentclass[preprintnumbers,prd,superscriptaddress,preprint,aps,10pt]{revtex4-1}

\usepackage{graphicx}
\usepackage{dcolumn}
\usepackage{bm}
\usepackage{xcolor}
\usepackage{comment}
\usepackage{hyperref}
\usepackage{cancel}
\usepackage{amsmath} 
\usepackage{amssymb} 
\usepackage{amsfonts} 
\usepackage{slashed}
\usepackage{multirow}
\usepackage{amsfonts} 
\usepackage{mathtools}
\usepackage{ulem}
\newcommand{\CP}{{CP}}

\newcommand{\CPV}{{\cancel{\rm CP}}}
\newcommand{\bea}{\begin{eqnarray}}
\newcommand{\eea}{\end{eqnarray}}
\newcommand{\be}{\begin{equation}}
\newcommand{\ee}{\end{equation}}
\newcommand{\Slash}[1]{{\ooalign{\hfil/\hfil\crcr$#1$}}}
\newcommand{\ba}{\begin{array}}
\newcommand{\ea}{\end{array}}

\newcommand{\mcP}{{\mathcal P}}
\newcommand{\mcF}{{\mathcal F}}
\newcommand{\mcA}{{\mathcal A}}

\newcommand{\mcD}{{\mathcal D}}
\newcommand{\mcE}{{\mathcal E}}

\newcommand{\mcO}{{\mathcal O}}

\newcommand{\mcN}{{\mathcal N}}
\newcommand{\mcH}{{\mathcal H}}

\newcommand{\Tr}{\mathrm{Tr}}

\newcommand{\lp}{\left}
\newcommand{\rp}{\right}
\newcommand{\la}{\langle}
\newcommand{\ra}{\rangle}
\newcommand{\vep}{\varepsilon}
\newcommand{\pd}[2]{\frac{\partial #1}{\partial #2}}

\begin{document}


\title{The Neutron Electric Dipole Moment from Lattice QCD
using a Background Electric Field }

\author{Thomas Blum}%
\affiliation{Physics Department, University of Connecticut, Storrs, Connecticut 06269, USA
}
\author{Fangcheng He}
\affiliation{Center for Nuclear Theory, Department of Physics and Astronomy, Stony Brook University, Stony Brook, New York 11794-3800, USA}
\affiliation{Department of Physics, New Mexico State University, Las Cruces, NM 88003, USA}
\affiliation{Nuclear Science Division, Lawrence Berkeley National Laboratory, Berkeley, CA 94720, USA}

\author{Taku~Izubuchi} 
\affiliation{Physics Department, Brookhaven National Laboratory, Upton, New York 11973, USA
}
\affiliation{RIKEN-BNL Research Center, Brookhaven National Laboratory, Upton, NY 11973, USA}

\author{Luchang Jin}
\affiliation{Physics Department, University of Connecticut, Storrs, Connecticut 06269, USA
}

\author{Hiroshi Ohki}
\affiliation{Department of Physics, Nara Women's University, Nara 630-8506, Japan}
\author{Sergey Syritsyn}
\affiliation{Center for Nuclear Theory, Department of Physics and Astronomy, Stony Brook University, Stony Brook, New York 11794-3800, USA}

\date{\today}

%%%%%%%%%%%%%%%%%%%%%%%%%%%%%%%%%%%%%%%%%%%%%%%%%%%%%%%%%%%%%%%%%%%%%%%%%%%%%%%%
\begin{abstract}
We present the calculation of the neutron electric dipole moment (nEDM) $d_n$ using 2+1 flavor domain wall fermion
ensembles with fixed lattice spacing $a\approx 0.11\,\text{fm}$ and pion masses of 340, 420, and 576 MeV.
We show that the neutron electric dipole moment can be extracted from the energy shift induced by a static uniform
external background electric field in the presence of the CP-violating QCD theta-term, $\bar\theta Q_{top}$.
Motivated by the Feynman-Hellmann theorem, we employ sampling of the topological charge $q_\text{top}(t)$ on a
single time-slice rather than the global topological charge $Q_\text{top}=\int q_\text{top}(t) \, dt$, which
dramatically improves the statistical precision of the $\theta$-induced nEDM.
Key to our method is to calculate the forward matrix element of the topological charge density in the nucleon deformed
by a background electric field. 
We find that calculation with the traditional positive parity-projected nucleon operator is subject to large excited-state contamination.
To remove the contamination, we construct the ground state of the deformed nucleon by solving a non-Hermitian generalized eigenvalue
problem.
With this approach, we find consistent values for the nEDM when using different nucleon interpolating operators,
regardless of whether they are covariant or non-covariant under chiral transformations.
Finally, after extrapolating to the physical point, we obtain $d_n=-0.0050(4)^\text{stat}(8)^\text{sys}\bar{\theta}$ $e$ fm, where the systematic uncertainty includes  excited-state effects estimated as variation with the Euclidean-time fits and the dependence on the strength of the electric field applied to the neutron. Conventional systematic errors like discretization, finite-volume, and chiral extrapolation effects will be addressed in future work.

%%%%%%%%%%%%%%%%%%%%%%%%%%%%%%%%%%%%%%%%%%%%%%%%%%%%%%%%%%%%%%%%%%%%%%%%%%%%%%%%
\begin{comment}
\item[Usage]
Secondary publications and information retrieval purposes.
\item[Structure]
You may use the \texttt{description} environment to structure your abstract;
use the optional argument of the \verb+\item+ command to give the category of each item. 
\end{comment}
%%%%%%%%%%

%\end{description}
\end{abstract}

\maketitle

\tableofcontents

%%%%%%%%%%%%%%%%%%%%%%%%%%%%%%%%%%%%%%%%%%%%%%%%%%%%%%%%%%%%%%%%%%%%%%%%%%%%%%%%
%%%%%%%%%%%%%%%%%%%%%%%%%%%%%%%%%%%%%%%%%%%%%%%%%%%%%%%%%%%%%%%%%%%%%%%%%%%%%%%%
\section{\label{sec:intro}Introduction} 
Permanent electric dipole moments (EDMs) of nucleons, nuclei, atoms, and molecules are sensitive probes for measuring
the strength of CP violation in Nature.
In the Electroweak sector of the Standard Model (SM), all CP violation is  caused by a single parameter, the imaginary
phase of the Cabibbo-Kobayashi-Maskawa (CKM) quark mixing matrix~\cite{Kobayashi:1973fv}.
The corresponding predicted contributions to the neutron EDM ($d_n$) and proton EDM ($d_p$) are  $O(10^{-32})$ $e$-cm
\cite{Khriplovich:1981ca,Czarnecki:1997bu,Seng:2014lea}, $10^5-10^6\times$ smaller than the most direct recent experimental
bound, $d_n \sim  10^{-26}$ $e$-cm~\cite{Abel:2020gbr} and the indirect bounds $d_n<1.6 \times 10^{-26}$ $e$-cm and
$d_p<2.0\times 10^{-25}$ $e$-cm obtained from the experimental limit  on the $^{199}$Hg atom EDM~\cite{Graner:2016ses}.
The sensitivity of experimental measurements to the nEDM is expected to improve by up to two orders of magnitude in the next
10-20 years to $d_n \approx 3 \times 10^{-28}$ $e$-cm at the 90\% confidence level~\cite{Alarcon:2022ero,Ito2019}.

In addition to the CKM matrix, the only other source of CP violation within the SM is the ``$\Theta$-term'' of the
strong interactions (QCD).
Beyond the Standard model, there may also be quark chromo–electric interactions, the Weinberg operator~\cite{Weinberg:1989dx}, 
four-quark and other higher-dimensional operators induced by yet undiscovered particles and interactions.
The difficulty of calculating the contributions of these interactions to the neutron EDM (nEDM) is due to the non-perturbative nature of QCD at the hadronic scale.
Calculations based on the chiral Lagrangian~\cite{Crewther:1979pi,Pich:1991fq,Cho:1992rv,Borasoy:2000pq,Hockings:2005cn,Narison:2008jp,Ottnad:2009jw,deVries:2010ah,Mereghetti:2010kp}, QCD sum rules~\cite{Pospelov:1999ha,Pospelov:2000bw,Lebedev:2004va,Pospelov:2005pr,Fuyuto:2012yf,Haisch:2019bml,Ema:2024vfn} and the instanton liquid model~\cite{Faccioli:2004jz,Liu:2025kuc}
have provided useful estimates of these matrix elements albeit with large model uncertainties.
In contrast to these phenomenological approaches, Lattice QCD provides, in principle, an {\it ab initio} framework for
calculating CP-violating matrix elements with systematically controllable uncertainties.
Numerous early lattice studies reported large values of the $\Theta$-induced
term~\cite{Shintani:2005xg,Berruto:2005hg,Shindler:2014oha,Guo:2015tla,Shindler:2015aqa,Alexandrou:2015spa,Shintani:2015vsx}
which were eventually attributed to an incorrect definition of the electric dipole form factor $F_3$ and mixing with the
Pauli form factor~\cite{Abramczyk:2017oxr}.
After subtracting the mixing term, those nEDM results became significantly smaller, and while comparable with
phenomenology, they were universally dominated by statistical noise even at unphysical heavy quark masses.
More recent Lattice calculations have addressed this mixing problem and now employ the correct definition of the electric
dipole form factors(EDFF)~\cite{Dragos:2019oxn,Alexandrou:2020mds,Bhattacharya:2021lol,Liang:2023jfj}.
However, it is still difficult to obtain a good statistical signal for $\Theta$-induced nEDM in direct calculations at
the physical point~\cite{Alexandrou:2020mds,Bhattacharya:2021lol}.
A nonzero value has been obtained by calculating the EDM using heavier quark masses and extrapolating to the physical
point, albeit with significant systematic uncertainties~\cite{Dragos:2019oxn,Liang:2023jfj}.
On the other hand, it remains controversial even whether the QCD $\Theta$-term can generate a nonzero nucleon 
EDM~\cite{Ai:2024cnp,Schierholz:2024var,Khoze:2025auv,Benabou:2025viy,Bhattacharya:2025qsk,Aghaie:2026pkf}.
In addition to the $\Theta$-induced EDM, nucleon EDMs induced by the quark EDM~\cite{Bhattacharya:2015esa}, 
quark chromo–EDM~\cite{Abramczyk:2017oxr,Bhattacharya:2018qat,Kim:2018rce,Bhattacharya:2022whc}, and the Weinberg
term~\cite{Dragos:2017wms,Bhattacharya:2022whc} have also been studied in lattice QCD.
Overall, a precise lattice calculation of nucleon EDMs from all phenomenologically relevant sources remains a demanding task.

Currently, most lattice calculations of nEDMs use the traditional form factor method, in which they are extracted from
the electric dipole form factor, $F_3(Q^2)$.
This form factor appears as a $\CP$-odd contribution to the nucleon matrix elements of the quark vector current due to the
$\theta$-angle or other sources of CP violation, and it has to be extrapolated to the forward limit ($Q^2\to0$) to obtain
the nEDM.
An alternative method, pursued here, is to calculate the EDM from the nucleon energy shift 
$\Delta E\propto d_N(2\vec S\cdot\vec E)$ in a uniform background electric 
field~\cite{Aoki:1989rx,Shintani:2006xr,Shintani:2008nt,Izubuchi:2007rmy,Abramczyk:2017oxr,Izubuchi:2020ngl,He:2023gwp}. In this approach, the external electric field is treated perturbatively and should be sufficiently weak for the higher-order corrections to be neglected.
Significant advantages of this method are that no forward-limit extrapolation is required (although the interpolation to zero external electric field is required) and there is no parity
mixing between $CP$-even and $CP$-odd form factors $F_2$ and $F_3$ (although the nucleon ground state is naturally parity-mixed due to its polarizability).
In addition, in the case of the $\Theta$-induced nEDM, one only needs to calculate the $\CP$-odd part of a simpler nucleon
two-point correlation function.

Also, all lattice calculations mentioned in the previous paragraphs used the global topological charge summed over the
entire lattice volume, which produces large statistical fluctuations in nucleon correlation functions.
In this work, we present our novel method to use local (summed over a single time-slice) topological charge to
compute the $\Theta$-induced neutron EDM, which requires imposing a uniform electric background field in Euclidean space.
According to the Feynman-Hellmann theorem, the resulting energy shift induced by the QCD ``$\Theta$-term" is
proportional to the matrix element of the local topological charge sandwiched between ground-state nucleon states deformed by the electric field, which can be isolated by solving a generalized eigenvalue problem (GEVP) in the space of
the nucleon, its parity partner, and their excited states~\cite{Hackl:2024whw}.
This method enables a more statistically precise and systematically reliable determination of the nEDM.
Furthermore, motivated by the suggestion that one should compute the nEDM using nucleon operators that transform covariantly
under chiral transformations~\cite{Ema:2024vfn}, we compare how such a choice may affect the nEDM result.
We find that nEDM values extracted from the non-covariant and covariant nucleon operators are comparable within statistical errors, provided that
the operators have good overlap with relevant nucleon eigenstates in the background field.

Finally, we compare nEDM values obtained with topological charge determined from gradient-flowed gluon fields and,
inspired by the ABJ anomaly, from the pseudoscalar quark density, with and without axial-vector quark density correction terms, and with varying quark masses~\cite{Smit:1986fn, Itoh:1987iy, Alexandrou:2017hqw}.
The results based on the gluon topological charge with varying gradient flow times are consistent with each other,
while those based on the pseudoscalar quark density converge to the same value as the mass of the quark probing the
topology is increased.
For smaller quark mass, we find the nEDM obtained from the pseudoscalar quark density shows stronger $N\pi$ contamination.

This paper is organized as follows: Section~\ref{sec:EDMext} introduces the electric dipole form factor and background
field methods, we discuss the extraction of the EDM both in the absence and presence of excited states, present the method
for isolating the ground state of the nucleon in the presence of a background field, and describe how to extract the
neutron EDM using the local topological charge.
In Section~\ref{sec:comp}, we describe the computational setup, including the details of our lattice ensembles and the
implementation of the background electric field on the lattice.
In Subsection~\ref{sec:top}, we compare two different definitions of topological charge, one constructed from the
gradient-flowed gluon field, and the other defined using the pseudoscalar quark density constructed with the
low eigenmodes of the preconditioned Möbius Domain Wall Fermion (MDWF) Dirac operator.
Numerical results  are presented in Sec.~\ref{sec:Numres}, in which we compare values of the nEDM obtained using
different nucleon interpolating operators.
Our conclusions are summarized in Section~\ref{sec:conc}.

%%%%%%%%%%%%%%%%%%%%%%%%%%%%%%%%%%%%%%%%%%%%%%%%%%%%%%%%%%%%%%%%
%%%%%%%%%%%%%%%%%%%%%%%%%%%%%%%%%%%%%%%%%%%%%%%%%%%%%%%%%%%%%%%%
\section{Extraction of the EDM from the nucleon correlation function
  \label{sec:EDMext}}

%%%%%%%%%%%%%%%%%%%%%%%%%%%%%%%%%%%%%%%%%%%%%%%%%%%%%%%%%%%%%%%%
\subsection{Electric dipole moment from background field method}\label{sec:theory}
The QCD $\Theta$-term is the only dimension-four operator allowed by the symmetries of the strong interactions, and it breaks P and CP symmetries. 
In the  presence of the $\Theta$-term, and under a Wick rotation to imaginary time, the Boltzmann weight of path integral reads
\begin{equation}\label{eq:actionME}
\exp\left[iS_{QCD,M}+i\frac{\bar\theta}{32\pi^2} \int d^4x_M  G^a_{\mu\nu,M}  {\widetilde G^a_{\mu\nu,M}}\right]
\quad\longrightarrow\quad
\exp\left[-S_{QCD,E}-i\frac{\bar\theta}{32\pi^2} \int d^4x_E  G^a_{\mu\nu,E}  {\widetilde G^a_{\mu\nu,E}}\right]\,,
\end{equation}
where $\bar{\theta}$ is the physical theta angle, defined as $\bar{\theta} = \theta_{QCD} + {\rm Arg\ Det}M_q$, which
incorporates both the original QCD angle $\theta_{QCD}$ and the phase of the quark mass matrix $M_q$.
The subscripts M and E denote Minkowski and Euclidean space, respectively. More details on the notation in  Minkowski
space and Euclidean space can be found in  Appendix~\ref{sec:appenx_EM}. 
The $\Theta$-term becomes an imaginary phase in Euclidean space, making it difficult to  perform direct Monte Carlo
simulations.

The form factor method has been described in multiple publications (see, e.g.,
Ref.\cite{Abramczyk:2017oxr} and references therein).
The electric dipole form factor (EDFF) is defined as
\begin{equation}
\label{eqn:ff_cpviol}
\la p^\prime,\sigma^\prime |J^\mu|p,\sigma
\ra_{\CPV}
= \bar{u}_{p^\prime,\sigma^\prime} 
   \lp[F_1(Q^2) \gamma^\mu_M + \lp(F_2(Q^2) 
  + iF_3(Q^2)\gamma^5_M\rp) \lp(\frac{i\sigma^{\mu\nu}q_\nu}{2M_N}\rp)_M \rp] u_{p,\sigma}\,
\end{equation}
where $Q^2=-(p^\prime-p)^2$ is the momentum transfer to the nucleon from the external photon, 
and $F_{1,2,3}$ denote the Dirac, Pauli, and electric dipole form factors.
The forward limit of the latter yields the nEDM, $F_3(0)=2m_nd_n$. 
Although nucleon states in the QCD vacuum with $\CP$-violation are no longer parity eigenstates,
it is crucial to ensure that their spinors satisfy the positive-parity Dirac equation with real-valued mass,
$(\Slash{p}-m_N) u_{p,\sigma} = 0$,
otherwise $F_3$ receives a spurious contribution from $F_2$~\cite{Abramczyk:2017oxr},
\begin{equation}\label{eq:FFrel}
\begin{aligned}
\tilde F_2(Q^2) &= F_2\cos(2\alpha) + F_3\sin(2\alpha), \nonumber\\
\tilde F_3(Q^2) &= F_3\cos(2\alpha) - F_2\sin(2\alpha),
\end{aligned}
\end{equation}
where $\alpha$ is the parity mixing of the nucleon spinor due to $\CP$-violation, and $\tilde F_2(Q^2)$ and $F_3(Q^2)$
are the form factors defined using the $\CPV$ Dirac spinor basis,
\begin{equation}
\label{eqn:ff_cpviolutilde}
\la p^\prime,\sigma^\prime |J^\mu|p,\sigma
\ra_{\CPV} = \bar{\tilde{u}}_{p^\prime,\sigma^\prime} 
   \left[\tilde{F}_1(Q^2) \gamma^\mu_M + \big(\tilde{F}_2(Q^2) 
  + i\tilde{F}_3(Q^2)\gamma_M^5\big) \left(\frac{i\sigma^{\mu\nu}q_\nu}{2M_N}\right)_M \right] \tilde{u}_{p,\sigma}\,
\end{equation}
with $\tilde{u}=e^{i\alpha\gamma_5}u$ and $\bar{\tilde{u}}=\bar{u}e^{i\alpha\gamma_5}$.  

In Euclidean space, the definition of the EDFF can be written as
\begin{equation}
\label{eqn:ff_cpviol_E}
\la p^\prime,\sigma^\prime |J^\mu|p,\sigma
\ra_{\CPV}
= \bar{u}_{p^\prime,\sigma^\prime} 
   \left[F_1(Q^2) \gamma^\mu_E + \big(F_2(Q^2) 
  - iF_3(Q^2)\gamma^5_E\big) \left(\frac{\sigma^{\mu\nu}q_\nu}{2M_N}\right)_E \right] u_{p,\sigma}.
\end{equation}
The details about the transformation from Minkowski to Euclidean space can be found in Appendix~\ref{sec:appenx_EM}.
Since we calculate the nEDM on the Euclidean lattice, we use the Euclidean setup as the default and omit the subscript E
in the following.

Alternatively, the nEDM can be extracted using the background electric field method.
The idea is that the nucleon energy shift in the presence of a background field is proportional to the strength of the electric field.
Such an approach was introduced in \cite{Aoki:1989rx, Shintani:2006xr, Shintani:2008nt, Izubuchi:2007rmy} and also has been extended
to analyze CP-even properties, such as the electric polarizability and magnetic moment of the
nucleon~\cite{Detmold:2009dx, Detmold:2010ts}.
In this work, we use this method to calculate only the neutron EDM, because the proton is more complicated due to its
acceleration in the electric field. 

After imposing a uniform background electric field, the Dirac equation for the parity-positive neutron spinor $u_N$ in the rest frame
($p_N=(iE_N,\vec0)$) becomes in Euclidean space~\cite{Abramczyk:2017oxr}\footnote{
  The explicit matrix form on the right hand side is obtained in the Weyl $\gamma$-matrix basis, following the convention adopted in Ref.~\cite{Abramczyk:2017oxr}.
  For most of the discussion below, we use spin-parity (Dirac) basis.
}
\begin{equation}
\label{eqn:neutron_dslash}
\lp[ i \slashed{p} + m_N 
  -\big(\frac12 \mathcal{F}_{\mu\nu}\sigma^{\mu\nu}\big)\frac{F_2(0) - i\zeta\gamma_5}{2m_N} \rp] \, u_{N}
  = \lp(\ba{cc} 
    m_N - \frac{(F_2(0) - i\zeta)\vec\mcE\cdot\vec\sigma}{2m_N} &   -E_N \\ 
    -E_N &   m_N + \frac{(F_2(0)+i\zeta)\vec\mcE\cdot\vec\sigma}{2m_N}
  \ea\rp) \, u_N  = 0\,, 
\end{equation}
where $E_N$ is the neutron energy, $\zeta=F_{3}(0)$ is the electric dipole
moment, and $\vec\mcE$ is the Euclidean electric field.
To linear order in the electric field strength $\mcE$, the nucleon energy $E_N$ is
\begin{equation}
E_N^2 = m_N^2 + i \zeta (\vec\Sigma \cdot\vec\mcE) + O(\mcE^2) \,,
\quad\text{ or }\quad
E_N = m_N + \frac{i\zeta}{2m_N}\,(\vec\Sigma \cdot\vec\mcE) + O(\mcE^2)
\end{equation}
where $\zeta/(2m_N)=d_N$ is the electric dipole moment and
$\vec\Sigma=\text{diag}[\vec\sigma,\vec\sigma]$ is the spin operator.
Note that the linear part of the energy shift 
$\delta E=\frac{i\zeta}{2m_N}\,(\vec\Sigma \cdot\vec\mcE)$ 
is imaginary because of the analytic continuation of the  electric field.
Naively one might think that it results in imaginary contributions to the eigen-energies and complex phases in
correlation functions, but in fact, this anti-Hermitian perturbation leads to non-orthogonality of eigenvectors and
resembles a so-called $PT$-even quantum-mechanical system~\cite{Bender:2023cem}.

%%%%%%%%%%%%%%%%%%%%%%%%%%%%%%%%%%%%%%%%%%%%%%%%%%%%%%%%%%%%%%%%
\subsection{EDM in the absence of excited states
  \label{sec:edm_point}}

In this subsection, we briefly consider the hypothetical case when the neutron does not interact with any other QCD states.
This means that we ignore contributions from any nucleon internal excited states as well as neglect any meson-cloud effects. In other words, we regard it as a ``sterile'', non-deformable elementary spin-1/2 particle strictly described by the Dirac equation~(\ref{eqn:neutron_dslash}) from which its magnetic $\kappa=F_2(0)$ and electric $\zeta=F_3(0)$ dipole moments may be precisely extracted\footnote{In the context of non-relativistic quantum mechanics, the only excited states such a ``sterile'' neutron would have are particle-antiparticle pair excited states with energies $\Delta E=2m_N,\,4m_N,\,\ldots$, which connect it to the relativistic picture.}.
The goal of this exercise is to merely set the scene for the discussion in the following sections of how excited states affect extraction of $\kappa$ and $\zeta$ in the presence of Euclidean background electric field.

The zero-momentum neutron two-point correlation function in the presence of a background electric field can be written as~\cite{Detmold:2010ts}
\begin{equation}
\label{eq:2pte}
C_{2pt,\mcE} = |Z_N|^2 \left(\frac{1+\gamma_4}{2}+i\frac{\kappa}{4m_N^2}\gamma_3\gamma_4\mcE_z\right)e^{-m_Nt}+O(\mcE_z^2),
\end{equation}
where $\kappa=F_2(0)$ is the neutron anomalous magnetic moment as if it was a point-like, ``non-deformable'', particle.
In principle, $\kappa$ might be estimated from the following ratio in the limit of large source-sink separation~\cite{Detmold:2010ts}
\begin{equation}
\label{eq:ratiokappa}
\kappa^\text{est}(t_f)=i\frac{2m_N^2}{\mcE_z}\frac{\Tr[\gamma_3\gamma_4C_{2pt,\mcE}(t_f)]}{\Tr[T^+C_{2pt,\mcE}(t_f)]}
\quad \xRightarrow{t_f\to\infty} \quad \kappa \,,
\end{equation}
where $T^+=\frac12(1+\gamma_4)$ is the positive-parity projector. 
However, in Sec.~\ref{sec:edm_exc} below we show that this estimator for $\kappa$ receives a contribution from parity
mixing between the positive-parity nucleon and its heavier negative-parity partner, as well as the vacuum polarization, 
caused by the ordinary electric dipole operator $\mcD = (\sum_a q_a\vec r_a)$, which is also responsible for nucleon and
vacuum polarizability.

We can express its two-point correlator $C^{{\rm 2pt},\vec{E}}_{\CPV}(\vec{0},t)$ in the presence of \emph{both} the
background electric field \emph{and} the $\bar\theta$-term as 
\begin{equation}
\label{eq:eshift}
\begin{aligned}
C_{2pt,\mcE,\bar\theta} 
  &=|Z_N|^2 \sum_{\sigma=\pm1} \tilde{u}_{E_z,s} \bar{\tilde{u}}_{E_z,s} \frac{e^{-(m_N+i\sigma d_n\mcE_z)t}}{2m_N} 
\\&=|Z_N|^2 e^{i\alpha\gamma_5}\left(\frac{1+\gamma_4}{2} 
  \left[ 
  \frac{(1+\Sigma_z)}{2}e^{-id_n\mcE_zt} +\frac{(1-\Sigma_z)}{2}e^{id_n\mcE_zt}
  \right]+i\frac{\kappa}{4m_N^2}\gamma_3\gamma_4\mcE_z\right) e^{-m_Nt} e^{i\alpha\gamma_5}
  +O(\mcE_z^2),
\\&=|Z_N|^2 \left(\frac{1+\gamma_4}{2}+i\frac{\kappa}{4m_N^2}\gamma_3\gamma_4\mcE_z\right)e^{-m_Nt}
\\&+|Z_N|^2\left(i\alpha\gamma_5
  -i\frac{1+\gamma_4}{2}\Sigma_zd_n\mcE_z t+i\alpha\frac{\kappa}{2m_N^2}\Sigma_z\mcE_z\right)
  e^{-m_Nt}+O(\mcE_z^2,\bar\theta^2)\,,
\end{aligned}
\end{equation}
where $\tilde{u}_{E_z,s} = e^{i\alpha\gamma_5} u_{E_z,s}$ is the $\CPV$ Dirac spinor and $\alpha$ is the parity-mixing angle that
leads to spurious mixing between the electric dipole moment, $d_n$, and the magnetic dipole moment, $\kappa$~\cite{Abramczyk:2017oxr}.
On the other hand, based on the Euclidean action defined in Eq.~(\ref{eq:actionME}) one can expand the path integral for the correlation function in $\bar{\theta}\ll1$ as
\begin{equation}\label{eq:thetaexp}
C_{2pt,\mcE,\bar{\theta}} 
\approx C_{2pt,\mcE}  -i\bar{\theta} \la Q_{top} \, N(t) \bar{N}(0)\ra_\mcE \,.
\end{equation}
By comparing Eq.~(\ref{eq:eshift}) with Eq.~(\ref{eq:thetaexp}), the first-order nEDM coefficient $d_n/\bar\theta$ can
in principle be extracted from nucleon correlators in an electric field $\vec\mcE = \mcE_z\hat z$ as 
\begin{equation}
\label{eqn:edm_estimator_sum}
\frac{d_n^G(t_f)}{\bar{\theta}} 
= \frac{1}{\mcE_z} \frac{d}{dt_f} \,
  \lp[\frac {\mathrm{Tr}\big[T^+_{S_z} \la Q_{top} \, N(t_f)\bar{N}(0)\ra_{\mcE_z}\big]}
            {\mathrm{Tr}\big[T^+ \la N(t_f)\bar{N} (0)\ra_{\mcE_z}\big]}\rp] 
  \xRightarrow{t_f\rightarrow\infty}d_n/\bar{\theta}\,,
\end{equation}
where $T^+_{S_z} =
T^+\cdot(1+\Sigma_z)$ is the  spin-$\hat z$ projector with $\Sigma_z=-i\gamma_1\gamma_2$. We use superscript G to denote the nEDM results obtained using global topological charge.
Although the above ratio approaches the nEDM in the large $t_f$ limit, the numerator in
Eq.~(\ref{eqn:edm_estimator_sum}) contains a $t_f$-independent contamination term,
\begin{equation}
\label{eq:nume}
\mathrm{Tr}\big[T^+_{S_z} \la Q_{top} \, N(t_f)\bar{N}(0)\ra_{\mcE_z}\big]=2|Z_N|^2\left(\frac{d_n}{\bar{\theta}}\mcE_z
t_f-\frac{\alpha}{\bar{\theta}}\frac{\kappa}{2m_N^2}\mcE_z\right) \,.
\end{equation}
This contamination term can be removed altogether in an alternative estimator using the $\gamma_4$ nucleon spin
projector,
\begin{equation}
\label{eqn:edm_estimator_sum_g4xy}
\frac{d_n^{'G}(t_f)}{\bar{\theta}} 
= \frac{1}{\mcE_z} \frac{d}{dt_f} \,  
  \frac {\mathrm{Tr}\big[\gamma_4\Sigma_z \la Q_{top} \, N(t_f)\bar{N}(0)\ra_{\mcE_z}\big]}
        {\mathrm{Tr}\big[T^+ \la N(t_f)\bar{N}(0)\ra_{\mcE_z}\big]}\xRightarrow{t_f\rightarrow\infty}d_n/\bar{\theta}\,.
\end{equation}
When the source-sink separation $t_f$ is sufficiently large and excited state contamination is negligible, the EDM
values extracted using Eq.~(\ref{eqn:edm_estimator_sum}) and Eq.~(\ref{eqn:edm_estimator_sum_g4xy}) should be
equivalent.

%%%%%%%%%%%%%%%%%%%%%%%%%%%%%%%%%%%%%%%%%%%%%%%%%%%%%%%%%%%%%%%%
\subsection{EDM in presence of excited states 
  \label{sec:edm_exc}}
So far, we have ignored excited and negative-parity states that can couple to the nucleon in presence of electric field.
This is equivalent to treating the nucleon as a nondeformable point-like particle.
Presence of excited states has two implications. 
First, one has to separate ground-state matrix elements from the excited-state contaminations, which may in principle be
achieved by fitting the time dependence of the EDM estimator~(\ref{eqn:edm_estimator_sum}) to a multi-state model.
Second and more problematic, the point-like nucleon model fails to account for its deformation in the background
electric field and mixing of positive- and negative-parity states, which generates its electric dipole moment~\cite{Baym:2016lyf}.
In fact, it is this deformation (more precisely, its correlation with the nucleon spin) that allows us to extract
$\bar\theta$-induced EDM and thus it must be properly examined.
It produces nucleon polarizability and, as we show below, also contributes to the estimator of the anomalous magnetic
dipole moment~(\ref{eq:ratiokappa}), which hinders its extraction.
Our discussion here is very similar to Ref.~\cite{Baym:2016lyf}, but we discuss it in Euclidean-lattice framework.

First we note that the formulas~(\ref{eqn:edm_estimator_sum},\ref{eqn:edm_estimator_sum_g4xy}) resemble the
``summation'' method of computing ground-state matrix elements.
This relation is made apparent by the Feynman-Hellmann theorem (recently discussed in Ref.~\cite{Bouchard:2016heu} in
the context of lattice QCD).
The two-point nucleon correlation function in the presence $\CPV$ interaction and background electric field in Euclidean
space can be written as
\begin{equation}
C_{2pt,\mcE,\bar\theta} =\la O(t_f) \, O^\dag(0) \, e^{ -i\bar{\theta} Q_{top} -i\int d^4x \mcA^\mu(x)J^\mu_{EM} }\ra_{S_{QCD}}
\end{equation}
where, to be specific, we set the electromagnetic vector potential to $\mcA^\mu=(0,0,0,z\mcE_z)$.
The combined effect of the $\CP$-odd $\bar\theta$-term and parity-violating dipole interaction\footnote{
  Throughout the paper, the lattice electric field $\vec\mcE$ contains an additional factor of electron charge $|e|$
  compared to the standard units, while the electric dipole operator $\mcD_z$ is defined in terms of its fractions
  $Q_{u,d}=2/3$ and $(-1/3)$ and has dimension of length.}
\begin{equation}
\Delta H = i\bar\theta q_{top} + i \mcE_z \mcD_z \,,
\quad \text{with}\quad q_{top} = \int d^3r \, \rho_Q(\vec r) \,,
\quad \mcD_z = \int d^3r \, z \, \rho_{EM}(\vec r) = 
\int d^3r \, z \, \sum_\psi Q_\psi \psi^\dag \psi
\end{equation}
results in the energy shift that can be interpreted as the energy of permanent electric dipole
induced by the $\theta$-term in background electric field,
\begin{equation}
\label{eq:eneshift}
E_N = m_N + i \bar{\theta} (d_n/\bar{\theta})\,(\vec\Sigma \cdot \vec\mcE)
+ \mathcal O(\bar\theta^2,\mathcal \mcE^2)\,,
\end{equation} 
Note that the energy shift is imaginary because the induced EDM is real-valued but the electric field is Wick-rotated to
imaginary value in Minkowski space to make it real-valued on a Euclidean lattice.

On the other hand, according to the Feynman-Hellmann theorem, the energy shift of the $\hat z$ spin-up state can be related to
the matrix element of the interaction, in this case equal to the topological charge density
$q_{top} = \int d^3x \rho_Q(\vec x)$:
\begin{equation}
\label{eq:localq}
\frac{\partial E^\uparrow_N}{\partial \bar{\theta}}\Big{|}_{\bar{\theta}=0} 
= i \la \mathcal{N}^{L,\uparrow}_0|q_{top}(0)|\mathcal{N}^{R,\uparrow}_0\ra\Big|_{\mcE_z}\,, 
\quad\text{ where }\quad 
q_{top}(0) = \frac1{32\pi^2} \sum_{\vec y} G^a_{\mu\nu}(0,\vec{y}) \, {\widetilde G^a_{\mu\nu}(0,\vec{y})}
\end{equation}
We use $\la \mathcal{N}^{L,\uparrow}_0|$ and $|\mathcal{N}^{R,\uparrow}_0\ra$ to denote bra- and
ket-vectors of the nucleon ground state \emph{in presence of Euclidean-space background electric
field}, which introduces anti-Hermitian perturbation to the QCD Hamiltonian (see below).  
Note that in absence of the electric field, the nucleon matrix element 
$\la N_0 | q_{top}(0)| N_0\ra=0$ because of spatial symmetry.
Therefore, the $\bar\theta$-induced nEDM can be alternatively calculated from the matrix element of
the local topological charge density as
\begin{equation}
\label{eqn:edm_bgem_plateau}
d_n/\bar{\theta} = \frac{1}{\mcE_z} \la \mathcal{N}_0^{L,\uparrow} |q_{top}(0)| \mathcal{N}_0^{R,\uparrow}\ra_{\mcE_z} 
\approx \frac{1}{\mcE_z} 
  \frac{ \la O^{L,\uparrow}_{\mathcal{N}_0}(t_f) \, q_{top}(\tau) \,  O^{\dag R,\uparrow}_{\mathcal{N}_0}(0)\ra_{\mcE_z}}
       { \la O^{L,\uparrow}_{\mathcal{N}_0}(t_f) \, O^{\dag
R,\uparrow}_{\mathcal{N}_0}(0)\ra_{\mcE_z}}\Bigg|_{t_f\rightarrow\infty,
t_f-\tau\rightarrow\infty}\,,
\end{equation}
where $O^L_{\mathcal{N}_0}(0)$ and $O^R_{\mathcal{N}_0}(0)$ are operators optimized to
project directly onto the ground nucleon state in the presence of a background field. 
This expression is sensitive to using correct ground-state operators creating mixed-parity states, and below in
Sec.~\ref{sec:gevp_bgem} we show how these operators are obtained using the generalized eigenvalue problem (GEVP)
method.
Equation~(\ref{eqn:edm_bgem_plateau}) illuminates how the problem of computing EDM is reduced from computing a 4-point
correlation function of the nucleon fields $N,\,\bar N$, the global topological charge $Q=\int dt \, q_{top}(t)$, and
the vector current $J_\mu=\bar \psi\gamma_\mu \psi$ to a correlator only of $N,\,\bar N$ in uniform electric field
and single-timeslice topological charge $q_{top}(x)$ 
\begin{equation}
\label{eqn:edm_bgem_3pt}
C_3(t_f,\tau) =   \la O^{L,\uparrow}_{\mcN_0}(t_f) \, q_{top}(\tau) 
  O^{\dag R,\uparrow}_{\mcN_0}(0)\ra_{\mcE_z} \,.
\end{equation}
This point is crucial to reducing stochastic fluctuations in ``disconnected'' correlators involving topological charge,
as well as Weinberg 3-gluon interaction, isoscalar 2-quark and 4-quark interactions, and similar.
In this paper, we concentrate on the topological charge but the methodology can be readily extended to these other
$\CPV$ interactions.

Now we consider breaking of spatial parity symmetry by the background electric field, with which the nucleon and vacuum
eigenstates no longer have definite parity because of the dipole interaction 
$\Delta H = i\mcE_z \mcD_z$.
To the leading order in the electric field, the two lowest states $|N\ra$ and $|N^*\ra$ of opposite parity become
\begin{equation}
\label{eq:Npm}
\begin{aligned}
|\mcN^R_0\ra &= |N_0\ra
+ i\mcE_z \sum_i \frac{\la N^*_i|\mcD_z|N_0\ra}{m_{N_0} - m_{N^*_i}} |N^*_i\ra + O(\mcE_z^2) \,,\\
|\mcN^R_1\ra &= |N^*_0\ra 
+ i\mcE_z \sum_i \frac{\la N_i|\mcD_z|N^*_0\ra}{m_{N^*_0} - m_{N_i}} |N_i\ra + O(\mcE_z^2) \,,
\end{aligned}
\end{equation}
where $|N_i\ra$ and $|N^*_i\ra$, $i=0,1,\ldots$ are towers of positive- and negative-parity
states with spin $1/2$ at $\vec\mcE=0$, $\bar\theta=0$, respectively.
\footnote{
  The dipole operator can also mix spin-$\frac12$ and spin-$\frac32$
  states, the latter one can be neglected since the topological charge forbids the mixing between states with different spin, assuming good rotation symmetry on lattice.}.
The $R$ subscript indicates that these are the right-side eigenvectors of the full
Hamiltonian.
They are not orthogonal to each other,
\begin{equation}
\la\mcN^R_0|\mcN^R_1\ra = 2i\mcE_z \frac{\la N_0|\mcD_z|N^*_0\ra}{m_{N^*_0} - m_{N_0}} + O(\mcE_z^2) \ne 0\,,
\end{equation}
because the perturbation is not Hermitian (see App.~\ref{sec:app_antiherm} for detail).
Instead, they are bi-orthogonal to the left-side eigenvectors
\begin{equation}
\label{eq:Npm_L}
\begin{aligned}
\la\mcN^L_0| &= \la N_0| 
+ i\mcE_z \sum_i \frac{\la N_0|\mcD_z|N^*_i\ra}{m_{N_0} - m_{N^*_i}} \la N^*_i| + O(\mcE_z^2)\,, \\
\la\mcN^L_1| &= \la N^*_0| 
+ i\mcE_z \sum_i \frac{\la N^*_0|\mcD_z|N_i\ra}{m_{N^*_0} - m_{N_i}} \la N_i| + O(\mcE_z^2) \,,
\end{aligned}
\end{equation}
It is straightforward to check that $\la\mcN^L_0|\mcN^R_1\ra = O(\mcE_z^2)$.
The transition matrix elements $\la N_0|\mcD_z|N^*_i\ra$ contribute both to the nucleon electric
polarizability\footnote{
  This definition corresponds to $\alpha_E\approx1.26\cdot10^{-3}\,\mathrm{fm}^3$~\cite{ParticleDataGroup:2024cfk}, 
  equivalent to $d(E)\approx0.86\,\mathrm{e}\cdot\mathrm{cm} \cdot\frac{E}{10^4\mathrm{V/cm}}$~\cite{Baym:2016lyf} in
  more common units.}
\begin{equation}
\label{eqn:nucl_polariz_pert2}
4\pi\alpha_E = 2e^2 \sum_i \frac{|\la N^*_i|\mcD_z|N_0\ra|^2}{m_{N^*_i} - m_{N_0}} > 0 \,,
\end{equation}
as well as permanent EDM induced if $\CP$-violating interaction is present.
To the first order in $\bar\theta$ and $\mcE_z$, the EDM defined in Eq.~(\ref{eqn:edm_bgem_plateau}) can be expressed
using perturbative expansion~\cite{Baluni:1978rf, Baym:2016lyf},
\begin{equation}\label{eq:nEDMdef}
d_n/\bar{\theta} 
= -i \sum_i
  \frac{  \la N_0|q_{top}|N^*_i\ra \la N^*_i|\mcD_z |N_0\ra + 
          \la N_0|\mcD_z |N^*_i\ra \la N^*_i|q_{top}|N_0\ra}
       { m_{N^*_i}-m_{N_0} }
= 2 \sum_i 
  \frac{ \mathrm{Im}\lp[\la N_0|q_{top}|N^*_i\ra \la N^*_i|\mcD_z |N_0\ra\rp]}
       { m_{N^*_i}-m_{N_0}}
\end{equation}
where the resulting value is real because the Euclidean topological charge density, 
as evidenced by its expression in terms of Euclidean color-electric and -magnetic fields 
$q_{top} = \frac1{8\pi^2} \int d^3x \, (\vec\mcE^a\cdot\vec H^a)$, 
acts as an \emph{anti-Hermitian} operator\footnote{
  The above equation is consistent with Eq.~(19) in Ref.~\cite{Baluni:1978rf} with $q_{top,E}=iq_{top,M}$ following
  Eq.~(\ref{eq:tolog}).}

To connect the above equation to the discussion in Sec.~\ref{sec:edm_point}, one can set the energy gaps $(m_{N^*_i}-m_N)$ very large so that the corresponding excited-state contributions vanish. 
However, it is our understanding that the virtual-pair states $|N\,\bar N N\ra,\,|N\,\bar N N\,\bar N N\ra,\, \cdots$ could appear  
appear as intrinsic contributions to the neutron EDM at much higher energy scale.

%%%%%%%%%%%%%%%%%%%%%%%%%%%%%%%%%%%%%%%%%%%%%%%%%%%%%%%%%%%%%%%%
\subsection{Effect on magnetic moment estimator in background electric field}
\label{sec:edm_mag}
Now consider the estimator~(\ref{eq:ratiokappa}), which depends on the relativistic nucleon correlator.
In this section, we show that this estimator is biased by contributions from nucleon and vacuum polarizabilities due to
mixing with opposite-parity excited states.
In background electric field $\mcE_z \hat z$ and without $\CP$ violation, the nucleon two-point correlation function is
diagonal in $S_z$ spin and can be expressed in the Dirac $\gamma$-matrix basis~(\ref{eqn:gammamatr_dirac}) as
\begin{equation}\label{eq:oobar}
\begin{aligned}
\la O(t_f) O^\dag(0)\ra_{\mcE_z} 
  &= C_{2pt,\mcE}(t_f)\gamma_4
\\&= \left(\begin{array}{cccc} 
  \la\psi^\uparrow_N(t_f) \psi^{\uparrow,\dag}_{N}(0)\ra_{\mcE_z} & 0 & \la\psi^\uparrow_N (t_f)\psi^{\uparrow,\dag}_{N^*}(0)\ra_{\mcE_z} & 0 \\
  0 & \la\psi^\downarrow_N(t_f) \psi^{\downarrow,\dag}_{N}(0)\ra_{\mcE_z} & 0 & \la\psi^\downarrow_N(t_f) \psi^{\downarrow,\dag}_{N^*}(0)\ra_{\mcE_z}\\
  \la\psi^\uparrow_{N^*}(t_f) \psi^{\uparrow,\dag}_{N}(0)\ra_{\mcE_z} & 0 & \la\psi^\uparrow_{N^*}(t_f) \psi^{\uparrow,\dag}_{N^*}(0)\ra_{\mcE_z} & 0\\
  0 & \la\psi^\downarrow_{N^*}(t_f) \psi^{\downarrow,\dag}_{N}(0)\ra_{\mcE_z} & 0 & \la\psi^\downarrow_{N^*}(t_f) \psi^{\downarrow,\dag}_{N^*}(0)\ra_{\mcE_z}
\end{array}\right)
\end{aligned}\end{equation}
where, for example, $\psi^{\uparrow,\downarrow}_N=\frac14(1+\gamma_4)(1\pm\Sigma_z)O$ and 
$\psi^{\uparrow,\downarrow}_{N^*}=\frac14(1-\gamma_4)(1\pm\Sigma_z)O$ 
denote the $S_z$-polarized positive- and negative-parity components of the nucleon interpolating operator $O$.
In contrast to the case without the background electric field, the parity off-diagonal correlators $\la\psi_N(t_f)
\psi^{\dag}_{N^*}(0)\ra_{\mcE_z}$ and $\la\psi_{N^*}(t_f) \psi^{\dag}_{N}(0)\ra_{\mcE_z}$ are nonzero, and they are related in such a way that 
$\la O(t_f) O^\dag(0)\ra^\dag_{\mcE_z}
= \la O(0) O^\dag(t_f)\ra_{\mcE_z}
= \gamma_4\la O(t_f) O^\dag(0)\ra_{\mcE_z} \gamma_4$, which is an important symmetry (see discussion following Eq.~(\ref{eq:nonHGEVP}), showing that our system is analogous to so-called PT-symmetric quantum-mechanical
systems~\cite{Bender:2023cem}. 
This observation is also consistent with the fact that the electric dipole interaction $\Delta_H=i\mcE_z\mcD_z$ satisfies the PT-symmetry condition $PT \Delta_H(PT)^{-1}= \Delta_H$. We further note that the corresponding interaction induced by an external magnetic field also satisfies PT symmetry.

To express all the nonzero elements shown in Eq.~(\ref{eq:oobar}),
first introduce their overlaps with zero-field, definite-parity states\footnote{These overlaps are spin-independent in
the $\CP$-even theory.}
\begin{equation}
Z_{N_i} = \la \Omega|\psi_N^\uparrow|N_i^\uparrow\ra\,,
\quad Z_{N^*_i} = \la \Omega|\psi_{N^*}^\uparrow|N^{*\uparrow}_i\ra
\end{equation}
where $|\Omega\ra$ denotes the parity even vacuum state without electric field.
However, when the field is present, the vacuum state is polarized and, similar to Eqs.~(\ref{eq:Npm},\ref{eq:Npm_L}), to
the first order in electric field
\begin{equation}
\begin{aligned}
|\Omega^R\ra &= |\Omega\ra
+ i\mcE_z \sum_{n} \frac{\la n^-|\mcD_z|\Omega\ra}{ - E_{n}} |n^-\ra + O(\mcE_z^2) ,\\
\la\Omega^L| &= \la \Omega| 
+ i\mcE_z \sum_{n} \frac{\la \Omega|\mcD_z|n^-\ra}{ - E_{n}} \la n^-| + O(\mcE_z^2)\,,
\end{aligned}
\end{equation}
where $|n^-\ra$ are parity-odd states with the same quantum numbers as the dipole operator $\mcD_z$, for example, 
$\hat z$-polarized $\rho$ and $\omega$ mesons.
Nucleon operators acting on such states produce polarized nucleons,
and then to the first order in $\mcE_z$, their overlaps with ket-states~(\ref{eq:Npm}) are\footnote{
  The nucleon field and state spins, where omitted, are presumed $(+1/2)$. 
  Formulas for the spin $(-1/2)$ case are analogous and would differ only due to Wigner-Eckart theorem, 
  for example, in $\protect\langle N^{*\protect\uparrow}|D_z|N^\protect\uparrow\protect\rangle
= -\protect\langle N^{*\protect\downarrow}|D_z|N^\protect\downarrow\protect\rangle$.}
\begin{equation}
\label{eq:proj}
\begin{aligned}
\la\Omega^L|\psi_{N}|\mcN^R_0\ra_{\mcE_z} 
  &= Z_{N_0} +O(\mcE_z^2) \,,\\
\la\Omega^L|\psi_{N^*}|\mcN^R_0\ra_{\mcE_z} 
  &=i\mcE_z \sum_i Z_{N^*_i} \frac{\la N^*_i|\mcD_z|N_0\ra}{m_{N_0}-m_{N^*_i}} -i\mcE_z \sum_{n} \frac{\la \Omega|\mcD_z|n^-\ra}{  E_{n}} \la n^-|\psi_{N^*}|N_0\ra + O(\mcE_z^2),\\
\la\Omega^L|\psi_{N}|\mcN^R_1\ra_{\mcE_z} 
  &=i\mcE_z \sum_i Z_{N_i} \frac{\la N_i|\mcD_z|N^*_0\ra}{m_{N^*_0}-m_{N_i}} -i\mcE_z \sum_{n} \frac{\la \Omega|\mcD_z|n^-\ra}{  E_{n}} \la n^-|\psi_{N}|N^*_0\ra + O(\mcE_z^2) \,,\\
\la\Omega^L|\psi_{N^*}|\mcN^R_1\ra_{\mcE_z} 
  &= Z_{N^*_0} +O(\mcE_z^2) \,,
\end{aligned}
\end{equation}
where matrix elements $\la n^-|\psi_{N}|N_0\ra=\la n^-|\psi_{N^*}|N^*_0\ra=0$ due to parity.
To shorten notation, we will also introduce matrix elements
\begin{equation}
Z^{n^-}_{N^*_i} = \la n^- | \psi_N^\uparrow |N^{*\uparrow}_i\ra\,,
\quad Z^{n^-}_{N_i} = \la n^- | \psi_{N^*}^\uparrow | N_i^\uparrow\ra\,.
\end{equation}

Using analogous expressions for the bra-state overlaps $\la \mcN^L_{0,1}|\psi^\dag|\Omega^R\ra_{\mcE_z}$ and taking into
account only two lowest states, the (mostly-positive parity) ground $\mcN_0$ and the (mostly-negative parity) first
excited $\mcN_1$, we can find the spin- and parity- projected components of the correlator matrix~(\ref{eq:oobar})
\begin{equation}
\la\psi(t_f) \psi^\dag(0)\ra_{\mcE_z} 
  \approx \la \Omega^L|\psi|\mcN^R_0\ra_{\mcE_z}\la \mcN^L_0|\psi^\dag|\Omega^R\ra_{\mcE_z} e^{-M_{\mcN_0}t_f} 
  + \la \Omega^L|\psi|\mcN^R_1\ra_{\mcE_z}\la \mcN^L_1|\psi^\dag|\Omega^R\ra_{\mcE_z} e^{-M_{\mcN_1}t_f} 
\end{equation}
the entries in Eq.~(\ref{eq:oobar}) can be expressed as the spectral decompositions
\begin{equation}
\label{eq:NNstar}
\begin{aligned}
\la\psi_N^\uparrow(t_f) \psi_N^{\uparrow,\dag}(0)\ra_{\mcE_z} 
&= |Z_{N_0}|^2 e^{-M_{\mcN_0}t_f} + O(\mcE_z^2)\,,
\\
\la \psi_N^\uparrow(t_f)\psi_{N^*}^{\uparrow,\dag}(0)\ra_{\mcE_z} 
&= i\mcE_z \Bigg[Z_{N_0}
    \lp( \sum_i Z^\dag_{N^*_i}\frac{\la N_0|\mcD_z|N^*_i\ra}{m_{N_0}-m_{N^*_i}}  
        -\sum_n Z^{n^- \dag}_{N_0}
                \frac{\la n^-|\mcD_z|\Omega\ra}{ E_{n}} 
    \rp) e^{-m_{\mcN_0}t_f} 
\\
&\phantom{= i\mcE_z} + Z^\dag_{N^*_0} 
    \lp( \sum_i Z_{N_i}\frac{\la N_i|\mcD_z|N^*_0\ra}{m_{N^*_0}-m_{N_i}}
        -\sum_n Z^{n^-}_{N^*_0}
                \frac{\la \Omega|\mcD_z|n^-\ra}{  E_{n}} 
    \rp) e^{-m_{\mcN_1}t_f} 
  \Bigg] \,,
\\
\la \psi_{N^*}^\uparrow(t_f)\psi_{N}^{\uparrow,\dag}(0)\ra_{\mcE_z} 
&= i\mcE_z \Bigg[ Z^\dag_{N_0} 
    \lp( \sum_i Z_{N^*_i}\frac{\la N^*_i|\mcD_z|N_0\ra}{m_{N_0}-m_{N^*_i}} 
        -\sum_n Z^{n-}_{N_0}
                \frac{\la \Omega|\mcD_z|n^-\ra}{  E_{n}} 
    \rp) e^{-m_{\mcN_0}t_f} 
\\&\phantom{= i\mcE_z \Bigg[} +Z_{N^*_0} 
    \lp( \sum_i Z^\dag_{N_i}\frac{\la N^*_0|\mcD_z|N_i\ra}{m_{N^*_0}-m_{N_i}}
        -\sum_n Z^{n^- \dag}_{N^*_0} 
                \frac{\la n^-|\mcD_z|\Omega\ra}{ E_{n}}  
    \rp) e^{-m_{\mcN_1}t_f}
\Bigg]\, \,,
\\
\la\psi_{N^*}^\uparrow(t_f) \psi_{N^*}^{\uparrow,\dag}(0)\ra_{\mcE_z}  
&= |Z_{N^*_0}|^2 e^{-m_{\mcN_1}t_f} + O(\mcE_z^2)\,.
\end{aligned}
\end{equation}
Note that the parity-mixed correlation between $\psi_N$ and $\psi_{N^*}$ is anti-Hermitian, 
$\la \psi_N^\uparrow(t_f)\psi_{N^*}^{\uparrow,\dag}(0)\ra_{\mcE_z} = - \la \psi_{N^*}^\uparrow(t_f)\psi_{N}^{\uparrow,\dag}(0)\ra^\dag_{\mcE_z}$, 
and these formulas apply to both spin states.
Combining Eqs.~(\ref{eq:oobar},\ref{eq:NNstar}), we obtain the following expression for the estimator of 
the nucleon anomalous magnetic moment~(\ref{eq:ratiokappa}) in large $t_f$ limit, including nucleon excited-state contributions,
\begin{equation}
\begin{aligned}
\label{eq:kappa_est_bias}
\kappa^\text{est}(t_f)
  &= 2\frac{m_N^2}{\mcE_z}\frac{\la \psi_N^\uparrow(t_f)\psi_{N^*}^{\uparrow,\dag}(0)\ra_{\mcE_z}-\la \psi_{N^*}^\uparrow(t_f)\psi_{N}^{\uparrow,\dag}(0)\ra_{\mcE_z}}{\la \psi_N^\uparrow(t_f)\psi_{N}^{\uparrow,\dag}(0)\ra_{\mcE_z}}
\\&= 2m_N^2\mathrm{Im}\lp[ \sum_i \frac{Z_{N^*_i}}{Z_{N_0}}\frac{\la N^*_i | \mcD_z|N_0\ra}{m_{N_0}-m_{N^*_i}}
                          -\sum_n \frac{Z^{n^-}_{N_0}}{Z_{N_0}}\frac{\la \Omega | \mcD_z|n^-\ra}{E_n} \rp].
\end{aligned}
\end{equation}
Note that only the spin up-components are included in the above equation, since the corresponding spin-down matrix
elements differ by an overall minus sign according to the Wigner–Eckart theorem, i.e, $\la
N_i^{*,\uparrow}|D_z|N_0^\uparrow\ra=-\la N_i^{*,\downarrow}|D_z|N_0^\downarrow\ra$, $\la n^- | \psi^\uparrow_{N^*} |
N^\uparrow_i\ra =- \la n^- | \psi^\downarrow_{N^*} | N^\downarrow_i\ra$.
These equations clearly show that $\kappa^\text{est}(t_f)$ does not converge to the neutron magnetic moment
$\kappa$ even in the large-$t_f$ limit because of the parity-mixing contributions of the nucleon ($N\leftrightarrow
N_i^*$) and vacuum ($\Omega\leftrightarrow n^-$) polarizability induced by the background electric field.
Furthermore, this bias depends on the nucleon interpolating operators used in calculations, and is not straightforward
to remove.
Our estimates in Appendix~\ref{sec:appenx_polariz} indicate that $\kappa^\text{est}$ may be comparable or larger in
magnitude than the nucleon anomalous magnetic moment.
Computing the excited-states contribution accurately requires knowledge of the overlap factors $Z_{N^*_i}$.
To calculate $\kappa$ using the estimator~(\ref{eq:ratiokappa}) introduced in Ref.~\cite{Detmold:2010ts}, one has to use
``perfect'' nucleon ground-state operators with all $Z_{N^*_i}=0$, which is difficult in practice, and may still be
subject to the effects of vacuum polarizability.

As we have shown in Sec.~(\ref{sec:edm_exc}), nEDM in our method is related to the nucleon energy shift and is well defined using the nucleon eigenstate in the presence of the background field. Unlike the magnetic momentum, generating a non-vanishing nEDM requires the transition matrix element between positive parity $N$ and its negative parity partner $N^*$, as shown in Eq.~(\ref{eq:nEDMdef}). This implies that the nEDM arises from the interplay between CP even electric dipole operator and the CP odd topological charge operator.

%%%%%%%%%%%%%%%%%%%%%%%%%%%%%%%%%%%%%%%%%%%%%%%%%%%%%%%%%%%%%%%%
\subsection{Generalized eigenvalue method in presence of electric field
  \label{sec:gevp_bgem}}
To isolate the nucleon eigenstates in the presence of a background field, we  solve the GEVP using the following correlation matrix 
\begin{equation}\label{eq:2ptppbar}
C^{\psi\psi^\dag,\uparrow}_{2pt,\mcE}(t_f)
=\left(\begin{array}{cc} \la\psi^\uparrow_N(t_f) \psi^{\uparrow,\dag}_{N}(0)\ra_{\mcE_z}  & \la\psi^\uparrow_N (t_f)\psi^{\uparrow,\dag}_{N^*}(0)\ra_{\mcE_z}  \\
\la\psi^\uparrow_{N^*}(t_f) \psi^{\uparrow,\dag}_{N}(0)\ra_{\mcE_z} &  \la\psi^\uparrow_{N^*}(t_f) \psi^{\uparrow,\dag}_{N^*}(0)\ra_{\mcE_z}\end{array}\right)
\end{equation}
We retain only the spin-up components for simplicity, the analysis for the spin-down state is similar. 
As discussed in sec~\ref{sec:edm_mag}, the off-diagonal part of the matrix shown in Eq.~(\ref{eq:2ptppbar}) 
is anti-hermitian.
For such a system, one can define the right generalized eigenstate and left generalized eigenstate through 
\begin{equation}
\label{eq:nonHGEVP}
\begin{aligned}
C^{\psi\psi^\dag,\uparrow}_{2pt,\mcE}(t_f)v_{R,n} 
  &= \lambda_n C^{\psi\psi^\dag,\uparrow}_{2pt,\mcE}(t_0)v_{R,n} \,,
\\
v_{L,n}^{*,T}C^{\psi\psi^\dag,\uparrow}_{2pt,\mcE}(t_f) 
  &= \lambda_n v_{L,n}^{*,T}C^{\psi\psi^\dag,\uparrow}_{2pt,\mcE}(t_0)
\quad\Leftrightarrow\quad
[C^{\psi\psi^\dag,\uparrow}_{2pt,\mcE}(t_f)]^\dag v_{L,n} 
  &= \lambda_n^* [C^{\psi\psi^\dag,\uparrow}_{2pt,\mcE}(t_0)]^\dag v_{L,n}\,,
\end{aligned}
\end{equation}
Furthermore,
$C^{\psi\psi^\dag,\uparrow}_{2pt,\mcE}(t_f)$  satisfies
\begin{equation}
\label{eq:psuedo hermiticity}
[C^{\psi\psi^\dag,\uparrow}_{2pt,\mcE}(t_f)]^\dag=S\,C^{\psi\psi^\dag,\uparrow}_{2pt,\mcE}(t_f)\,S^{-1}, 
~~~~~~
S=\mathrm{diag}\{1,-1\}
\end{equation}

i.e., it is similar to its Hermitian conjugate, or $S-$pseudo-Hermitian.
It implies that the eigenvalues $\lambda_n$ are either real or appear in complex-conjugate pairs (see App.~\ref{sec:app_antiherm}). For the real generalized eigenvalues observed in our analysis, the left eigenvector can be chosen as $v_{L,n}=S v_{R,n}$. If a complex-conjugate eigenvalue pair appears, $S$ maps the right eigenvector of one member of the pair to the left eigenvector of the conjugate partner.

This relation is consistent with the definition of left and right eigenstates shown in Eq.~(\ref{eq:Npm}), in which
$|\mcN^R_0\ra$ and $\la \mcN^L_0|$ differ by a sign in the term involving the operator $\mcD_z$. This sign difference arises because the matrix element $\la N^*_i|\mcD_z|N_0\ra$ is purely imaginary, as we shown in Appendix~\ref{sec:appenx_tmax}. An analogous argument applies to $|\mcN^R_1\ra$ and $\la \mcN^L_1|$.

Since the correlation matrix is not hermitian, we solve this non-Hermitian generalized eigenvalue problem by diagonalizing
\begin{equation}
[C^{\psi\psi^\dag,\uparrow}_{2pt,\mcE}(t_0)]^{-1} \, C^{\psi\psi^\dag,\uparrow}_{2pt,\mcE}(t_f) \, V_R
= V_R \, \Lambda\,,
\end{equation}
where the columns of $V_R$ are eigenvectors $v_{R,n}$ and $\Lambda=\mathrm{diag}\{\lambda_n\}$.
We observe in practice that the eigenvalues $\lambda_n$ are real and correspond to the exponential decay rates for these eigenstates in the correlation matrix.
This matches the observation that the PT symmetric non-Hermitian Hamiltonian (or $S-$pseudo-Hermitian Hamiltonian) 
would have real eigenvalues~\cite{Bender:2023cem} (see
also App.~\ref{sec:app_antiherm}) as long as the anti-Hermitian component remains smaller than the gaps between the
energy levels of the hermitian component, so that the former can be treated as perturbation to the latter\footnote{
  If the anti-Hermitian part becomes nonperturbative, i.e., larger than the energy gaps between unperturbed coupled
  states, then these two states ``merge'' into a pair of states with complex-conjugate eigenvalues, see
  App.~\ref{sec:app_antiherm}.
}.
In our case, the anti-Hermitian neutron electric dipole interaction with the background field has to remain small
compared to the mass gaps between the ground and excited states of the nucleon and its parity partner.
Our data suggest that it is indeed the case, because we find no significant disagreements from expected perturbative
behavior with $\mcE_z$ in all observables that we calculate.
Also, our estimates in Appendix~\ref{sec:appenx_polariz} indicate that this anti-Hermitian interaction has matrix
elements $\lesssim 100\,\mathrm{MeV})$, well below the mass gaps between the nucleon and its parity partners, as well as any $N\pi$ states.

The ground-state nEDM can be obtained using the following ratio 
\begin{equation}
\label{eqn:edm_bgem_GEVP}
d_n/\bar{\theta} = \frac{1}{\mcE_z} \la \mathcal{N}_0^{L,\uparrow} |q_{top}(0)| \mathcal{N}_0^{R,\uparrow}\ra_{\mcE_z}  
=\frac{1}{\mcE_z} 
  \frac{ v^\dag_{L,0}C^{\psi\psi^\dag,\uparrow}_{3pt,\mcE}(t_f,\tau)v_{R,0}}       {v^\dag_{L,0}C^{\psi\psi^\dag,\uparrow}_{2pt,\mcE}(t_f)v_{R,0}}\Bigg|_{t_f\rightarrow\infty,t_f-\tau\rightarrow\infty}\,,
\end{equation}
where $v_{L,0}$ and $v_{R,0}$ are the generalized eigenvectors of the ground state. 
The three-point correlation matrix $C^{\psi\psi^\dag,\uparrow}_{3pt,\mcE}(t_f,\tau)$ is defined as
\begin{equation}
\label{eq:3ptoriQtop}
C^{\psi\psi^\dag,\uparrow}_{3pt,\mcE}(t_f,\tau)
=\left(\begin{array}{cc} 
    \la\psi^\uparrow_N(t_f) q_{top}(\tau)\psi^{\uparrow,\dag}_{N}(0)\ra_{\mcE_z}  
  & \la\psi^\uparrow_N (t_f)q_{top}(\tau)\psi^{\uparrow,\dag}_{N^*}(0)\ra_{\mcE_z}  
\\  \la\psi^\uparrow_{N^*}(t_f)q_{top}(\tau)\psi^{\uparrow,\dag}_{N}(0)\ra_{\mcE_z} 
  &  \la\psi^\uparrow_{N^*}(t_f) q_{top}(\tau)\psi^{\uparrow,\dag}_{N^*}(0)\ra_{\mcE_z}
\end{array}\right)\,.
\end{equation}

%%%%%%%%%%%%%%%%%%%%%%%%%%%%%%%%%%%%%%%%%%%%%%%%%%%%%%%%%%%%%%%%%%%%%%%%%%%%%%%%
%%%%%%%%%%%%%%%%%%%%%%%%%%%%%%%%%%%%%%%%%%%%%%%%%%%%%%%%%%%%%%%%%%%%%%%%%%%%%%%%
\section{\label{sec:comp}Computational details}

%%%%%%%%%%%%%%%%%%%%%%%%%%%%%%%%%%%%%%%%%%%%%%%%%%%%%%%%%%%%%%%%
In this section, we present the details of the lattice computation, including the implementation of the background
electric field, the definition of the topological charge, and the all mode averaging (AMA) and low mode averaging (LMA)
techniques that are employed to improve the signal. 

We study the nEDM on three gauge ensembles generated with 2+1 flavors of domain wall fermions (DWF) and the Iwasaki gauge action ($\beta=2.13$), generated by the RBC/UKQCD collaborations~\cite{Allton:2008pn}. 
The 4d lattice size is $L^3\times T=24^3\times64$ and has inverse lattice spacing $a^{-1}=1.785$ GeV ($a=0.1105$ fm). 
The bare light quark masses are 0.005, 0.01, and 0.02 in lattice units and the corresponding pion masses are
approximately 340, 420, and 576 MeV, respectively. 
The extra 5th dimension has size $L_s=16$ with domain wall height $M_5=1.8$. 
We use more than 1000 configurations from each of the two ensembles with lighter quark masses, and 523 configurations for the ensemble with the heavier mass. 
The ensemble details are summarized in Tab.~\ref{Tab:ensembles}.
%------------------------------------------------------------------------------
\begin{table}[ht!]
\caption{\label{Tab:ensembles}
  Gauge ensembles parameters. 
  The pion mass $m_\pi$ corresponds to the $u,d$ sea quark mass parameter, residue mass $m_{res}$
  and $N_\text{cfg}$ is the number of configurations used in the calculation. We perform one exact (ex) inversion and 64 sloppy (sl) inversions per configuration for the AMA correction described in Sec.~\ref{sec:LMA}.
}
\begin{ruledtabular}
\begin{tabular}{c c c c c c c c}
\text{ensemble} & $L^3 \times T$   &  $a$ (fm)   & $m_\pi$ (MeV) & $m_l/m_s$  & $m_{res}$ &  $N_\text{cfg}$ & ex/sl\\
\hline
24I-005             & $24^3 \times 64$ & 0.1105(3)   & 340       & 0.005/0.04  &0.00308  & 1400  & 1/64\\
24I-010             & $24^3 \times 64$ & 0.1105(3)   & 420       & 0.01/0.04   &0.00308 & 1100   & 1/64 \\
24I-020             & $24^3 \times 64$ & 0.1105(3)   & 576       & 0.02/0.04   &0.00308 & 523  & 1/64
\end{tabular}
\end{ruledtabular}
\end{table}

\begin{figure}[ht!]
\centering
\includegraphics[width=.5\textwidth]{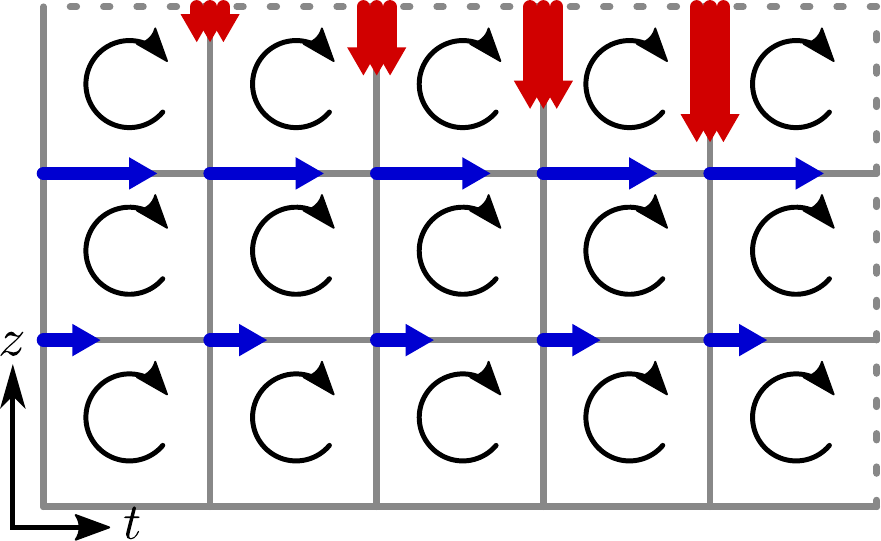}
\caption{Constant background electric field on a periodic lattice, following
  Ref.~\cite{Detmold:2009dx}.
  \label{fig:bgelectric}}
\end{figure}

The electric field is introduced following the method in Ref.~\cite{Detmold:2009dx}, which preserves the (anti)periodic
boundary condition in Euclidean time.
To achieve this, the electric field is analytically continued to an imaginary value.
If the particle's electric dipole moment is finite and real-valued, the energy shift will be imaginary, which might
present a problem in the analysis of the corresponding lattice correlators.
However, in our methodology, the $\CP$-odd interaction ($\bar\theta$-term) is an infinitesimal perturbation. 
Accordingly, the induced electric dipole moments and the corresponding energy shifts are determined from the first-order
Taylor expansion of the nucleon correlation functions with $\CP$-odd interactions (see above).
In this paper, we study only neutral particles; analysis of charged particle propagators
is more complicated~\cite{Detmold:2010ts}.
The rest energy $E_0$ is modified from the mass $m$ due to electric and magnetic dipole
interactions.
To avoid any confusion, we imply no relation between the Minkowski $\vec E,\vec H$ and Euclidean
$\vec\mcE,\vec\mcH$ electric and magnetic fields. 
Instead, we introduce \emph{ad hoc} uniform Abelian fields on a lattice (see Fig.~\ref{fig:bgelectric}) preserving
boundary conditions in both space and time~\cite{Detmold:2009dx} that probe the MDM and EDM: 
the magnetic field is $\epsilon^{ijk}\mcH^k = (\partial_i \mcA_j - \partial_j \mcA_i) = n_{ij} \Phi_{ij}$
(no summation over $i,j$) with vector potential
\begin{equation}
\label{eqn:euc_const_bgmag}
\left\{\begin{array}{ll}
  \mcA_{j}(x) &= n_{ij}\,\Phi_{ij} \, x_i \\
  \mcA_{i}(x)|_{x_i=L_i-1} &= -n_{ij}\,\Phi_{ij}\, L_i x_j 
\end{array}\right.
\end{equation}
and the electric field is $\mcE^k = (\partial_k \mcA_4 - \partial_4 \mcA_k) = n_{k4} \Phi_{k4}$ with
\begin{equation}
\label{eqn:euc_const_bgel}
\left\{\begin{array}{ll}
  \mcA_{4}(x) &= n_{k4}\,\Phi_{k4} \, x_k \\
  \mcA_{k}(x)|_{x_k=L_k-1} &= -n_{k4}\,\Phi_{k4}\, L_k x_4
\end{array}\right.\,,
\end{equation}
where $\Phi_{\mu\nu} = \frac{6\pi}{L_\mu L_\nu}$ is the quantum of flux through a plaquette $(\mu\nu)$ and $n_{\mu\nu}$
is the corresponding number of quanta.
The fractional quark charges $Q_u=2/3$, $Q_d=-1/3$ and periodic boundary conditions require that the flux through the
edge of the lattice is $L_\mu L_\nu \cdot \Phi_{\mu\nu} = 3\cdot2\pi$.
From potentials~(\ref{eqn:euc_const_bgmag},\ref{eqn:euc_const_bgel}), the Euclidean field strength tensor is
\begin{equation}
\label{eqn:fmunu_euc_lat}
\mcF_{\mu\nu} 
= \left(\begin{array}{r|rrrr}
  & 1 & 2 & 3 & 4\\
\hline
    1 &        0  &   \mcH^3  &  -\mcH^2  & \mcE^1 \\
    2 &  -\mcH^3  &        0  &   \mcH^1  & \mcE^2 \\
    3 &   \mcH^2  &  -\mcH^1  &        0  & \mcE^3 \\
    4 &  -\mcE^1  &  -\mcE^2  &  -\mcE^3  &      0 
\end{array}\right)
\end{equation}
with
$\vec\mcH=(n_{23}\Phi_{23}, n_{31}\Phi_{31}, n_{12}\Phi_{12})$ and
$\vec\mcE=(n_{14}\Phi_{14}, n_{24}\Phi_{24}, n_{34}\Phi_{34})$.

In this calculation, we are only interested in the electric field, and we choose the following four values,
\begin{equation}
\vec\mcE=(0, 0, \frac{6\pi}{L_z L_t}n_{z}), \quad n_{z}=\pm1,\pm2.  
\end{equation}
To implement the background field on the lattice, the $SU(3)$ gauge link $U_\mu(x)$ is multiplied by the $U(1)$ gauge
link $U^{(\mcE)}_\mu(x)$ associated with the background electric field,
\begin{equation}
\begin{aligned}
U_4(x) &\rightarrow U_4(x)U^{(\mcE)}_4(x)\,,
\\ U_{1,2,3}(x) &\rightarrow U_{1,2,3}(x)
\end{aligned}
\end{equation}
with $U^{(\mcE)}_4(x)=e^{i Q \mcA_4(x)}=e^{i Q \mcE_z x_3}$ for the lattice sites satisfy $0\le x_3\le L_z-2$, where $Q$ is
the quark electric charge and $\mcA_4(x)$ is defined in Eq.~(\ref{eqn:euc_const_bgel}) with $k=\hat{3}$.
To make all the $U(1)$ gauge plaquettes identical, the gauge links on the boundary $x_3=L_z-1$ must be modified as
\begin{equation}
\begin{aligned}
U_4(x)_{x_3=L_z-1} &\rightarrow U_4(x)U^{(\mcE)}_4(x) \,,\\
U_3(x)_{x_3=L_z-1} &\rightarrow U_3(x)U^{(\mcE),\perp}_3(x)\,,\\
U_{1,2}(x)_{x_3=L_z-1} &\rightarrow U_{1,2}(x)
\end{aligned}
\end{equation}
with $U^{(\mcE),\perp}_3=e^{-iQ\mcE_z L_z x_4}$. 
The setup is depicted in Fig.~\ref{fig:bgelectric}. Note that we employ the electro-quenched approximation in this calculation, such that the background electric field couples only to the valence quark. Given that the sea quark contribution to the nucleon magnetic momentum is approximately 1\%~\cite{Sufian:2017osl}, one may expect that the electro-quenched approximation to provide a reasonable description for the present nEDM study.
%------------------------------------------------------------------------------

%%%%%%%%%%%%%%%%%%%%%%%%%%%%%%%%%%%%%%%%%%%%%%%%%%%%%%%%%%%%%%%%
\subsection{Definition of the topological charge}\label{sec:top}
To obtain well-defined  topological charge density on the lattice, we employ the gradient
flow~\cite{Luscher:2010iy,Luscher:2011bx} to tame the short distance fluctuations of the gauge fields.
The differential equation for gradient flow can be written as~\cite{Luscher:2010iy,Luscher:2011bx}
\begin{equation}
\frac{d}{dt_{gf}}B_\nu(t_{gf},x)=D_\mu G_{\mu\nu}(t_{gf},x)\,, 
\quad\text{with } B_\mu(0,x)=A_\mu(x)\,,
\end{equation}
where $A_\mu$ is the original gauge field at $t_{gf}=0$ flow time, $B_\mu(t_{gf})$ is the gauge field at flow time $t_{gf}$,
and $G_{\mu\nu}(t_{gf})$ is the gauge field strength corresponding to $B_\mu(t_{gf})$.
At the leading order, the solution of the gradient flow equation is
\begin{equation}
\label{eq:diffu}
B_{\mu}(t_{gf},x)=\int d^4y \frac{e^{-(x-y)^2/(4t_{gf})}}{(4\pi t_{gf})^2}A_\mu(y),
\end{equation}
indicating that the flow equation is equivalent to four-dimensional Gaussian smearing with r.m.s. radius $\sqrt{8t_{gf}}$.

The topological charge at flow time $t_{gf}$ can be written as
\begin{equation}
Q_{top}(t_{gf}) 
  = \int d^4x \, \rho_Q(t_{gf}, x)
  = \frac{g^2}{32\pi^2} \int d^4x \, \epsilon^{\mu\nu\alpha\beta} \,
    \mathrm{Tr}\lp[G_{\mu\nu}(t_{gf},x)G_{\alpha\beta}(t_{gf},x)\rp].
\end{equation}
In practice we use the so-called five-loop improved (5LI) definition of the topological charge~\cite{deForcrand:1997esx}.

\begin{figure}[tp]
\begin{center}
\includegraphics[scale=0.6]{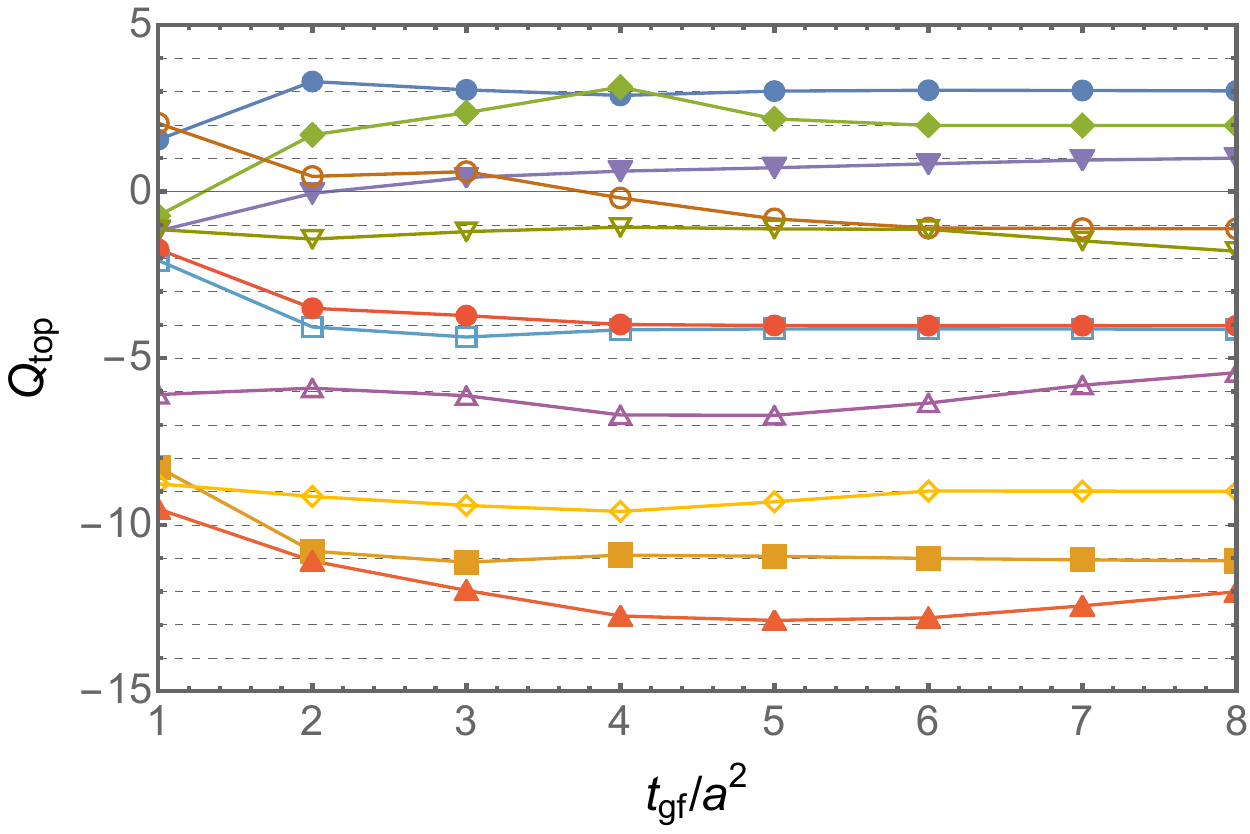}
\caption{The dependence of topological charge on gradient flow time. The result is obtained from 11 configurations on ensemble 24I-005.} 
\label{fig:gf_topocharge}
\end{center}
\end{figure}

The topological (instanton) sectors are resolved after several steps of the gradient flow.
The dependence of topological charge on gradient flow time is shown in Fig.~(\ref{fig:gf_topocharge}).
Here we present the results for 11 configurations from ensemble 24I-005.
Before employing the gradient flow, the topological charge exhibits large fluctuations and is generally not integer-valued.
For flow times  $t_{gf}>4a^2$, topological charges tend to be integer-valued and remain (mostly) stable as gradient flow time increases.

%------------------------------------------------------------------------------

\begin{figure}[ht!]
\centering
\includegraphics[scale=0.35]{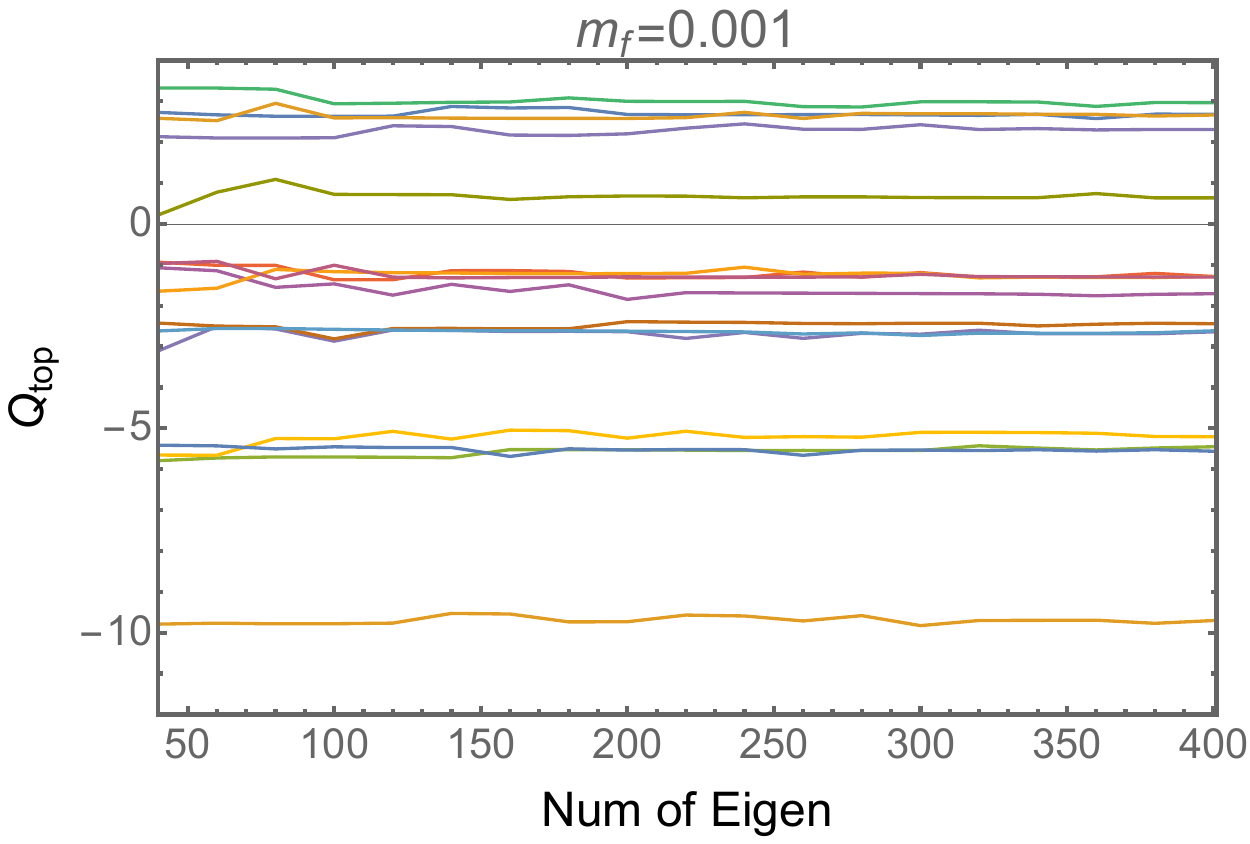}
\includegraphics[scale=0.35]{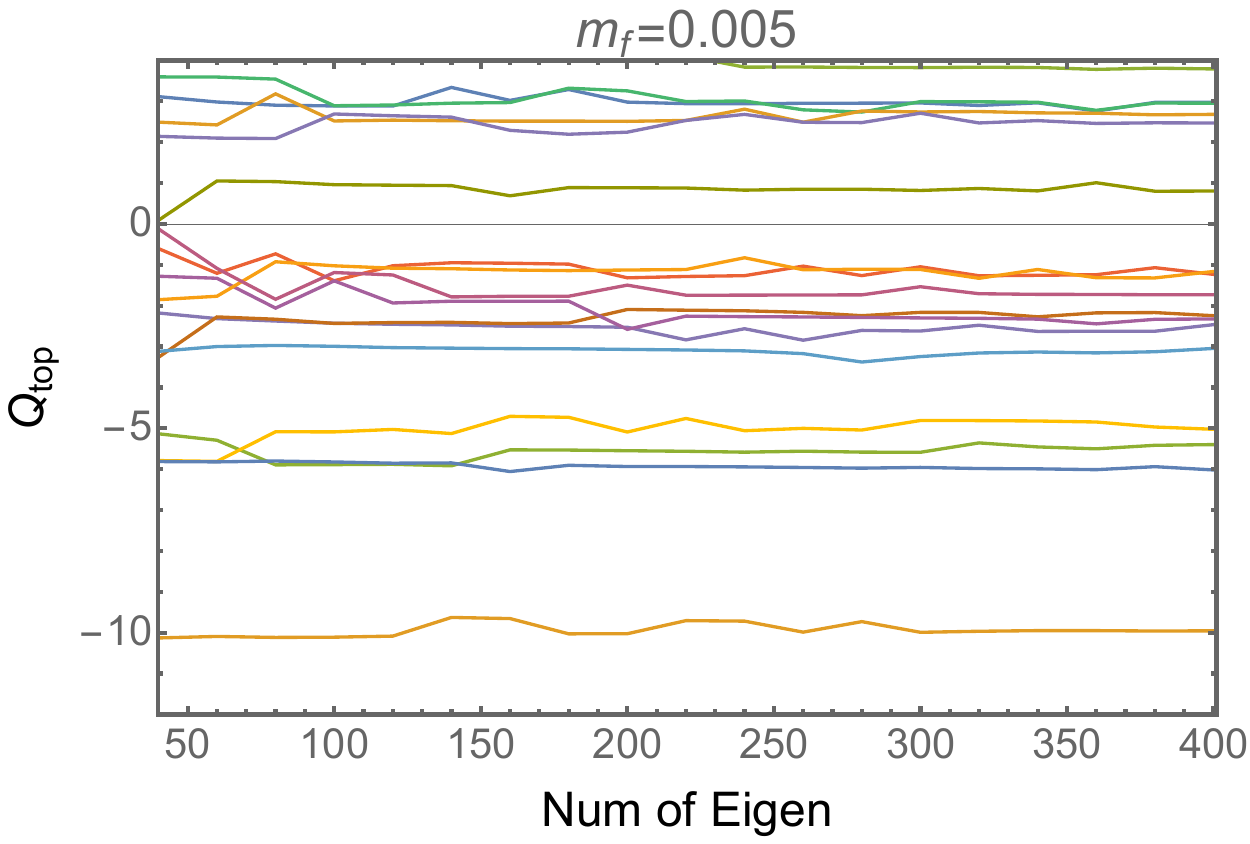}
\\
\includegraphics[scale=0.35]{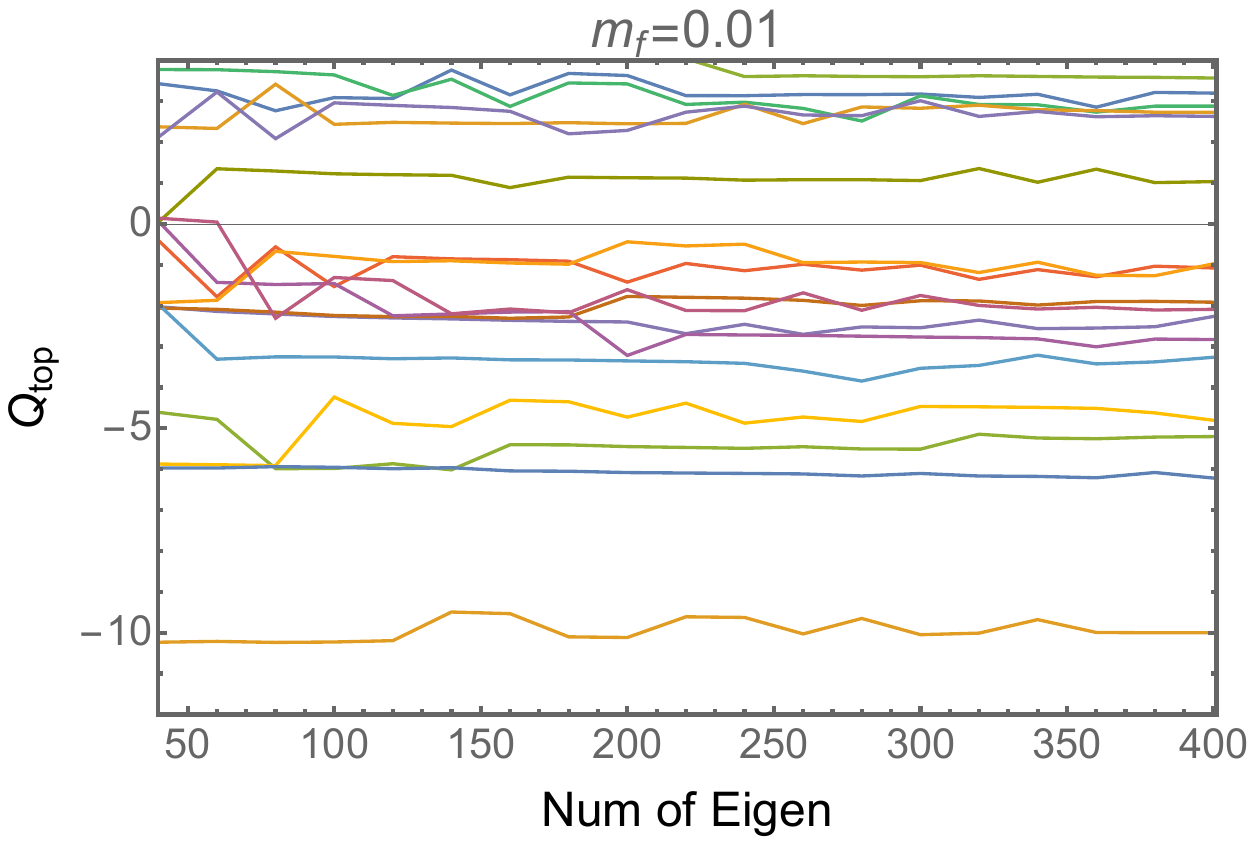}
\includegraphics[scale=0.35]{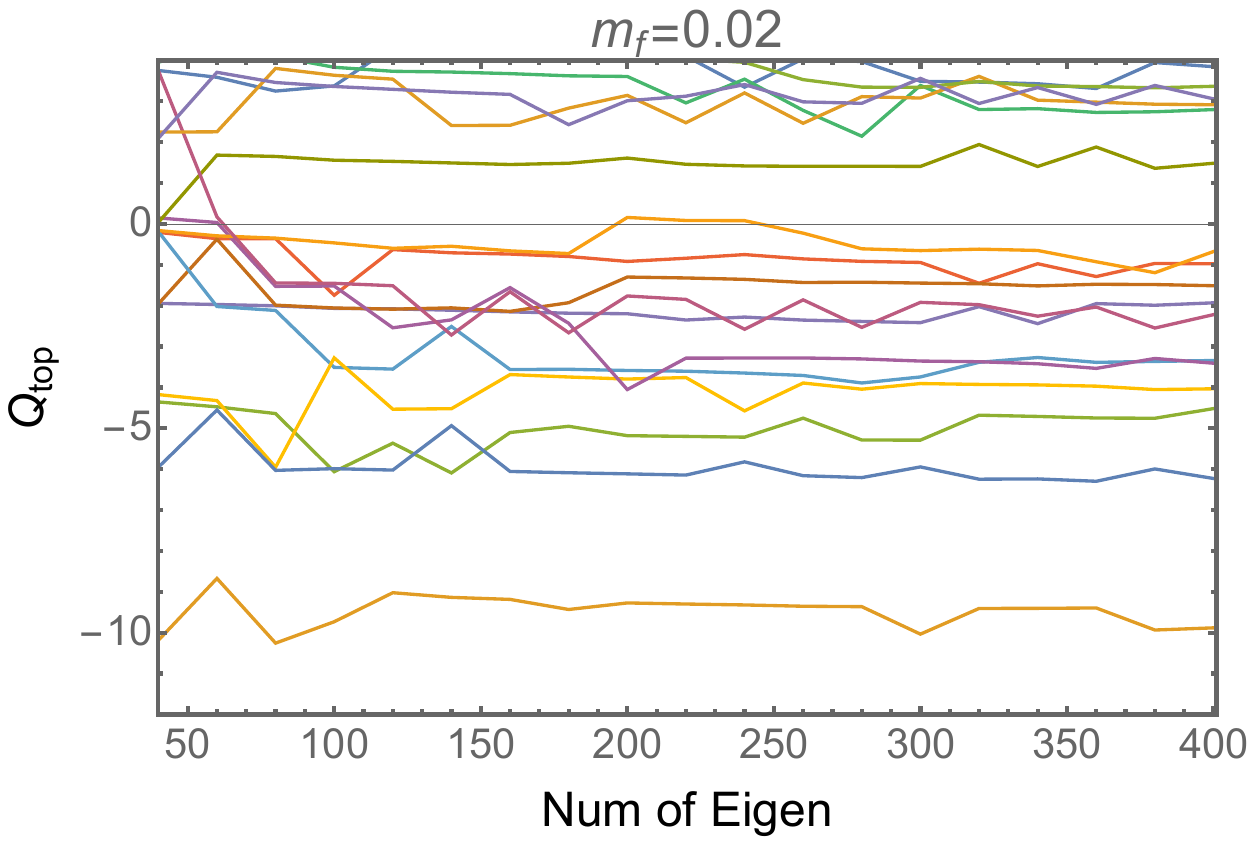}
\caption{
Comparison of global topological charge constructed from pseudoscalar quark density using varying numbers of lowest
modes preconditioned Domain-Wall fermion (ensemble 24I-005).
We examine $O(10)$ configurations denoted by the different color lines, with varying quark masses used to compute the
quark ``loops''. 
\label{fig:eigendep}
}
\end{figure}

Another way to determine the topological charge is via the Adler-Bell-Jackiw (ABJ) anomaly and the pseudoscalar quark
density.
The ABJ relation in Euclidean space with quarks satisfying the Dirac equation $(\slashed D + m_\psi)\psi=0$ can be
written as
\begin{equation}
\label{eq:ABJE}
Z_A\partial^\mu(\bar{\psi}\gamma^\mu\gamma^5\psi)
=2m\bar{\psi}\gamma^5\psi-2\rho_Q
=2m\bar{\psi}\gamma^5\psi-\frac{g^2}{16\pi^2}\epsilon^{\mu\nu\alpha\beta}\mathrm{Tr}[G_{\mu\nu}G_{\alpha\beta}],
\end{equation}
where $Z_A$ is the renormalization constant of the (local) axial vector current (see~\cite{Allton:2008pn} for
numerical values relevant to this work).
The term on the left hand side of (\ref{eq:ABJE}) vanishes after integration over the four-dimensional volume, and the
global topological charge is given by~\cite{Aoki:1990ix}
\begin{equation}
Q_{top} = m \, \int d^4x \, \la \bar{\psi}\gamma^5\psi\ra 
= - m \, \mathrm{Tr}\lp[\gamma_5\lp(\slashed D + m\rp)^{-1}\rp]
\end{equation}
(the minus sign arises from the Wick contraction of the quark fields).
In practice we calculate the $\gamma_5$-trace of the inverse quark propagator on a lattice using the (M\"obius) Domain
Wall operator $D_{DW}$ with varying current quark mass $m_f$,
\begin{equation}
\label{eq:tp_ql}
\begin{aligned}
\lp[m\int d^4x\, \bar{\psi}(x) \gamma^5\psi(x) \rp]^\text{lat}
  &= -(m_f+m_{res}) \sum_x \text{Tr} \lp[\gamma_5 D_{DW}^{-1}(x,x)\rp] 
\\&\approx -(m_f+m_{res}) \sum_x \text{Tr} \sum_n^{N_\psi}[\gamma_5 v_n(x)w^\dag_n(x)]  \,
\end{aligned}
\end{equation}
which is estimated using $N_{\psi}$ vector pairs $v_n, w_n$ reconstructed from eigenmodes of the preconditioned
Dirac operator, the same as those used to construct the all-to-all quark propagators and nucleon correlators
(see Sec.~\ref{sec:LMA}).
The residual mass $m_{res}$ equals $0.00308$~\cite{Mawhinney:2019cuc} for the gauge ensembles used in this calculation.
In Fig.~\ref{fig:eigendep}, we show the dependence of the global topological charge on the number of eigenvectors of the
Dirac operator for bare quark masses in the range $m_f=0.001$ to $m_f=0.02$.
The results obtained with smaller quark mass exhibit smaller fluctuations and saturate with $N_{\psi}$ more rapidly
compared to the those obtained with heavier quark mass. 
It may be argued that the effective quark mass $(m_f + m_{res})$ plays the role of a smearing parameter, with lighter
mass corresponding to larger smearing r.m.s radius and gradient flow time (see also Ref.~\cite{Alexandrou:2017hqw}).

\begin{figure}[tp]
\begin{center}
\includegraphics[scale=0.4]{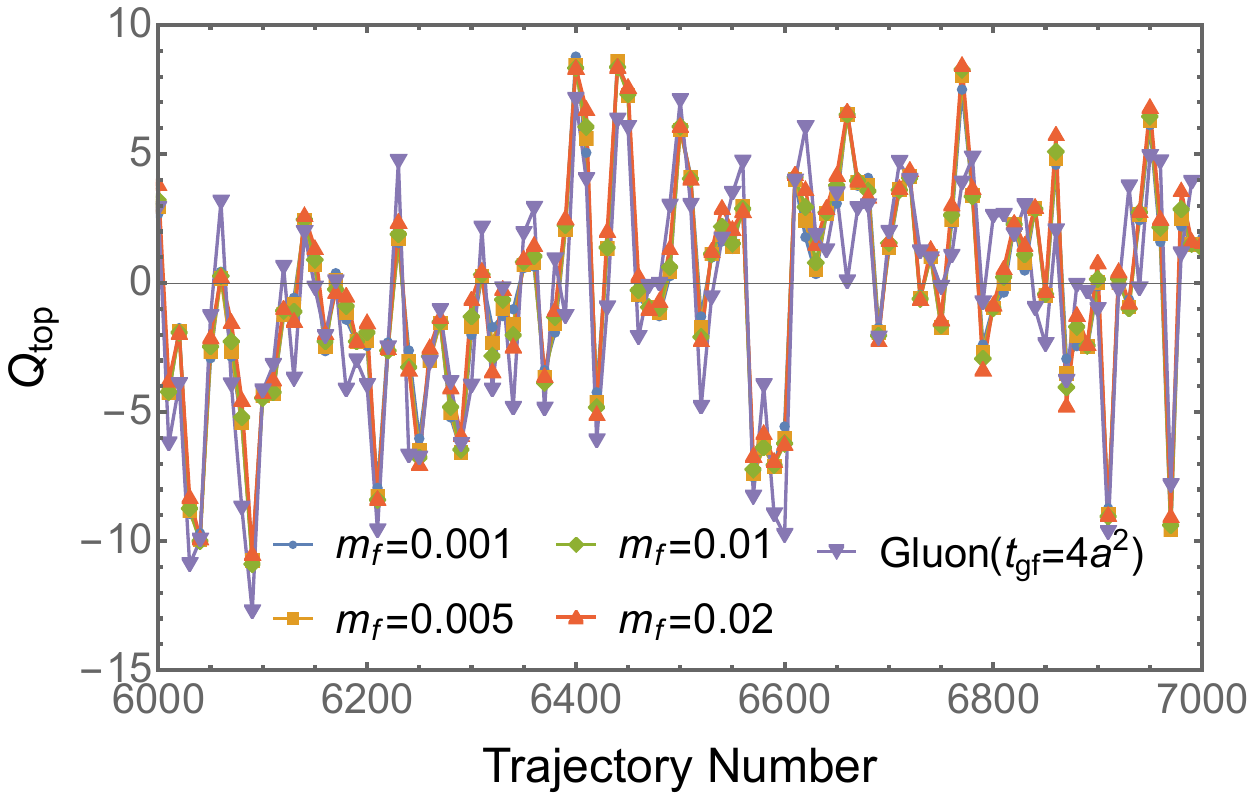}
\caption{
Comparison of global topological charge computed from gluon field and pseudoscalar quark density with varying quark mass
$m_f$ (ensemble 24I-005).
The gluon field is gradient-flowed to $t_{gf}=4a^2$. 
\label{fig:tp_comp}
}
\end{center}
\end{figure}

We compare the topological charge constructed from the quark loop at different quark mass with that obtained from the
gluon fields on 100 configurations in Fig.~(\ref{fig:tp_comp}).
The $x$ axis denotes the corresponding trajectory number of the configuration.
The topological charges computed from the pseudoscalar density with different quark masses are consistent with each other,
and are strongly correlated with that computed from gradient-flowed gauge links.

In our calculation, the nEDM is calculated from matrix elements of the local topological charge summed over a time slice, 
\begin{equation}
q^G_\text{top}(t_{gf}, t) = \frac1{16\pi^2}\sum_{\vec y}\mathrm{Tr} 
  \lp[ G_{\mu\nu}(t_{gf},t,\vec{y}) \, \widetilde G_{\mu\nu}(t_{gf},t,\vec{y}) \rp]\,.
\end{equation}
We also study nEDM using the alternative local topological charge definition constructed using the pseudoscalar density,
\begin{equation}
q^\psi_{top}(t) = -(m_f+m_{res}) \sum_{\vec x, n} \mathrm{Tr}[\gamma_5 v_n(t,\vec x)w_n^\dag(t,\vec x)]\,.
\end{equation}

\begin{figure}[ht!]
\centering
\includegraphics[scale=0.35]{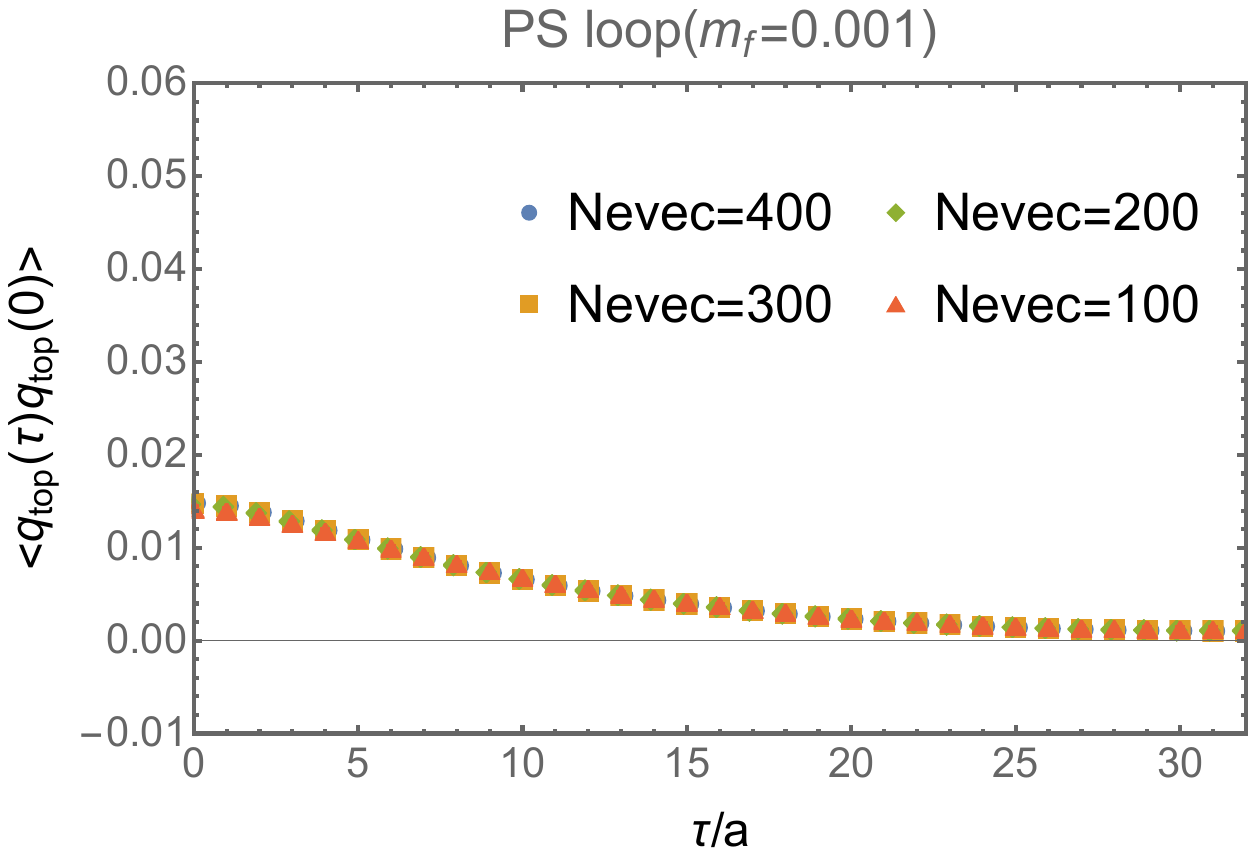}
\includegraphics[scale=0.35]{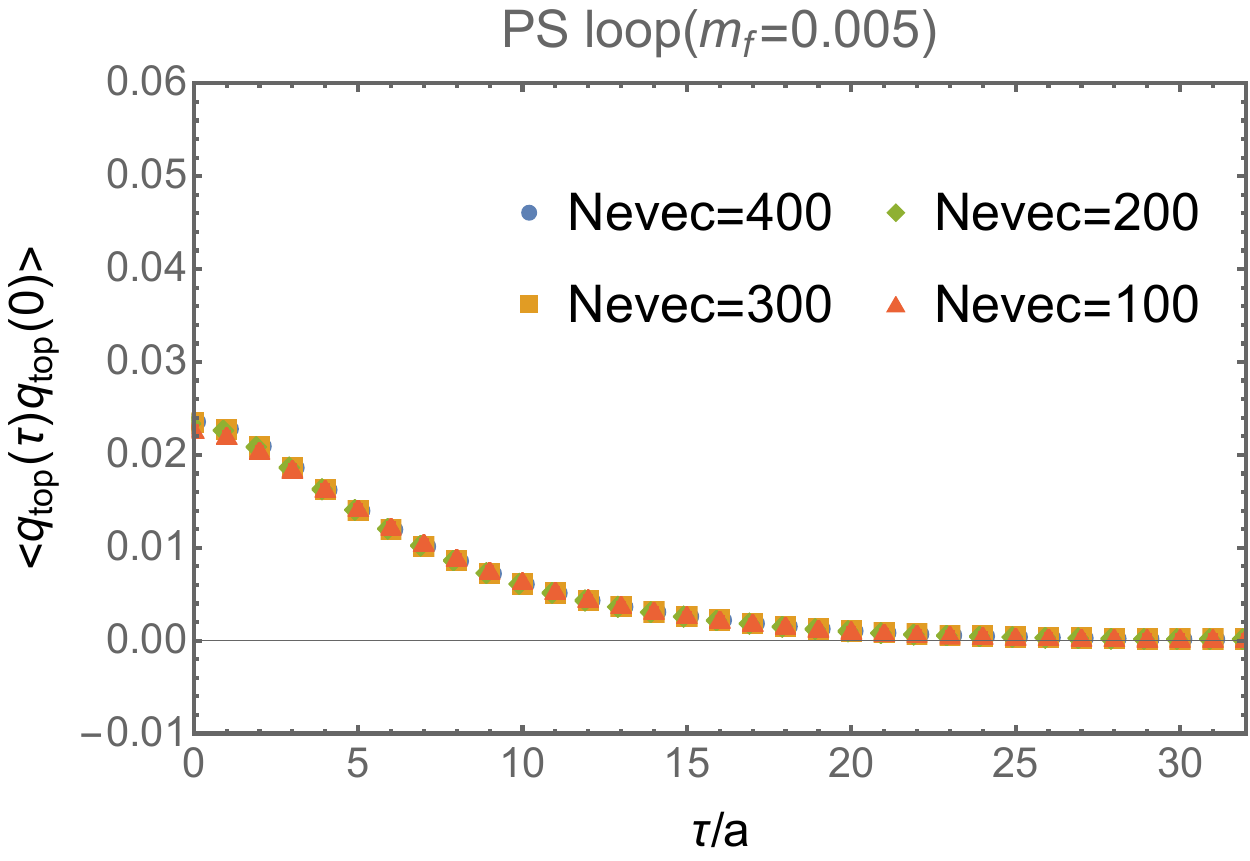}
\\
\includegraphics[scale=0.35]{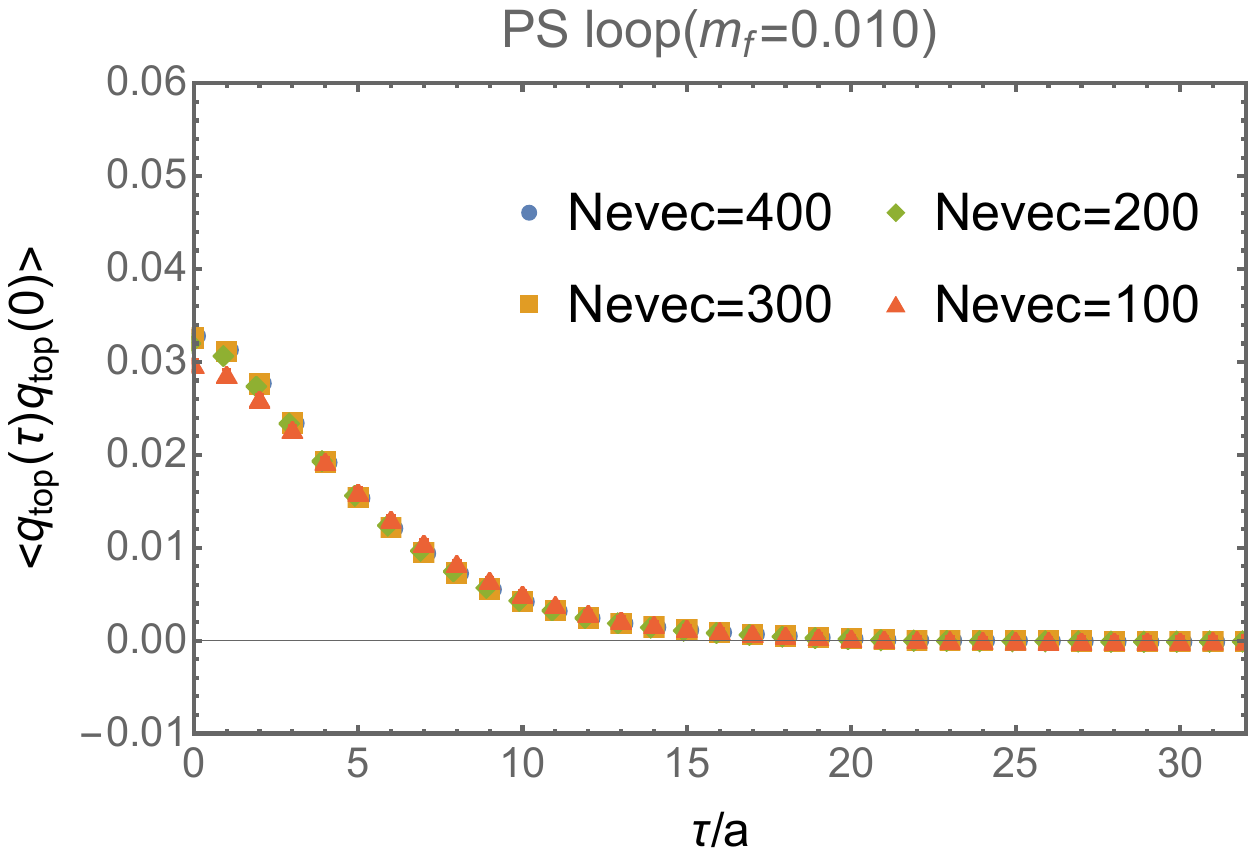}
\includegraphics[scale=0.35]{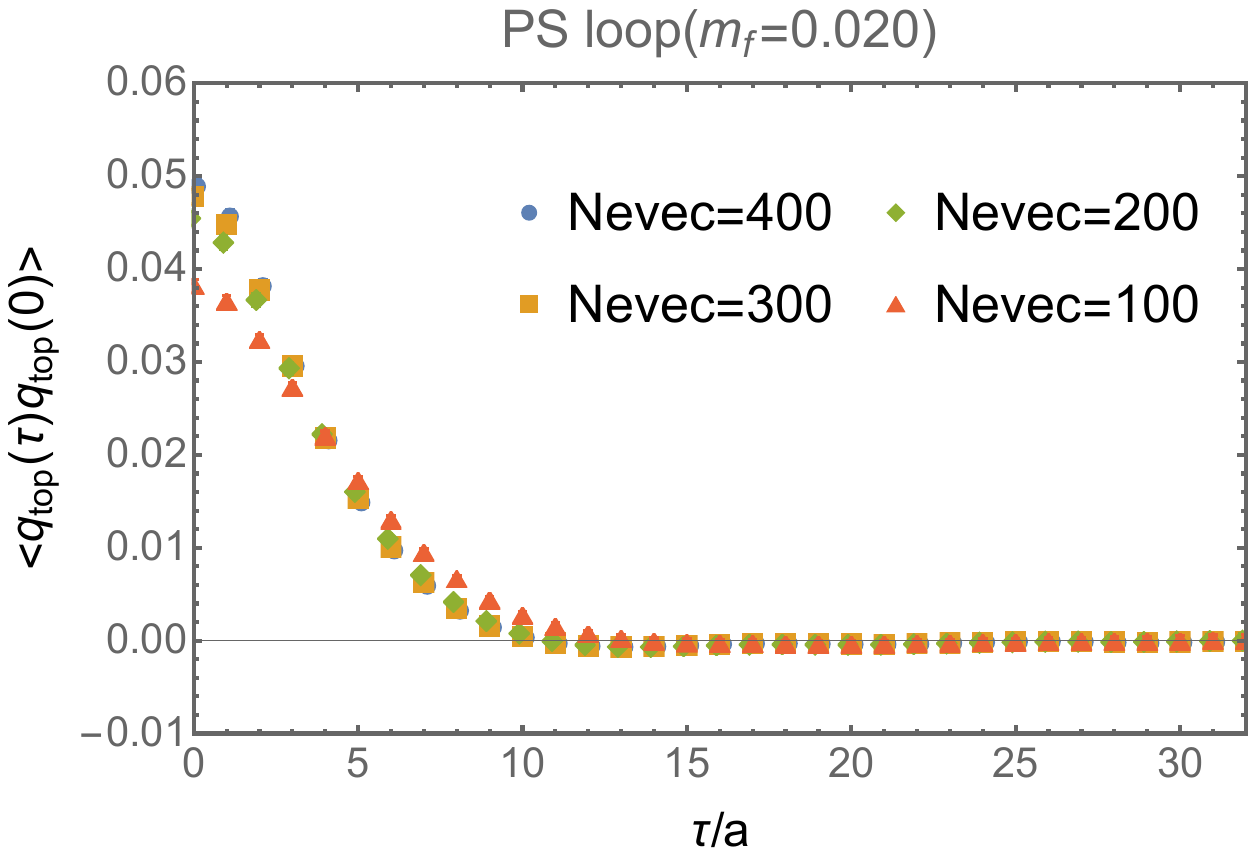}
\\
\includegraphics[scale=0.35]{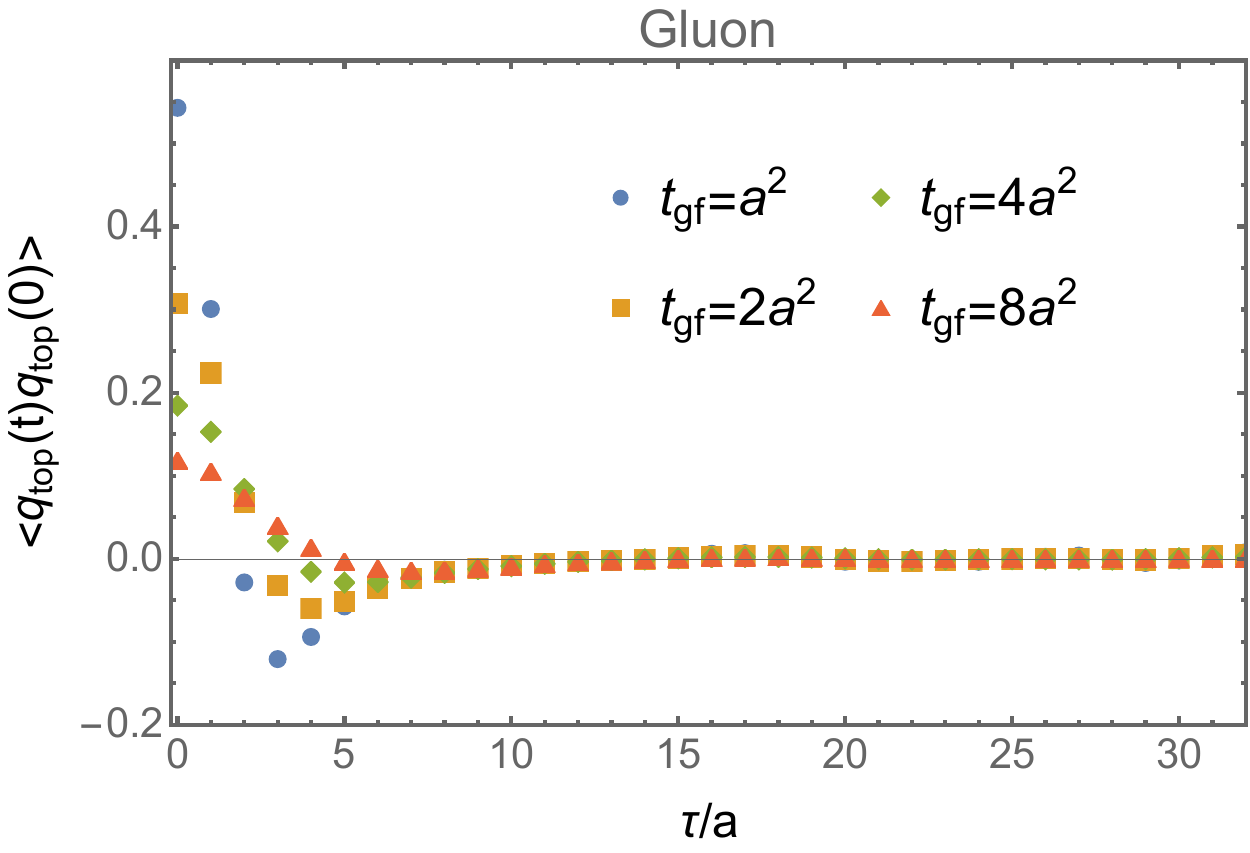}
\caption{
Correlation functions of local topological charge~(\ref{eq:qtopcorre}) computed from pseudoscalar quark density (PS
loop) with varying quark mass and the number of eigenvectors. 
The last figure is the same but constructed from gluon fields with varying gradient-flow time.}
\label{fig:qcoreeeigendep}
\end{figure}

To compare locality of these topological charge density definitions, we calculate their two-point correlation functions,
\begin{equation}
\label{eq:qtopcorre}
C_\text{top}(\tau)=\la q_\text{top}(t + \tau)q_\text{top}(t)\ra
\end{equation}
Although the global topological charge defined by the pseudoscalar density is consistent for different quark masses,
differences in their local behavior become apparent from their correlation functions as shown in Fig.~\ref{fig:qcoreeeigendep}.
The first four panels show the topological charge correlator constructed from the pseudoscalar density of quarks with different
masses, with varying numbers of eigenvectors used to estimate the trace.
The last panel shows the gluon topological charge correlator at varying gradient flow time.
In contrast to the gluonic definition, the topological correlators from pseudoscalar density are always positive.
For smaller quark masses, the results saturate rapidly with the number of eigenvectors.
Even with the heaviest quark mass, the correlation functions saturate at 300-400 eigenvectors within uncertainties.

\begin{figure}[tp]
\begin{center}
\includegraphics[scale=0.45]{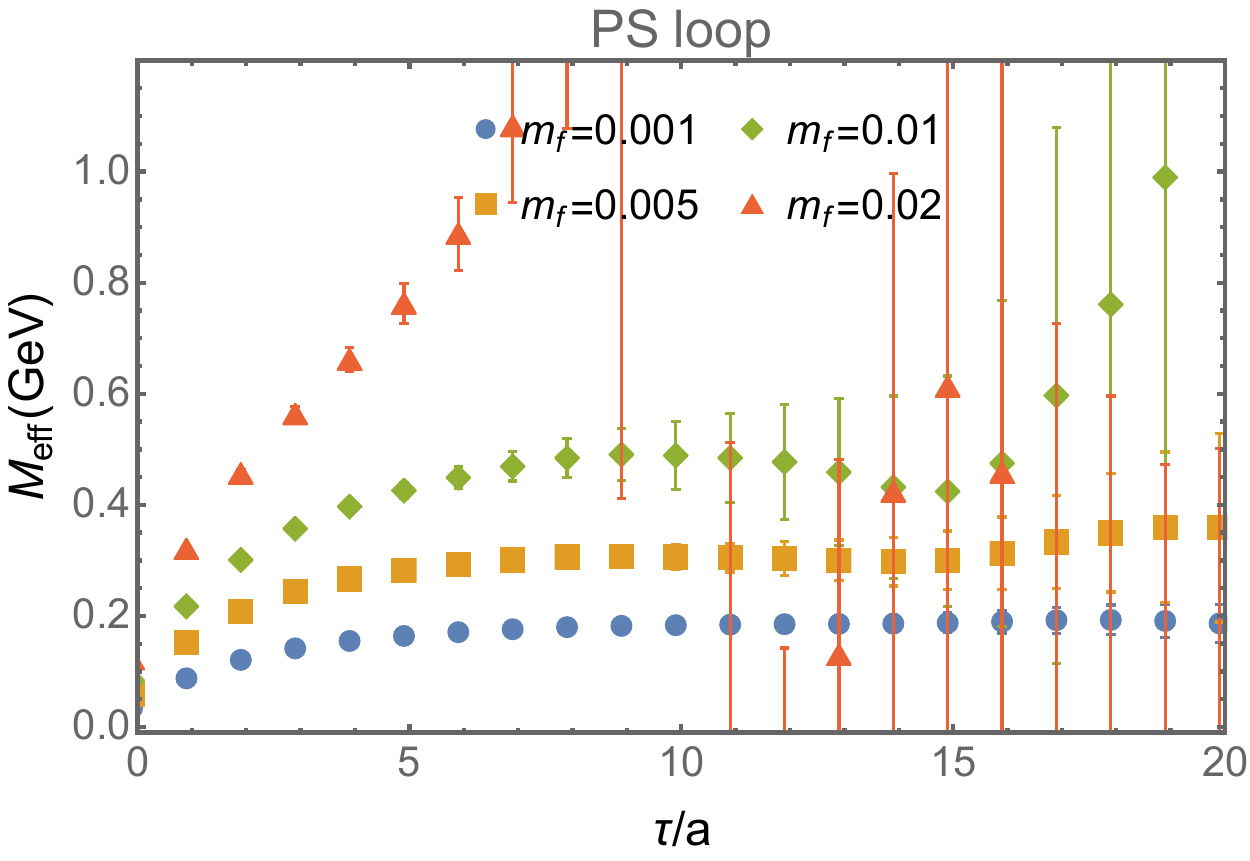}
\caption{The effective mass extracted from correlation functions of pseudoscalar quark density computed with varying
quark mass.} 
\label{fig:tp_eff}
\end{center}
\end{figure}

As the quark mass $m_f$ increases, the correlation function becomes larger at small time separation regions, and decays
more rapidly compared to the lighter quark mass cases.
In Figure ~\ref{fig:tp_eff} we show the effective mass of the topological charge correlation function 
$M^\text{eff}_\text{top}=\log\lp(C_\text{top}(\tau)/C_\text{top}(\tau+1)\rp)$,
which increases with increasing quark mass.
To extract the spectrum, we fit the correlation function using the Ansatz
\begin{equation}
C_\text{top}(\tau)=a_0[e^{-m_0 \tau}+e^{-m_0(T-\tau)}]+a_1[e^{-m_1 \tau}+e^{-m_1(T-\tau)}]\,,
\end{equation}
with fit results for $m_0$ and $m_1$ shown in Table.~\ref{tab:effm}.
Roughly speaking, the ground state mass $m_0$ is close to the pion mass corresponding to the input quark mass $m_f+m_{res}$.
The excited state mass is heavier, around $900$ MeV, and close to the $\eta'$ mass.
This is the mass that would dominate the full isoscalar correlator, in which connected and disconnected
diagrams are combined, and the pion-like isovector contribution is cancelled.

The correlation function constructed from the gluon topological charge operator is sensitive to the gradient flow time.
It exhibits a sharp peak at small $\tau$ for small flow time.
As the flow time increases, the peak broadens and becomes less localized, consistent with the diffusive nature of the
gradient flow, as shown in Eq.~(\ref{eq:diffu}). 
As shown in Fig.~\ref{fig:gf_topocharge}, the global topological charge  becomes stable at sufficiently large flow
times, which indicates that the nEDM results should be insensitive to the gradient flow evolution once the flow time is
large enough.
However, since we employ the local topological charge operator to extract the nEDM, one may still question whether the
resulting nEDM is sensitive to the choice of gradient flow time.
In Sec.~\ref{sec:Numer_ESC}, we will show that the nEDM results obtained using the local gluonic topological charge at
different gradient flow times are consistent with each other.

%------------------------------------------------------------------------------
\begin{table}[ht]
\centering
\caption{The fitting results of $m_0$ and $m_1$, obtained by fitting the correlation of topological charge density using 
  the Ansatz given in Eq.~(\ref{eq:qtopcorre}).
  \label{tab:effm}
}
\begin{tabular}{c|ccc}
\hline\hline
& $m_f=0.001$   & $m_f=0.005$  & $m_f=0.01$  \\
\hline
$m_0$(GeV)             & 0.180(6) & 0.318(13)   & 0.588(112)       \\
$m_1$(GeV)              & 0.907(27) & 0.923(37)   & 0.726(128)
\\
\hline\hline
\end{tabular}
\end{table}
%------------------------------------------------------------------------------
%------------------------------------------------------------------------------
%------------------------------------------------------------------------------
%------------------------------------------------------------------------------

According to Eq.~(\ref{eq:ABJE}), the local topological charge density constructed from the gluon field, $\rho_Q$, and
that obtained from the pseudoscalar density are related though 
\begin{equation}
\rho_Q=m\bar{\psi}\gamma^5\psi-Z_A\frac{\partial^\mu(\bar{\psi}\gamma^\mu\gamma^5\psi)}{2}\,.
\end{equation}
This leads to the identity for topological charge density correlators with other operators
\begin{equation}
\sum_{\vec{x}} \la { \rho_Q(\tau,\vec{x}) \mcO \ldots}\ra 
  = -Z_A\frac{\partial_\tau \sum_{\vec{x}}\la  A_4(\tau,\vec{x})  \mcO \ldots \ra}{2} 
  + m \sum_{\vec{x}} \la  J_5(\tau,\vec{x}) \mcO \ldots\ra.
\end{equation}
In our EDM calculation, such correlators with nucleon interpolating operators in a background electric field are
\begin{eqnarray}\label{eq:ABJ}
\la 
O_N(t_f) q_\text{top}^G(\tau) \bar O_N(0) \ra
  = -Z_A\frac{\partial_\tau \la  O_N(t_f) \sum_{\vec x} A_4(\tau,\vec{x}) \bar O_N(0)\ra}{2} 
  + m \sum_{\vec{x}} \la O_N(t_f) J_5(\tau,\vec{x})  \bar O_N(0)
 \ra.
\end{eqnarray}
where the spatial derivatives in the first term vanish by integration by parts and $m = (m_f + m_{res})$.
The first term on the RHS can be expressed using a spectral decomposition,
\begin{equation}
\la  O_N(t_f) \sum_{\vec{x}}A_4(\tau,\vec{x})  \bar O_N(0)\ra=
\sum_{i,j} \la 0 | O_N(t_f) | E_i\ra  \la E_i| A_4(0) |E_j\ra \la E_j | \bar O_N(0)|0\ra e^{-E_i (t_f-\tau) - E_j \tau}
\end{equation}
and is $\tau$-independent when $E_i=E_j$ and $0 < \tau < t_f$. 
This implies that this term contributes only to transition matrix elements between different eigenstates $i\ne j$ and
can be regarded as excited state contamination.

%%%%%%%%%%%%%%%%%%%%%%%%%%%%%%%%%%%%%%%%%%%%%%%%%%%%%%%%%%%%%%%%
\subsection{All mode Averaging and Low Mode Averaging}\label{sec:LMA}
Measurements of the correlation functions are done in an all-mode-average (AMA)~\cite{Blum:2012uh,Shintani:2014vja}
framework with a full-volume low-mode average (LMA)~\cite{Neff:2001zr,DeGrand:2004qw,Giusti:2004yp,Giusti:2005sx}.
These techniques are employed to enhance statistics with lower computational cost.
In addition, numerical emphasis on the low modes of the Dirac operator is crucial in the case of the $\theta$-induced
nEDM because of their close relationship to the topological charge density~\cite{tHooft:1976rip}.

The AMA and LMA techniques are combined to yield an improved estimator for a generic observable $\mathcal{O}$ (here,
$(\mcO\ldots)$ is the nucleon correlation function),
\begin{equation}
\label{eq:AMA}
\la (\mcO\ldots)_\text{AMA} \ra_{N_\text{sl}}
  = \la(\mcO\ldots)_\text{sl}\ra_{N_\text{sl}} + \la((\mcO\ldots)_\text{ex}- (\mcO\ldots)_\text{sl}) \ra_{N_\text{ex}}\,,
\end{equation}
where ``sl'' refers to cheaper approximate (sloppy) samples with full statistics $N_{sl}$ and 
``ex'' refers to exact samples with reduced statistics $N_{ex}\ll N_{sl}$. 
To accelerate the inversion of Dirac operator using the conjugate gradient (CG) algorithm, we employ low-mode deflation by first computing the lowest 200 eigenmodes of preconditioned Dirac operator. The exact solves use a CG residual tolerance of $\mathcal{O}(10^{-9})$, while the sloppy solves are obtained by performing a fixed 200 CG iterations, without requiring convergence to the target residual.
The first high-statistics term is designed to suppress the statistical noise ($N_{sl}\gg N_{ex}$),
and the second low-statistics term remove the bias of the approximation.
The key idea is that the approximate and exact values have strongly correlated fluctuations so that the bias correction
contributes very little to the total statistical error~\cite{Blum:2012uh}.
In practice the exact and approximate observables are constructed using conjugate gradient solvers for the quark
propagators that make up the nucleon correlation function.
The sloppy solvers are computed with relaxed tolerance or by executing a fixed (but significantly reduced) number of
iterations after deflation with precomputed low-lying modes of the intrinsic even-odd preconditioned operator.
Deflation is the crucial element of the AMA scheme because it dramatically reduces the bias correction and its
fluctuations.

In addition to deflation-enhanced AMA, we improve the statistical precision of nucleon correlators with a full-volume
($N_\text{V}=L^4$) average of the low mode-only samples (``low''), which are evaluated using the same modes as in the
deflation step,
\begin{equation}
\label{eq:LMA}
\la (\mcO\ldots)_\text{LMA} \ra 
  = \la(\mcO\ldots)_\text{low} \ra_{N_\text{V}}
  + \la((\mcO\ldots)_\text{AMA} - (\mcO\ldots)_\text{low}) \ra_{N_\text{sl}}\,,
\end{equation}
The LMA improvement results in a dramatic increase in statistics because the ``low'' samples are computed at
all the sites of the lattice. 
For correlation functions involving pions or other mesons, which are quickly saturated by a relatively small number of
low-modes, Eqs.~(\ref{eq:AMA}) and (\ref{eq:LMA}) are quite effective in reducing statistical errors in correlation
functions.
However, nucleon correlation functions are not dominated by the low-modes, so in principle a large number may be needed
for significant gains.
This can be prohibitively expensive because the cost scales as the power of the number of quark fields in the hadron
interpolating operator.

Nevertheless, our goal is to examine the nucleon EDM which is computed as the \emph{neutron energy shift} relative to
the zero-electric field case.
Further, the EDM is $\CP$-odd, and only configurations with non-zero topological charge contribute.
Therefore, EDM must be sensitive only to the lowest modes of the Dirac operator, which also contribute most of the noise
in the correlation functions.
In the energy shifts, however, such noise is unrelated to the $\CP$-odd interactions and is expected to cancel.
Since the low-mode samples span the entire volume, we expect our LMA procedure to be a precise probe of
$\bar\theta$-induced nEDM, even though the nucleon correlation functions themselves are quite noisy.

There is one caveat to the above discussion: in our study we work with the low-modes of a preconditioned Dirac operator
which are both easier to compute and can be used for deflation in the conjugate gradient solver, but the topology is
related to the low-lying modes of the Dirac operator itself.
The two types of eigenvectors are not equivalent but are related; for example if the gauge field supports an exact
zero-mode of the Dirac operator, the same is true for the preconditioned operator.
We expect that there is strong correlation between the near-zero modes as well, however there is no exact relation
between them. In this calculation, we construct the low mode-only samples using the lowest 200 eigenmodes of preconditioned Dirac operator. The comparison between the nEDM results obtained using AMA alone with those including the LMA correction is presented in Appendix~\ref{sec:appenx_error}. 

Specifically, we use the 4D even-odd preconditioning of the Domain Wall Dirac operator,
\begin{equation}
D_{DW} = \lp(\ba{cc} M_{ee} & M_{eo} \\ M_{oe} & M_{oo} \ea \rp)\,,
\end{equation}
where $M_{ij}$ ($i,j\in {e,o}$) are the matrices between 4D-even and 4D-odd site subsets.
We employ the ``SchurDiagTwo" preconditioned Dirac operator defined in Grid~\cite{Grid,Boyle:2016lbp},
\begin{equation}
\label{eqn:Doo_def}
D_{oo}  = 1 -  M_{oe} M_{ee}^{-1}  M_{eo} M_{oo}^{-1}\,,
\end{equation}
which can be inverted using block LDU decomposition
\begin{equation}
\label{eq:decomp}
\begin{aligned}
D_{DW}^{-1} 
  &=\lp(\ba{cc} 1 & -M_{ee}^{-1}  M_{eo} M_{oo}^{-1} \\ 0 & M_{oo}^{-1} \ea\rp)
    \lp(\ba{cc} M_{ee}^{-1} & 0 \\ 0 & D_{oo}^{-1} \ea\rp)
    \lp(\ba{cc} 1 & 0\\  -M_{oe} M_{ee}^{-1} & 1 \ea\rp)
\\&=\lp(\ba{cc}
        M_{ee}^{-1} +M_{ee}^{-1}  M_{eo} M_{oo}^{-1} D_{oo}^{-1}M_{oe} M_{ee}^{-1}
    &   -M_{ee}^{-1}  M_{eo} M_{oo}^{-1} D_{oo}^{-1}
    \\  -M_{oo}^{-1}D_{oo}^{-1} M_{oe} M_{ee}^{-1} 
    &   M_{oo}^{-1}D_{oo}^{-1}
  \ea\rp)\,.
\end{aligned}
\end{equation}
The computationally demanding part is the inverse of the preconditioned operator $D_{oo}^{-1}$\footnote{
  We actually calculate the inverse of $D_{oo}^\dag D_{oo}$ acting on a source multiplied by $D_{oo}^\dag$.} 
and the associated low-lying modes of $D_{oo}^\dag D_{oo}$ ($D_{ee}$ is defined similarly to $D_{oo}$~(\ref{eqn:Doo_def})). 
The all-to-all vectors~\cite{Foley:2005ac} $v_{n}$ and $w_{n}$ used to construct the baryons fields are given by
\begin{equation}
\begin{aligned}
v_{n} &= \frac{1}{\lambda_n}\lp(\ba{c} -M_{ee}^{-1}  M_{eo} M_{oo}^{-1} h_n \\ M_{oo}^{-1}h_n \ea\rp) \,,
\\
w_{n} &=\lp(\ba{c} -M_{ee}^{-1,\dag} M_{oe} ^\dag D_{oo} h_n \\ D_{oo}h_n \ea\rp)\,,
\end{aligned}
\end{equation}
where $(\lambda_n,h_n)$ are eigen-pairs the $D^\dag_{oo}D_{oo}$ operator.

%%%%%%%%%%%%%%%%%%%%%%%%%%%%%%%%%%%%%%%%%%%%%%%%%%%%%%%%%%%%%%%%%%%%%%%%%%%%%%%%
%%%%%%%%%%%%%%%%%%%%%%%%%%%%%%%%%%%%%%%%%%%%%%%%%%%%%%%%%%%%%%%%%%%%%%%%%%%%%%%%
\section{Numerical results}
\label{sec:Numres}
\begin{figure}[ht!]
\centering
\includegraphics[width=.32\textwidth]{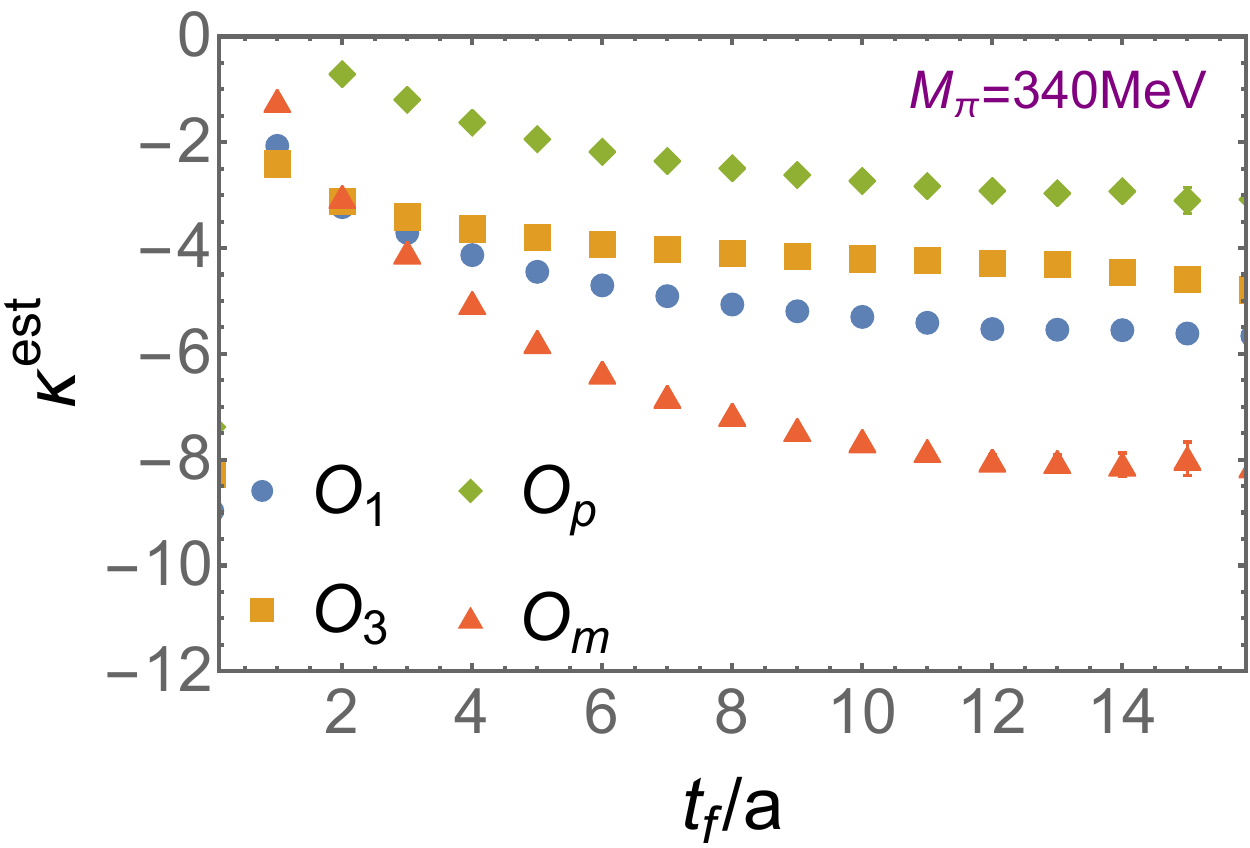}
\includegraphics[width=.32\textwidth]{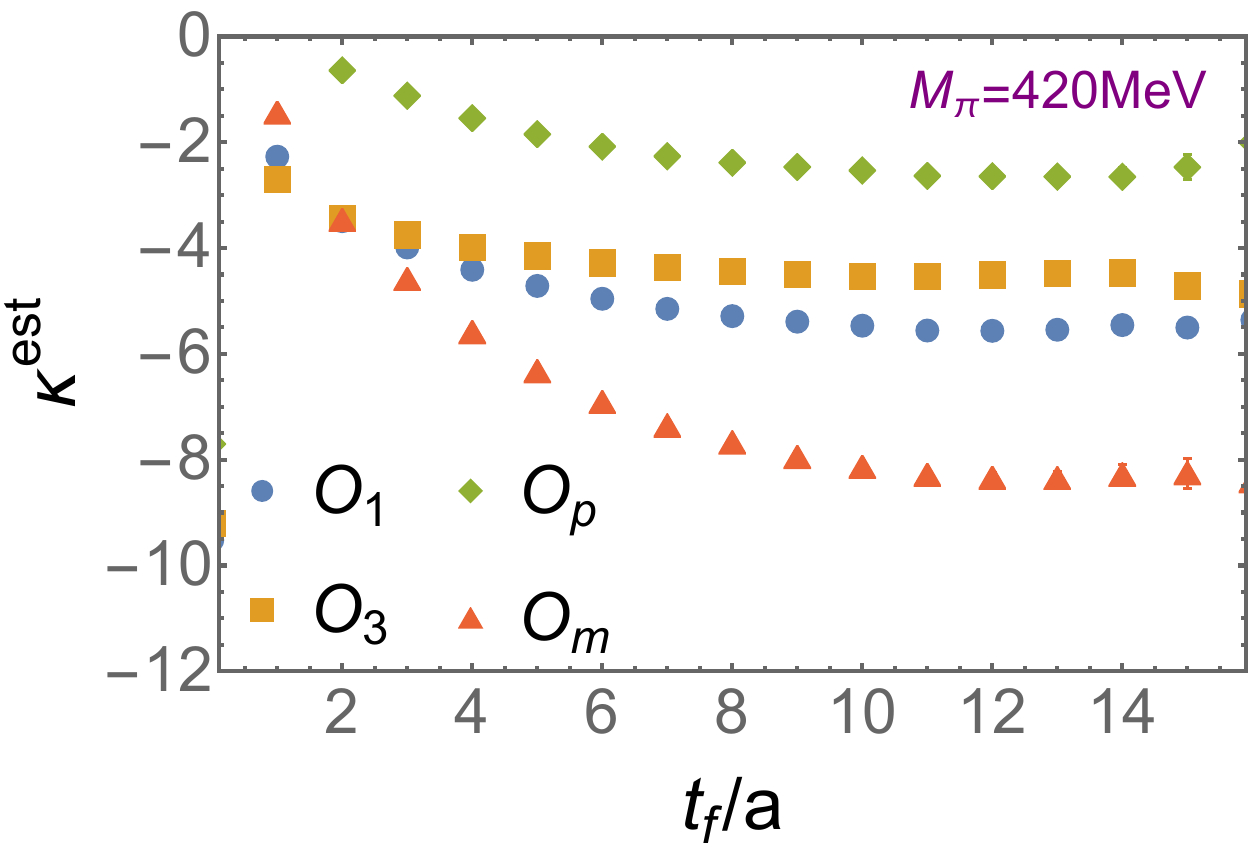}
\includegraphics[width=.32\textwidth]{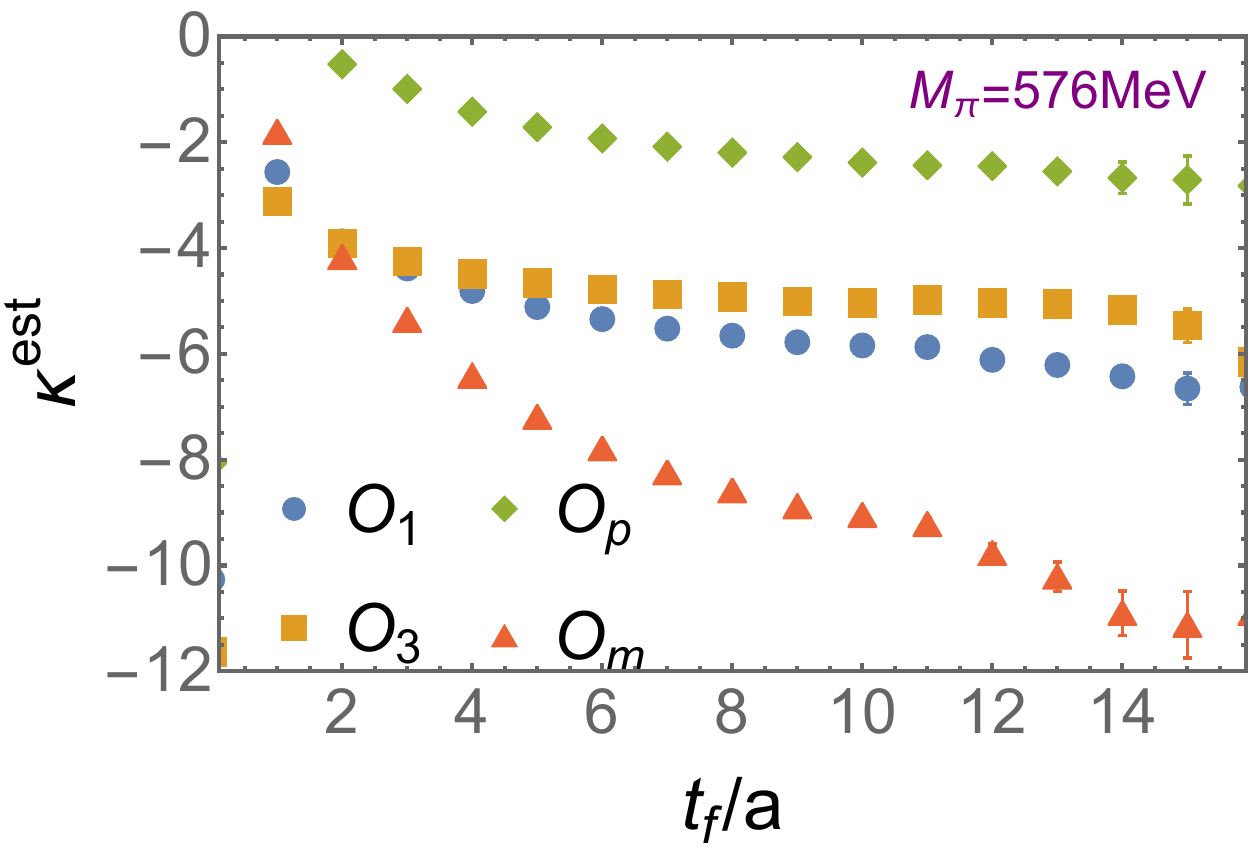}
\caption{Results for the neutron anomalous magnetic moment estimator $\kappa^\text{est}$ obtained using Eq.~(\ref{eq:ratiokappa}) on three different ensembles.
(Different colors show their dependence on the nucleon interpolating operator used.)
\label{fig:kappa_Ez2}
}
\end{figure}

In this calculation, we consider four nucleon interpolating operators
with spin-$\frac12$,
\begin{equation}
\label{eqn:nucleon_ops}
\begin{aligned}
O_1 &=\epsilon^{abc} (d^{aT}C\gamma_5u^b)d^c
\,,\qquad
&&O_2 =-\epsilon^{abc} (d^{aT}Cu^b)\gamma_5d^c
\,,\\
O_3 &=\epsilon^{abc} (d^{aT}C\gamma_4\gamma_5u^b)d^c
\,,\qquad
&&O_4 =\epsilon^{abc} (d^{aT}C\gamma_4u^b)\gamma_5d^c,
\end{aligned}
\end{equation}
According to Ref~\cite{Ema:2024vfn}, one can define chiral covariant operators $O_p=O_1+O_2$ and $O_m=O_1-O_2$, which can be rewritten as $O_p=\frac{1}{2}\epsilon^{abc}(d^{aT}C\gamma_\mu d^b)\gamma_\mu\gamma_5u^c$ and
$O_m=\frac{1}{2}\epsilon^{abc}(d^{aT}C\sigma_{\mu\nu} d^b)\sigma_{\mu\nu}\gamma_5u^c$~\cite{Ioffe:1981kw}.
The operators $O_p$, $O_m$, $O_3$ and $O_4$ are covariant under the chiral transformation $q\to e^{i\alpha\gamma_5}q$.
As we mentioned in Sec.~\ref{sec:EDMext}, the estimator $\kappa^\text{est}(t_f)$ defined in Eq.~(\ref{eq:ratiokappa})
contains parity mixing and vacuum polarization effects induced by the background electric field.
It does not converge to the neutron anomalous magnetic moment and includes the overlap factors of the nucleon
interpolating operators with nucleon excited states, as shown in Eq.~(\ref{eq:kappa_est_bias}).
The numerical results for $\kappa^\text{est}(t_f)$ are shown in Fig.~\ref{fig:kappa_Ez2}, where data points with
different colors represent and display dependence of the results obtained using different nucleon interpolating
operators.
Not only are the results nucleon-operator dependent, but they also have pronounced excited state contamination visible as
the lack of a plateau for long source-sink separation. 

%------------------------------------------------------------------------------

In the following  subsections, we first present the nEDM results obtained using global topological charge.
These results are very noisy and have strong excited state contamination, making it challenging to obtain a good
statistical signal for the nEDM.
We then show the results extracted with local topological charge operator and compare the nEDM results obtained using
different interpolating operators.
Eventually, we combine these operators to solve a multi-operator GEVP to extract the ground-state nEDM matrix element.

%%%%%%%%%%%%%%%%%%%%%%%%%%%%%%%%%%%%%%%%%%%%%%%%%%%%%%%%%%%%%%%%%%%%%%%%%%%%%%%%
\subsection{nEDM results using global topological charge}
\begin{figure}[ht!]
\centering
\includegraphics[width=.48\textwidth]{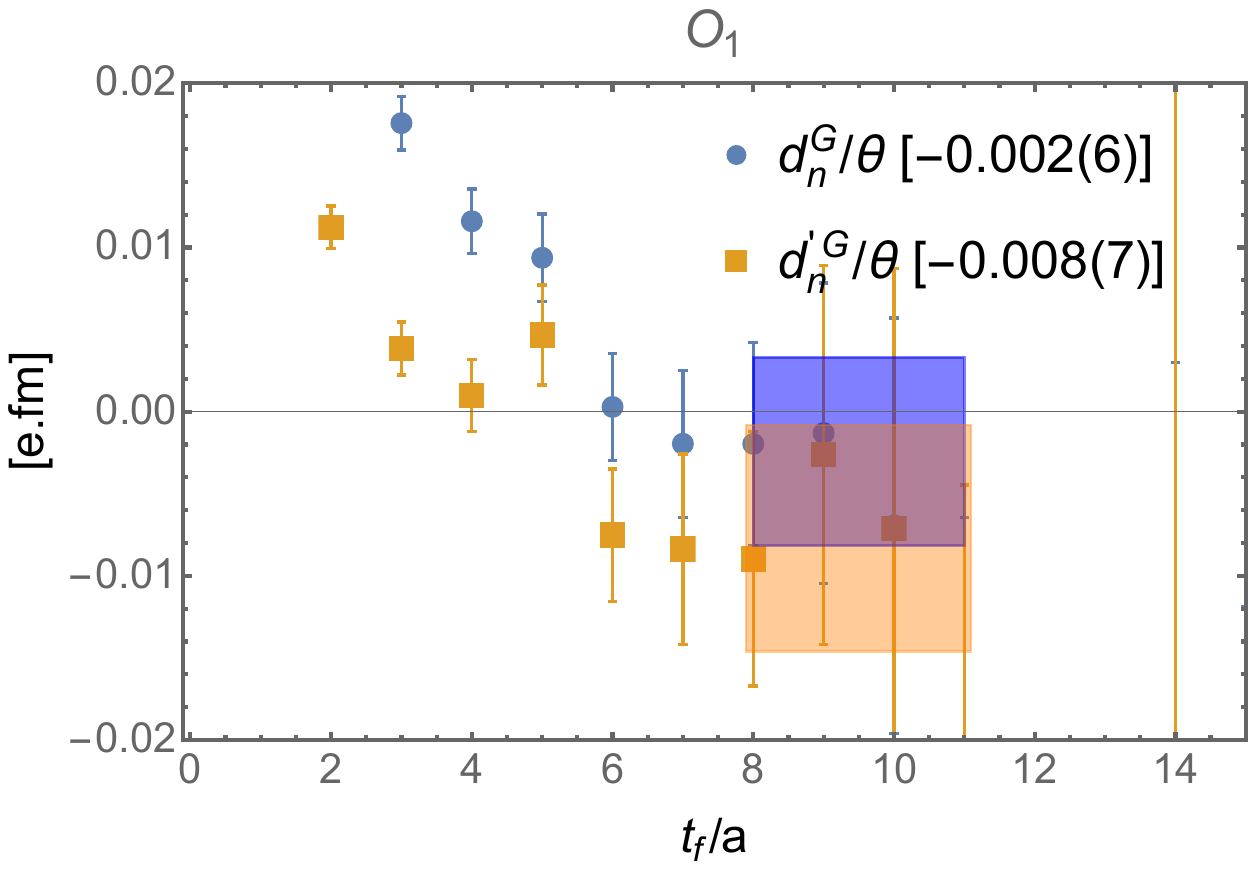}
\includegraphics[width=.48\textwidth]{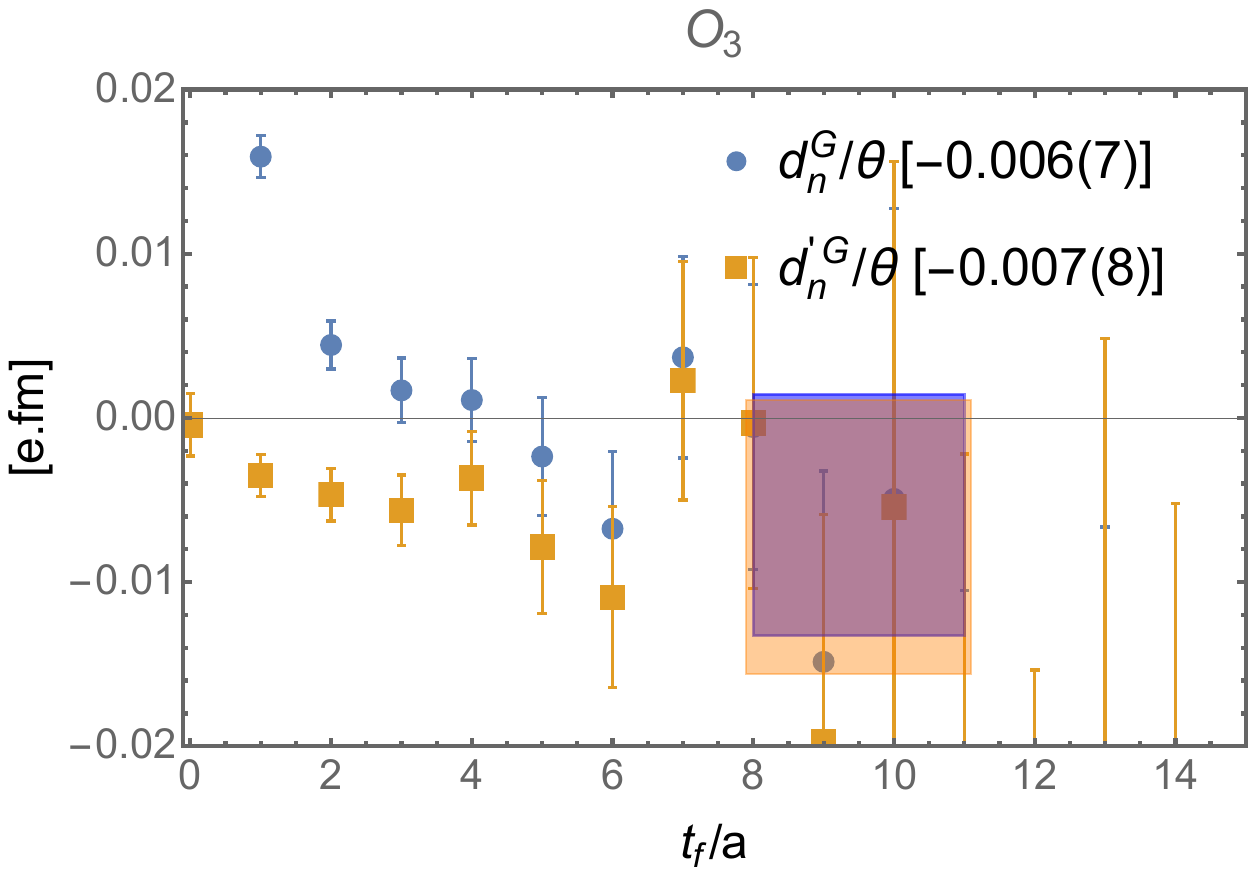}
\\
\includegraphics[width=.48\textwidth]{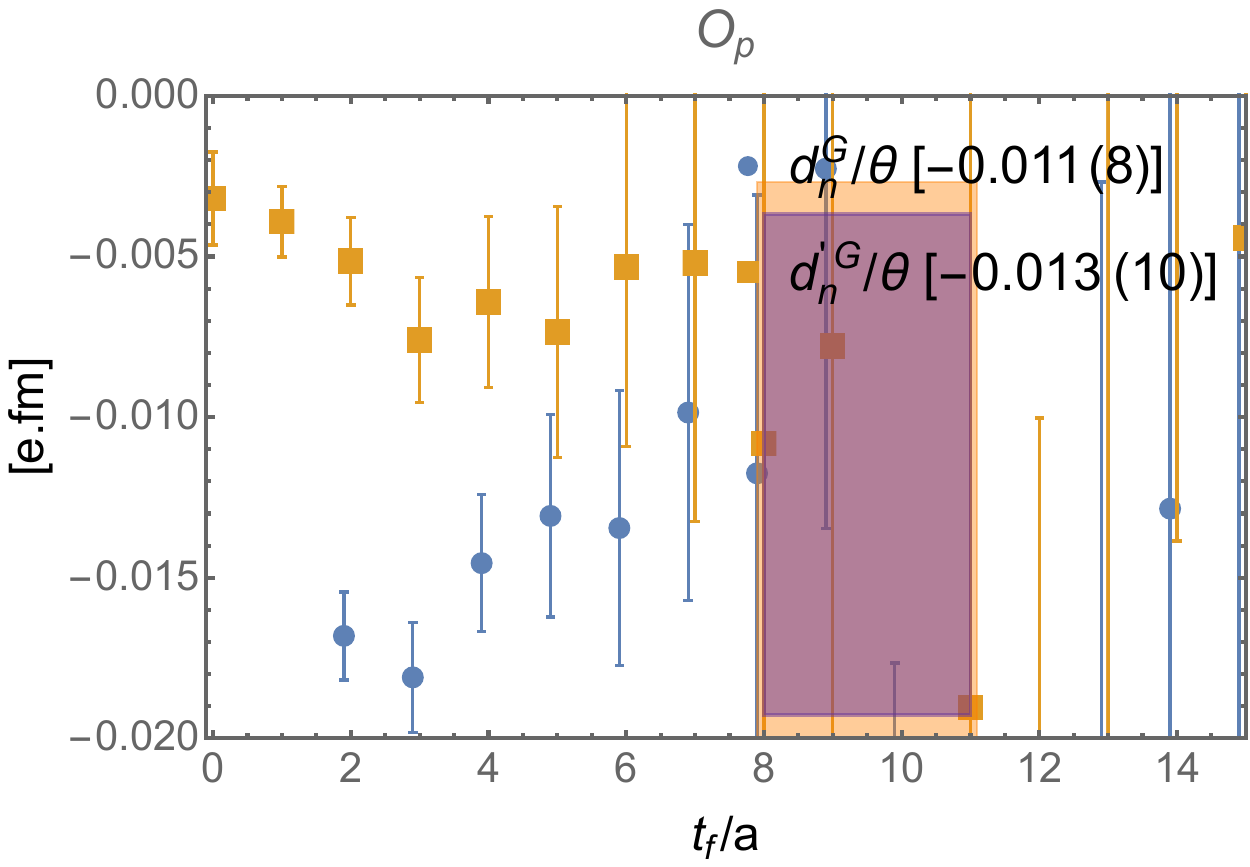}
\includegraphics[width=.48\textwidth]{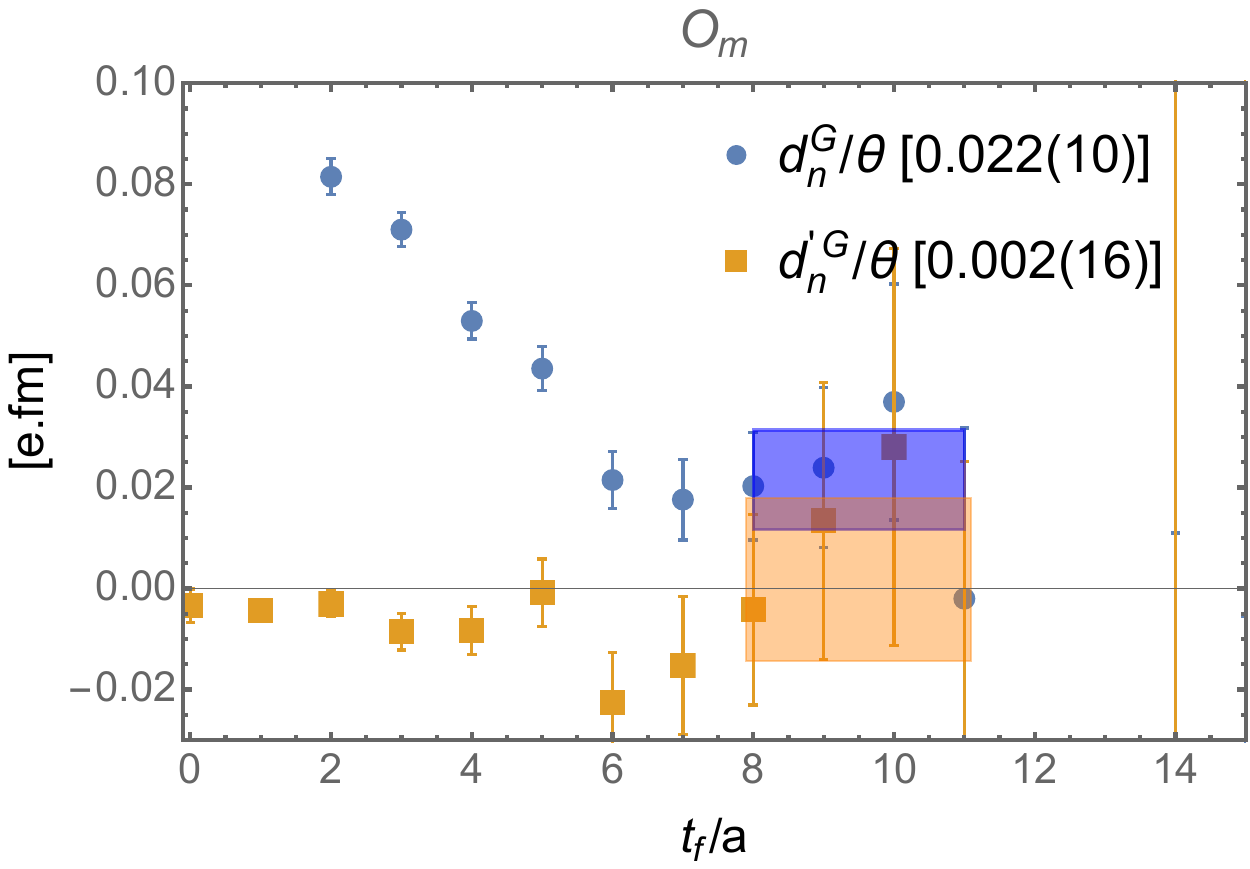}
\caption{Neutron EDM results obtained using global topological charge constructed from \emph{gluon field} and with
different nucleon interpolating operators.
The yellow and blue data points represent the results obtained using Eq.~(\ref{eqn:edm_estimator_sum}) and
Eq.~(\ref{eqn:edm_estimator_sum_g4xy}), respectively.
The bands are obtained using constant plateau fits with $8a \le t_f \le 11a$, and the fit results are given in
the square brackets.}
\label{fig:global_g}
\end{figure}

\begin{figure}[ht!]
\centering
\includegraphics[width=.48\textwidth]{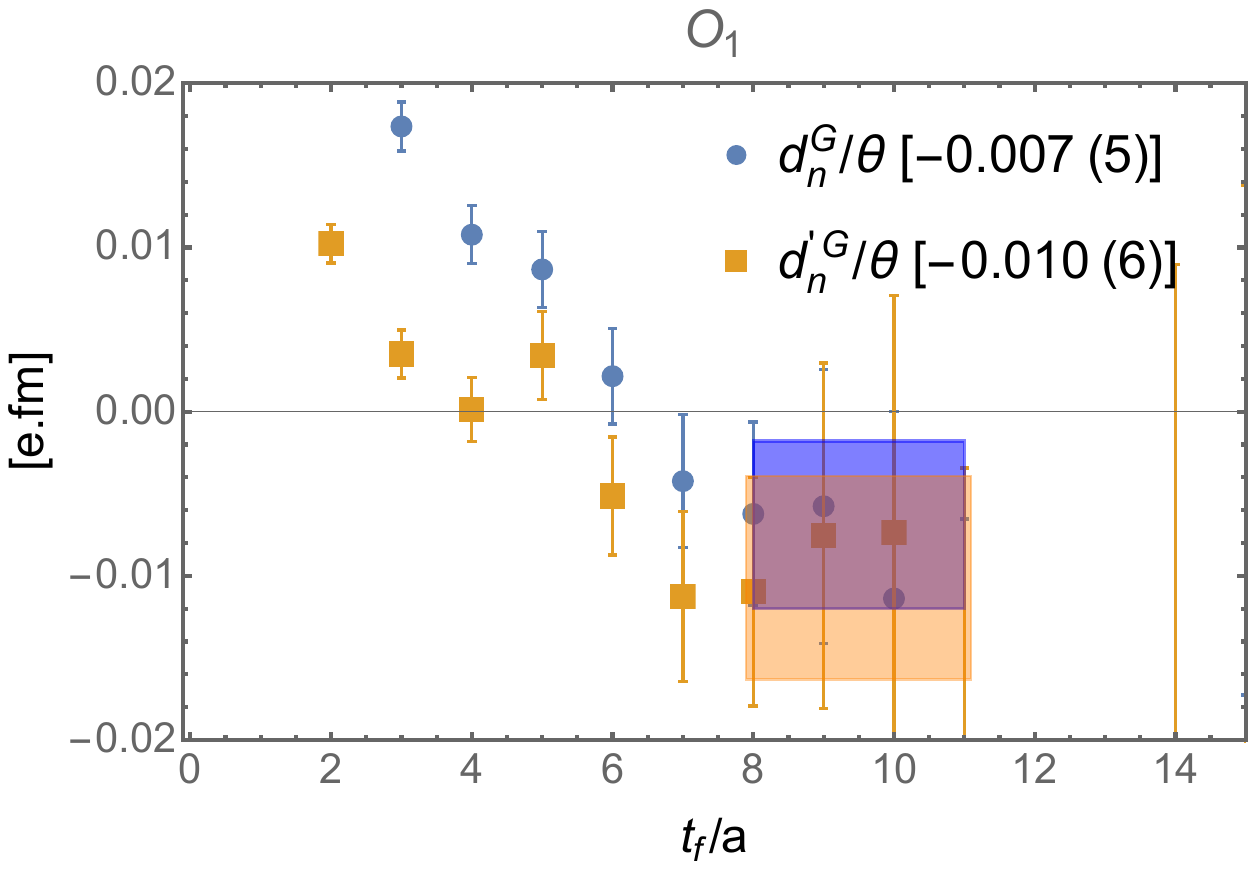}
\includegraphics[width=.48\textwidth]{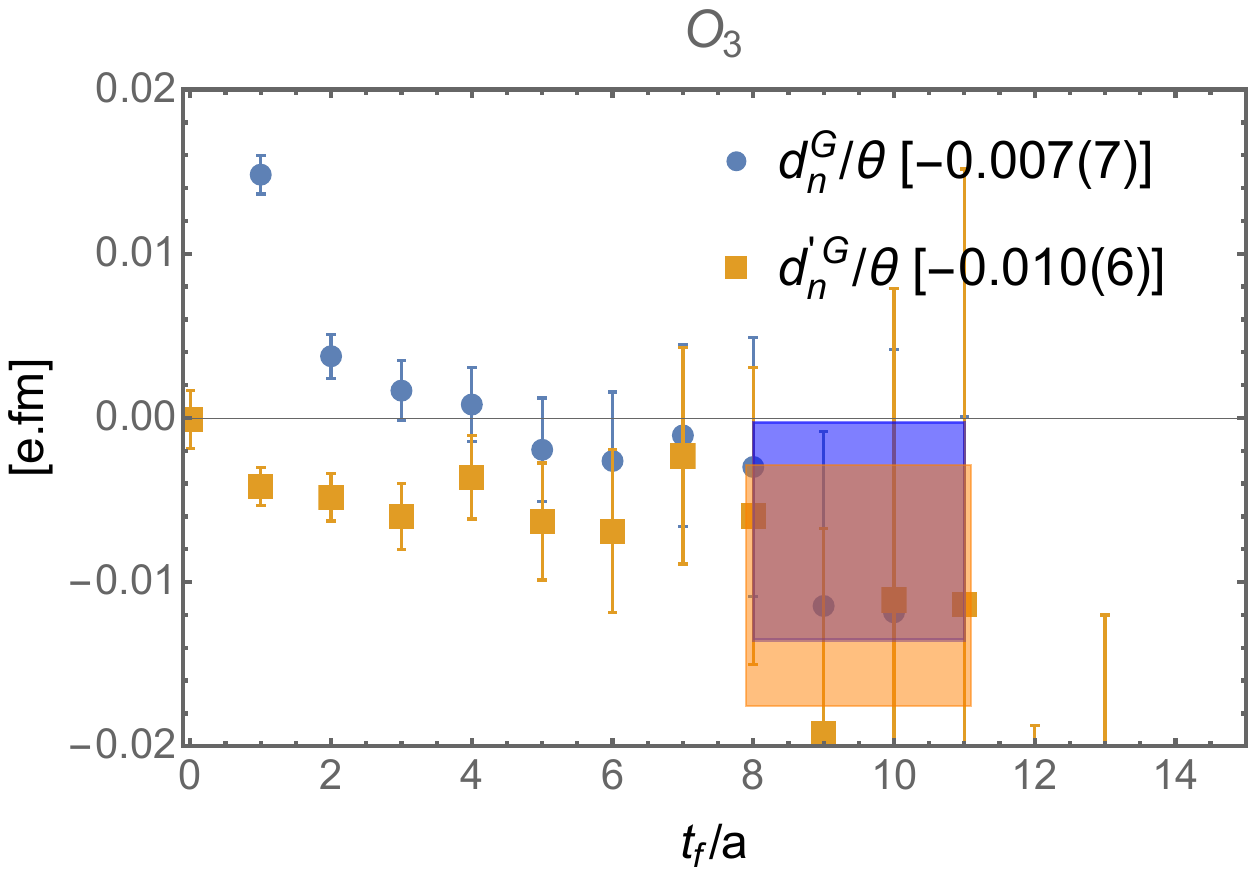}
\\
\includegraphics[width=.48\textwidth]{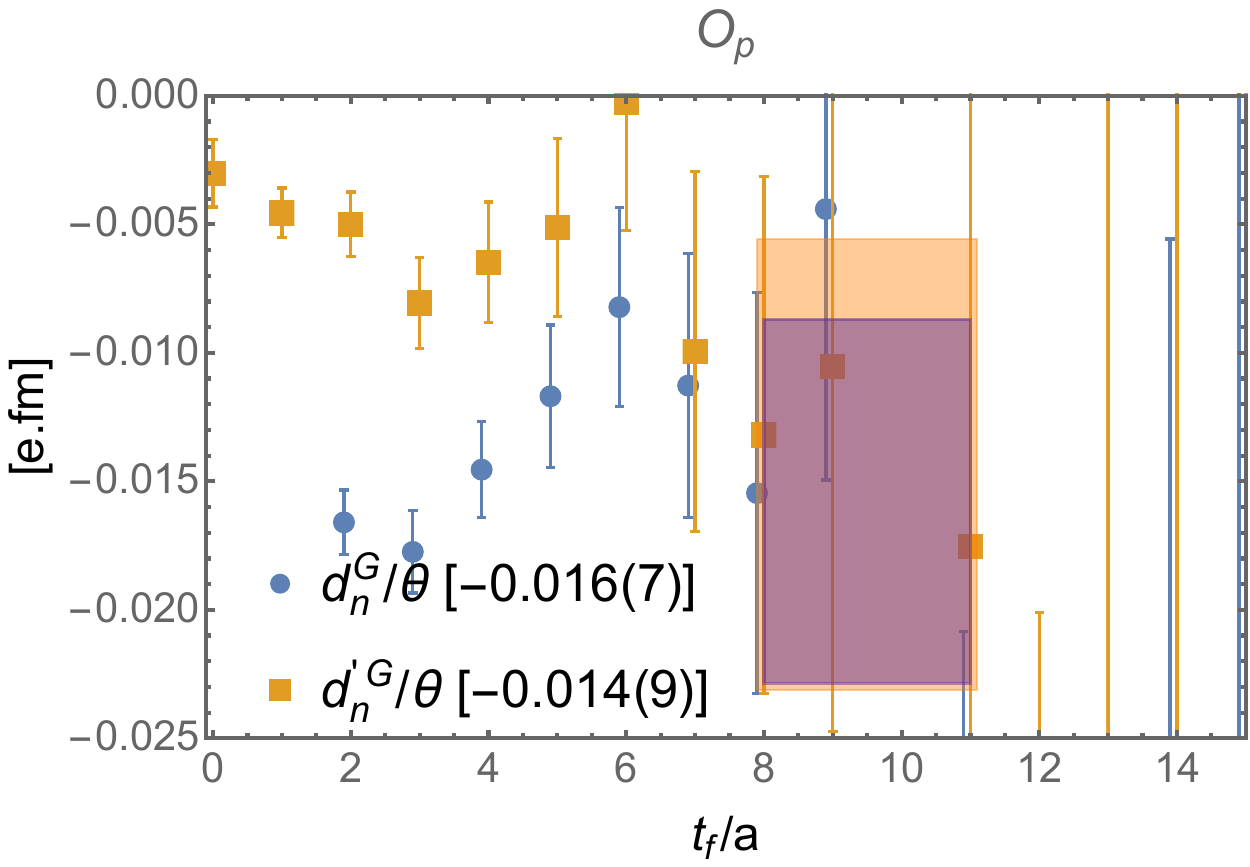}
\includegraphics[width=.48\textwidth]{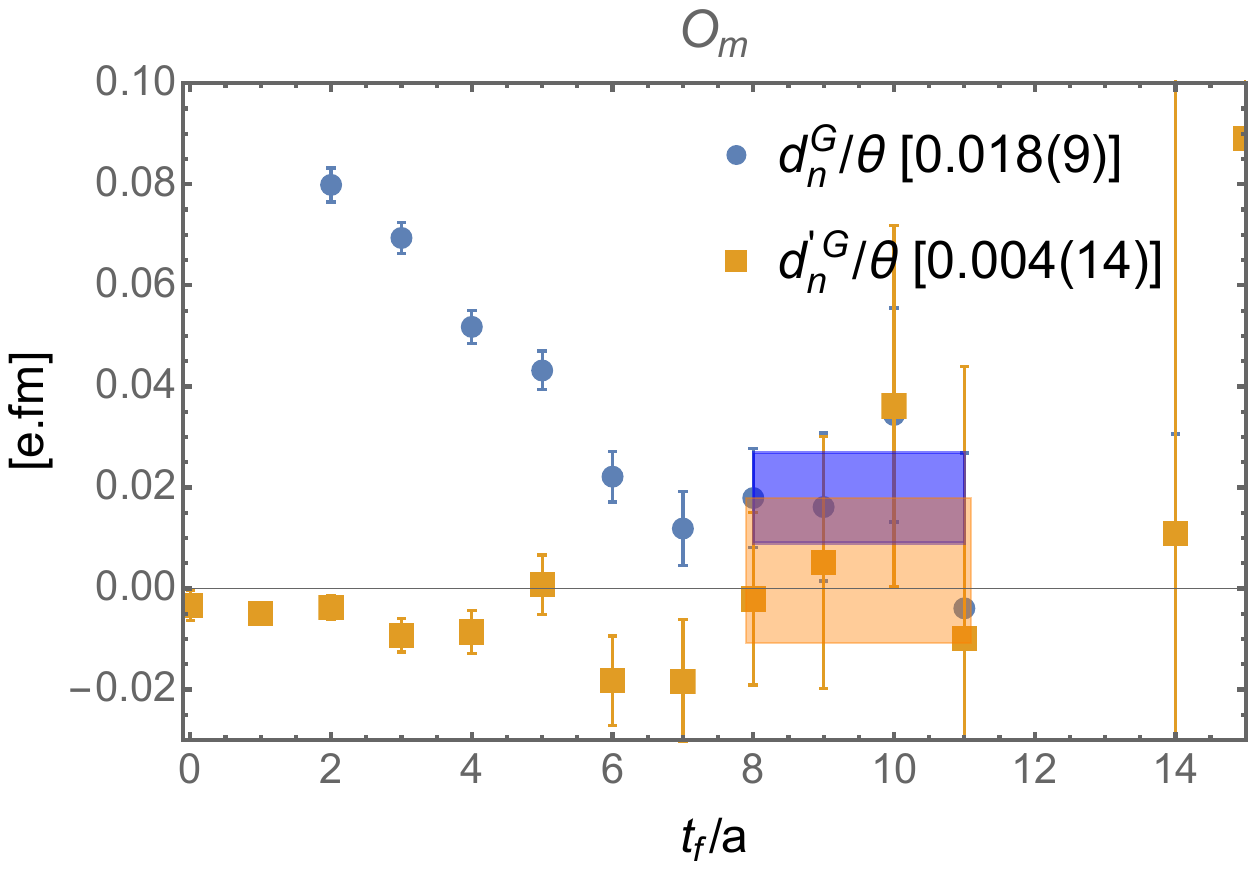}
\caption{Same as Fig.~\ref{fig:global_g} but the topological charge is defined using \emph{pseudoscalar quark density}
with quark mass $m_f=0.001$.}
\label{fig:global_q}
\end{figure}

In Section~\ref{sec:EDMext}, we introduced the extraction of the nEDM using the global topological charge.
The numerical results for $d^G_n/\theta$ and $d^{'G}_n/\theta$ defined in Eq.~(\ref{eqn:edm_estimator_sum}) and
Eq.~(\ref{eqn:edm_estimator_sum_g4xy}) are shown in Fig.~\ref{fig:global_g} and Fig.~\ref{fig:global_q} for the
topological charge constructed from the gluon field and from the pseudoscalar quark density, respectively.
The time derivatives are approximated by the differences of adjacent timeslices. 
The bands are obtained from constant fits to the data with source-sink separations ranging from $t_f=8a$ to $11a$.
The number within the square brackets in the figure is the fitted value. 
The results obtained using two different definitions of the topological charge are consistent with each other, but are
also consistent with zero in almost all cases.

Equations~(\ref{eqn:edm_estimator_sum}) and (\ref{eqn:edm_estimator_sum_g4xy}) have been derived assuming that the
$\CPV$ nucleon two-point correlation function in the background field $C_{2pt,\mcE,\theta}$ is due entirely to the ground state
contribution.
However, the difference between $d^G_n$ and $d^{'G}_n$ in the small $t_f$ region indicates the presence of excited state contamination. 
The results for $d^{'G}_n$ exhibit less dependence on $t_f$, indicating that it has less excited state contamination
compared to $d^G_n$.
As shown in Eq.~(\ref{eq:nume}), the numerator of Eq.~(\ref{eqn:edm_estimator_sum}) contains an
additional term proportional to $\alpha\kappa$.
Although this contribution can be removed by taking the time derivative, excited state terms can still contribute to 
the numerator of Eq.~(\ref{eqn:edm_estimator_sum}) if they are not negligible.
Further, the results obtained using different nucleon operators are noticeably different in the small-$t_f$ region.
At the same time, the statistical uncertainties become large at $t_f>6a$ making it challenging to extract the nEDM using the global
topological charge, and substantially higher statistics is needed to obtain reliable nEDM results with this method.

%------------------------------------------------------------------------------

%------------------------------------------------------------------------------

%%%%%%%%%%%%%%%%%%%%%%%%%%%%%%%%%%%%%%%%%%%%%%%%%%%%%%%%%%%%%%%%%%%%%%%%%%%%%%%%
\subsection{Nucleon eigenstates in the presence of Euclidean electric field}
\label{sec:NSE}
\begin{figure}[ht!]
\centering
\includegraphics[width=.96\textwidth]{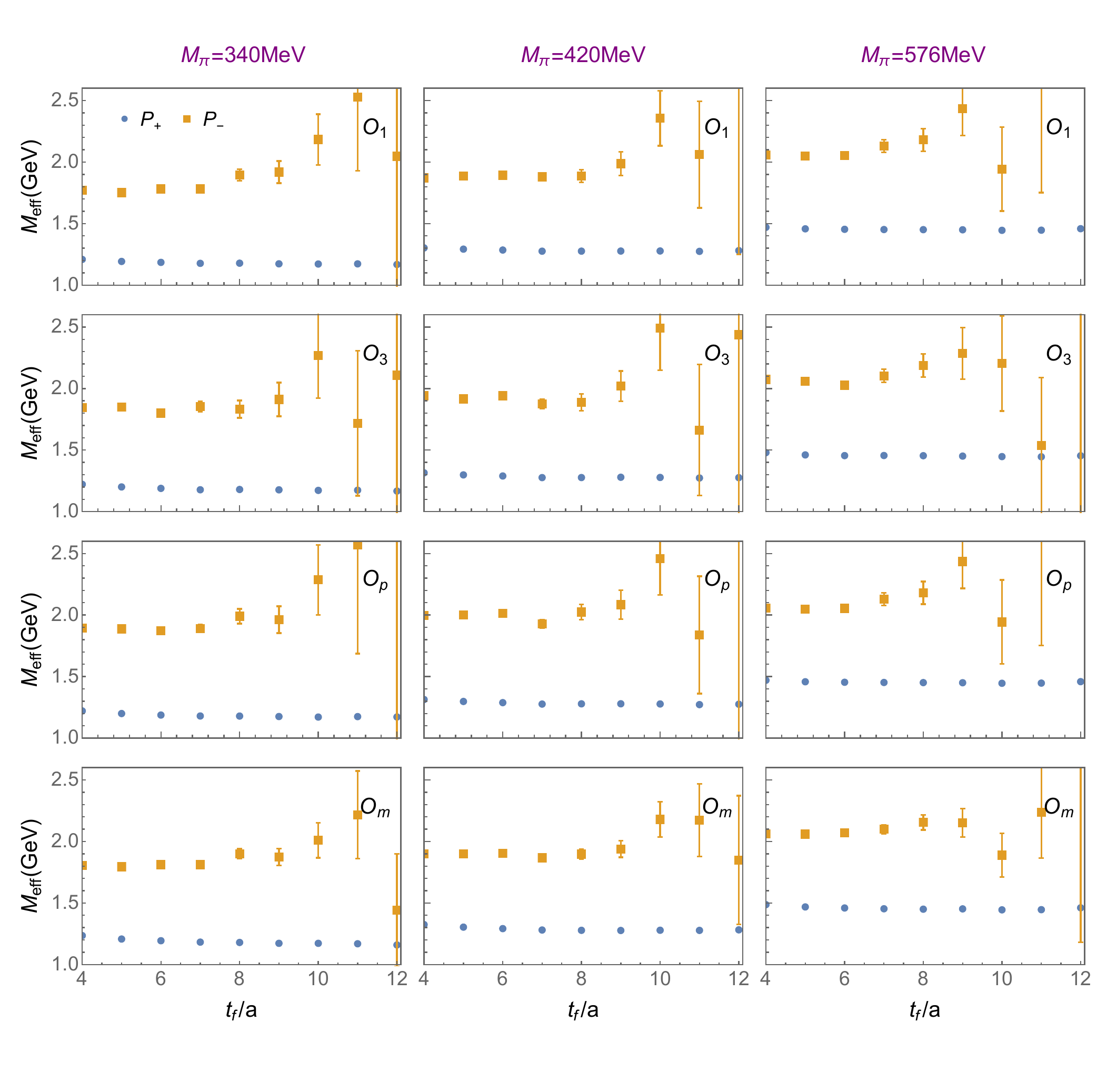}
\vspace{-12mm}
\caption{
  The effective mass of positive-parity ($P_+$) and negative-parity ($P_-$) nucleon correlator components, 
  computed with electric field strength $|n_z|=2$ and with different nucleon interpolating operators.
  Results are shown for ensembles 24I-005(left), 24I-010(middle), and 24I-020(right).
  \label{fig:ori_res_mass005Ez2}
}
\end{figure}

As Eqs.~(\ref{eq:NNstar}) show, correlators $\la\psi_N(t_f) \psi^{\dag}_{N}(0)\ra_{\mcE_z}$ and
$\la\psi^\uparrow_{N^*}(t_f) \psi^{\uparrow,\dag}_{{N^*}}(0)\ra_{\mcE_z}$ receive contributions from both
eigenstates $|\mcN_0\ra$ and $|\mcN_1\ra$ in the presence of an electric field.
More specifically, they can be expressed as
\begin{equation}
\label{eq:NN2pt}
\begin{aligned}
\la\psi^\uparrow_N(t_f) \psi^{\uparrow,\dag}_{N}(0)\ra_{\mcE_z}
  &= |Z_{N_0}|^2e^{-m_{\mcN_0}t_f}
\\&\quad-\mcE_z^2\left| \sum_{i}Z_{N_i}\frac{\la N_i|\mcD_z|N^*_0\ra}{m_{N^*_0}-m_{N_i}}
                       -\sum_{n} \frac{\la\Omega|\mcD_z|n^-\ra}{  E_{n}} Z^{n^-}_{N^*_0}\right|^2
        e^{-m_{\mcN_1}t_f} + O(\mcE_z^4)\,,
\\
\la\psi^\uparrow_{N^*}(t_f) \psi^{\uparrow,\dag}_{{N^*}}(0)\ra_{\mcE_z}
  &= |Z_{N^*_0}|^2e^{-m_{\mcN_1 }t_f}
\\&\quad-\mcE_z^2\left| \sum_{i}Z_{N^*_i}\frac{\la N^*_i|\mcD_z|N_0\ra}{m_{N_0}-m_{N^*_i}}
                       -\sum_{n} \frac{\la\Omega|\mcD_z|n^-\ra}{  E_{n}} Z^{n^-}_{N_0}\right|^2
      e^{-m_{\mcN_0}t_f} + O(\mcE_z^4)\,.
\end{aligned}
\end{equation}
The second terms in the these equations are proportional to $\mcE_z^2$. We find these effects are not apparent in  $\la\psi^\uparrow_N(t_f) \psi^{\uparrow,\dag}_{N}(0)\ra_{\mcE_z}$ but are visible in  $\la\psi^\uparrow_{N^*}(t_f) \psi^{\uparrow,\dag}_{{N^*}}(0)\ra_{\mcE_z}$ at large $t_f$ due to the mass gap
$m_{\mcN_1}-m_{\mcN_0}\approx m_{N^*_0}-m_{N_0} + O(\mcE_z^2)$.
To examine this parity breaking effect caused by the background field, we define the following effective masses for the correlators.
\begin{equation}
m_{P_+}(t_f) = \log\left[\frac{\la\psi^\uparrow_N(t_f) \psi^{\uparrow,\dag}_{N}(0)\ra_{\mcE_z}}
                              {\la\psi^\uparrow_N(t_f+1) \psi^{\uparrow,\dag}_{N}(0)\ra_{\mcE_z}}\right] \,,
m_{P_-}(t_f) = \log\left[\frac{\la\psi^\uparrow_{N^*}(t_f) \psi^{\uparrow,\dag}_{{N^*}}(0)\ra_{\mcE_z}}
                              {\la\psi^\uparrow_{N^*}(t_f+1) \psi^{\uparrow,\dag}_{{N^*}}(0)\ra_{\mcE_z}}\right] \,.
\end{equation}
The numerical results for $m_{P_+}(t_f)$ and $m_{P_-}(t_f)$ obtained using different nucleon interpolating operators on
three ensembles are shown in  Fig~\ref{fig:ori_res_mass005Ez2}.
These results are calculated with the strength of electric field fixed to $\mcE_z=6\pi |n_z|/(L_z L_t)$ with $|n_z|=2$,
and data with $+\mcE$ and $-\mcE$ are averaged to improve the signal.
The effective $m_{P_+}(t_f)$ remains stable as the source-sink separation $t_f$ increases, whereas $m_{P_-}(t_f)$ exhibits an 
upward trend. This behavior arises from the relative sign difference between the coefficients of the first and second terms in Eq.~(\ref{eq:NN2pt}). To illustrate this effect, in Appendix~\ref{sec:appenx_of} we show how these coefficients can be extracted and used to reconstruct the correlator $\la\psi^\uparrow_{N^*}(t_f) \psi^{\uparrow,\dag}_{{N^*}}(0)\ra_{\mcE_z}$ and the corresponding effective mass. The reconstructed effective mass reproduces the upward trend.
One expects that $m_{P_-}(t_f)$ approaches the ground-state mass in the sufficiently large $t_f$ limit. In practice, however, this behavior is difficult to observe because the signal to noise degrades rapidly with $t_f$.

\begin{figure}[ht!]
\centering
\includegraphics[width=.96\textwidth]{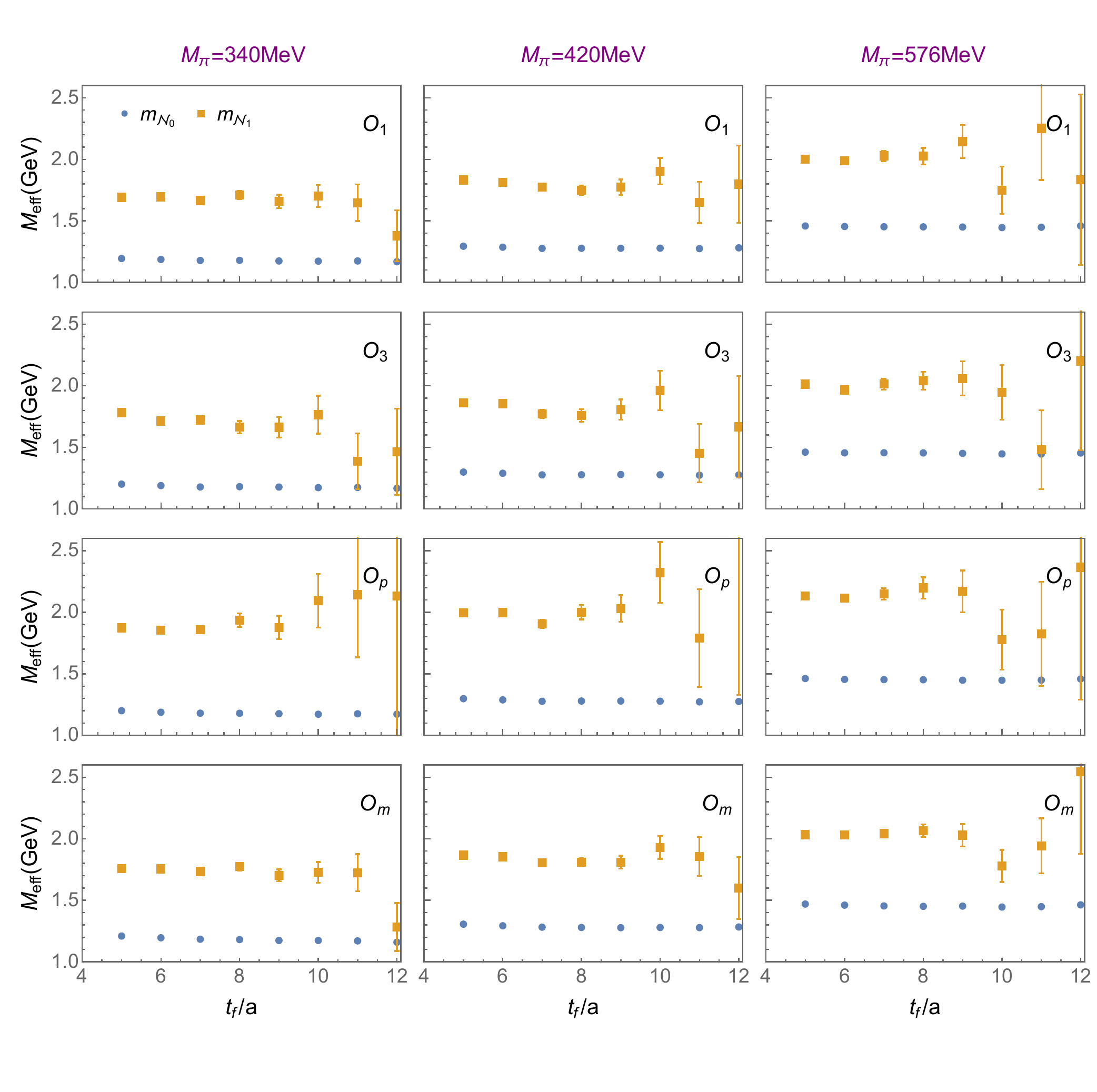}
\vspace{-12mm}
\caption{The ground and the first-excited $2\times2$ GEVP effective masses computed with electric field strength
  $|n_z|=2$ and with different nucleon interpolating operators.
  Results are shown for ensembles 24I-005(left), 24I-010(middle), and 24I-020(right).
  \label{fig:GEVP_res_mass005Ez2}
}
\end{figure}

To obtain the nucleon-like eigenstates in the presence of a background field, we solve the GEVP, 
Eq.~(\ref{eq:nonHGEVP}).
The effective mass for the eigenstates can be computed from the generalized eigenvalues,
\begin{equation}
m_{\mcN_{0,1}}(t_f)=\log\left[\frac{\lambda_{0,1}(t_f)}{\lambda_{0,1}(t_f+1)}\right],
\end{equation}
where $m_{\mcN_0}$ and $m_{\mcN_1}$ represent the effective mass of states $\mcN_0$ and $\mcN_1$, respectively.
$\lambda_0$ and $\lambda_1$ denote the ground state and first excited state eigenvalues obtained by solving the GEVP.
The numerical results of $m_{\mcN_0}(t_f)$ and $m_{\mcN_1}(t_f)$ are shown in Fig~.\ref{fig:GEVP_res_mass005Ez2}, where
$t_0=5a$ is used as the reference time.
The blue and yellow data points represent the effective mass for $|\mcN_0\ra$ and $|\mcN_1\ra$, respectively.
The ground state effective masses $m_{\mcN_0}$ do not show any $t$ dependence and agree for all four operators.
The effective masses of the $|\mcN_1\ra$ states are also largely consistent across the whole $t$ range.
We observe, however, that $m_{\mcN_1}$ obtained using operators $O_1$, $O_3$ and $O_m$ are consistent, while those using
operator $O_p$ are systematically higher by approximately $150\,\mathrm{MeV}$.
We interpret that as evidence that operator $O_p$ couples predominantly to the ``N(1650)'' state, whereas the other
operators couple predominantly with the lighter ``N(1535)'' state.

\begin{figure*}[t]
\centering
\includegraphics[width=.48\textwidth]{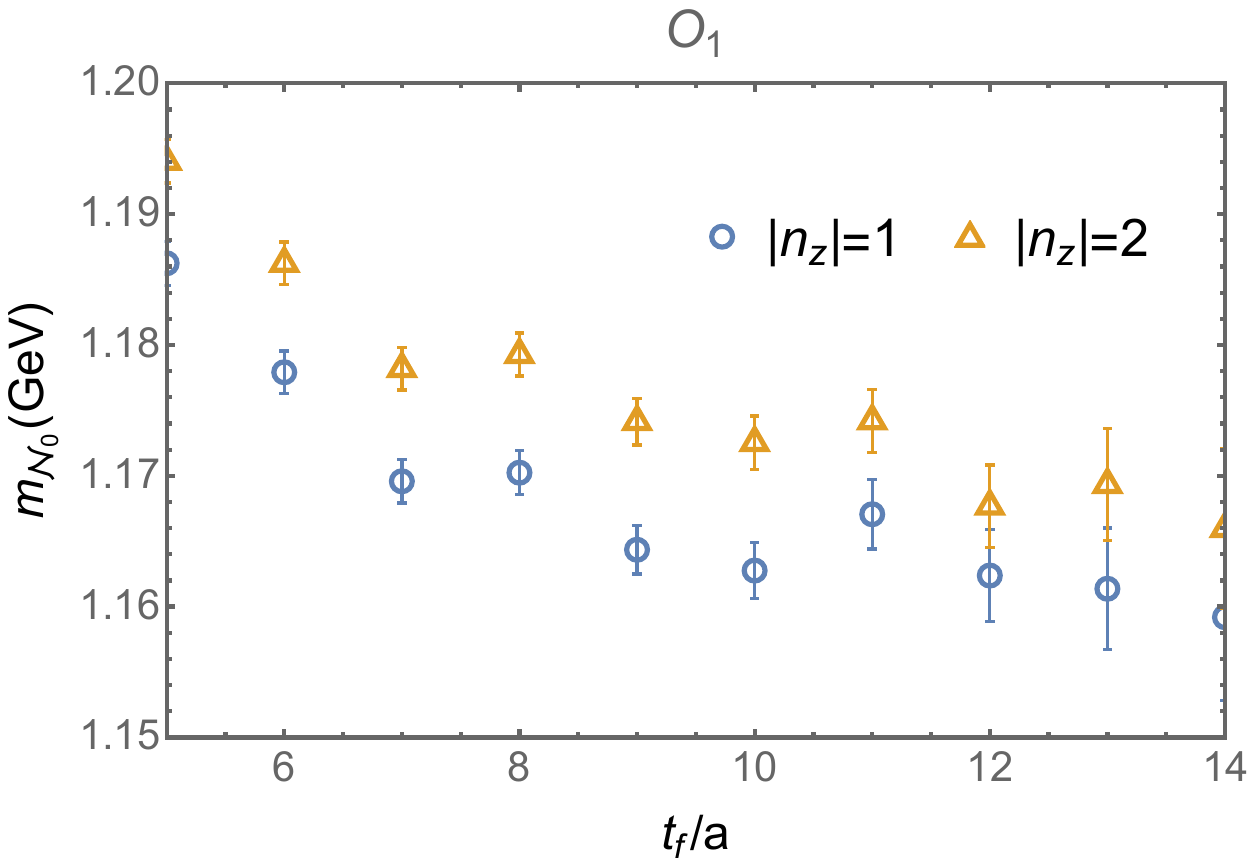}
\includegraphics[width=.48\textwidth]{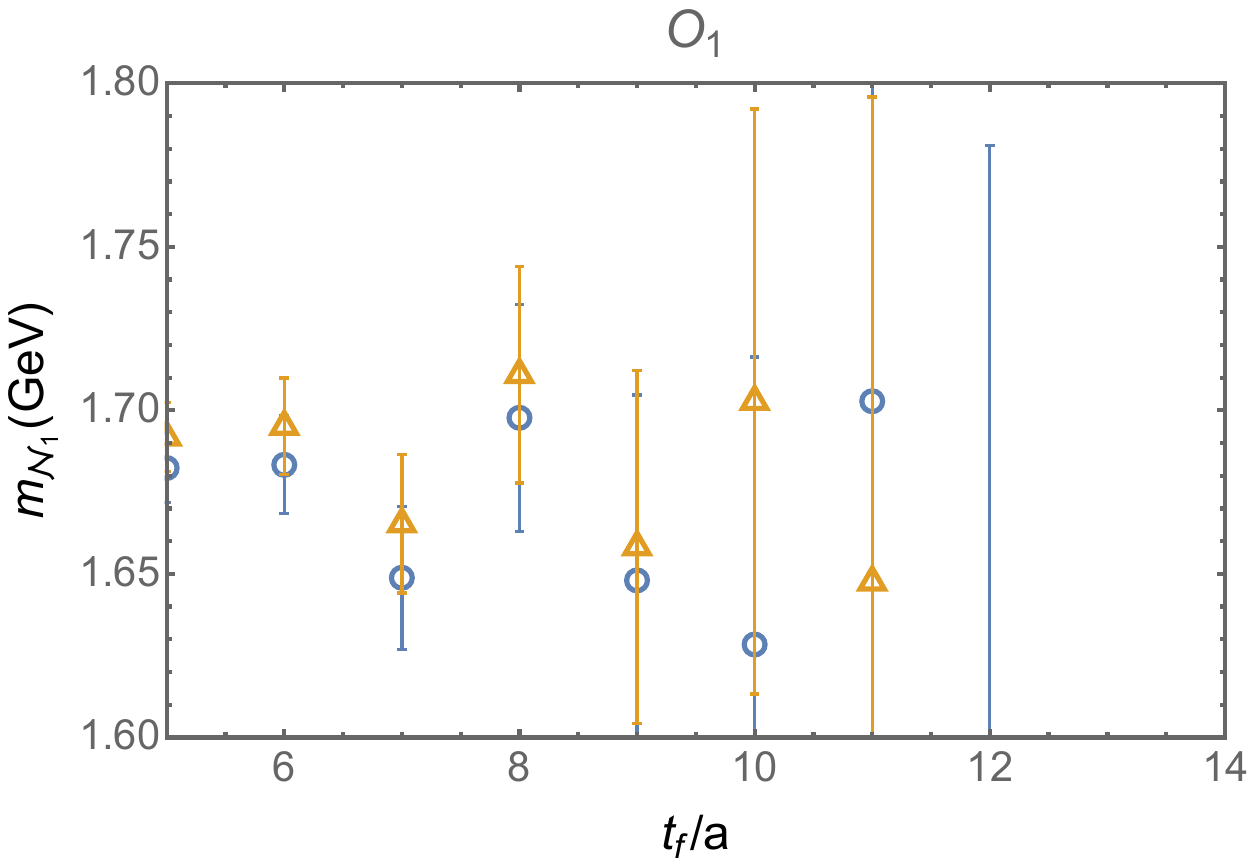}
\caption{The effective mass of the ground state (left panel) and the first excited state (right panel) with different electric field strength. Results are shown for ensemble 24I-005.}
\label{fig:effmass01}
\end{figure*}

In Fig.~\ref{fig:effmass01}, we present the effective masses of the ground state (left panel) and the first excited state (right panel) for different electric field strengths. One can see that the effective mass of ground state increases with increasing electric field strength, which is consistent with the expectation that the non-hermitian interaction leads to an attractive shift of the energy levels between parity partners, as demonstrated in Appendix.~\ref{sec:app_antiherm}. On the other hand, the effective mass of the first excited state is much noisier, making it difficult to resolve the dependence on the electric field strengths. Since the first excited state is predominately of negative parity, and it receives an attractive contribution not only from the ground state but also the higher positive-parity excited nucleon states. These contributions can partially cancel each other, rendering the corresponding energy level shift less pronounced. 
%------------------------------------------------------------------------------

%------------------------------------------------------------------------------

%%%%%%%%%%%%%%%%%%%%%%%%%%%%%%%%%%%%%%%%%%%%%%%%%%%%%%%%%%%%%%%%%%%%%%%%%%%%%%%%
\subsection{Excited state contamination}\label{sec:Numer_ESC}
\begin{figure}[ht!]
\centering
\includegraphics[width=.96\textwidth]{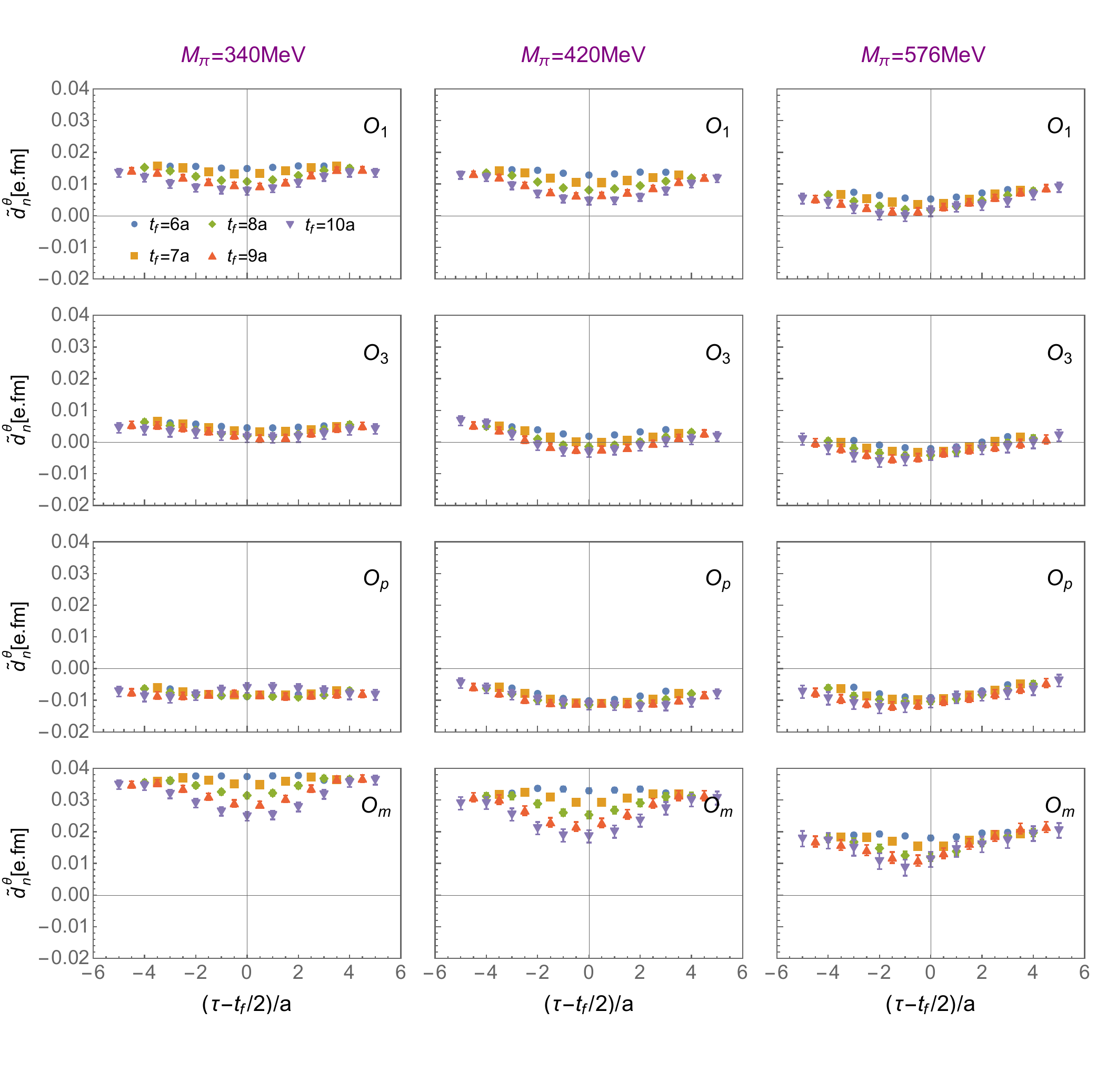}
\vspace{-12mm}
\caption{The EDM results obtained using definition $\tilde{d}_n^\theta(t_f,\tau)$ from Eq.~(\ref{eqn:edm_tilde}) on
  ensemble 24I-005(left), 24I-010(middle), and 24I-020(right) with electric field strength $|n_z|=2$. 
  The topological charge density is constructed from gluon field at gradient flow time $t_{gf}=4a^2$.
  \label{fig:Ori_res_mass005Ez2_igf4}
}
\end{figure}

Using ground-state nucleon GEVP vectors, we use Eq.~(\ref{eqn:edm_bgem_GEVP}) to calculate the matrix element for the nEDM.
To show the necessity of the GEVP method, we compare nEDM results obtained using Eq.~(\ref{eqn:edm_bgem_GEVP}) with those
obtained using the previous method defined in~\cite{Izubuchi:2020ngl,He:2023gwp},
\begin{equation}
\label{eqn:edm_tilde}
\tilde{d}_n^\theta(t_f,\tau)  = \frac{1}{\mcE_z} 
  \frac{\Tr \big[T^+_{S_z} \la N(t_f) \, q(\tau) \, \bar N(0)\ra_{\mcE_z}\big]}
       {\Tr \big[T^+       \la N(t_f) \, \bar N(0)\ra_{\mcE_z}\big]}
  \xRightarrow{t_f,\tau,(t_f-\tau)\rightarrow\infty}d_n/\bar{\theta}.
\end{equation}
which is inspired by Eq.~(\ref{eqn:edm_estimator_sum}) but probes a single-time slice of the topological charge.
This ratio should converge to the neutron EDM in the large source-sink separation limit.
However, as shown in Eq.~(\ref{eq:nume}), it contains the contamination term $\propto\alpha\kappa$.
This term vanishes in the $t_f$-derivative of the nucleon correlation function with global topological charge,
but it is not guaranteed to vanish when the local topological charge is used, particularly if  excited state
contamination is present.
One can instead consider extracting the nEDM using the following ratio that is inspired by
Eq.~(\ref{eqn:edm_estimator_sum_g4xy}) that does not contain the $\alpha\kappa$ term and is directly related to $d_n$,
\begin{equation}
\label{eqn:edm_bar}
\bar{d}_n^\theta(t_f,\tau)  = \frac{1}{\mcE_z} 
  \frac{\Tr \big[\gamma_4\Sigma_z \la N(t_f) \, q(\tau) \, \bar N(0)\ra_{\mcE_z}\big]}
       {\Tr \big[T^+       \la N(t_f) \, \bar N(0)\ra_{\mcE_z}\big]}
       \xRightarrow{t_f,\tau,(t_f-\tau)\rightarrow\infty}d_n/\bar{\theta}.
\end{equation}
In Figure~\ref{fig:Ori_res_mass005Ez2_igf4} we show the results of $\tilde{d}_n^\theta(t_f,\tau)$ obtained on the three
ensembles using the gluon topological charge operator at fixed gradient flow time $t_{gf}=4a^2$ and source-sink
separations $6a \le t_f \le 10a$, computed with $|n_z|=2$ electric field.
We combine the nEDM correlators with $n_z=\pm2$ to enhance the signal because they should be proportional to $\mcE_z$.
We observe significant discrepancies between the nEDM results obtained with different nucleon interpolating operators.
In particular, the results obtained using operator $O_m$ are significantly larger than those obtained from other nucleon
interpolating operators, while the result from $O_p$ has the opposite sign. 

\begin{figure}[ht!]
\centering
\includegraphics[width=.96\textwidth]{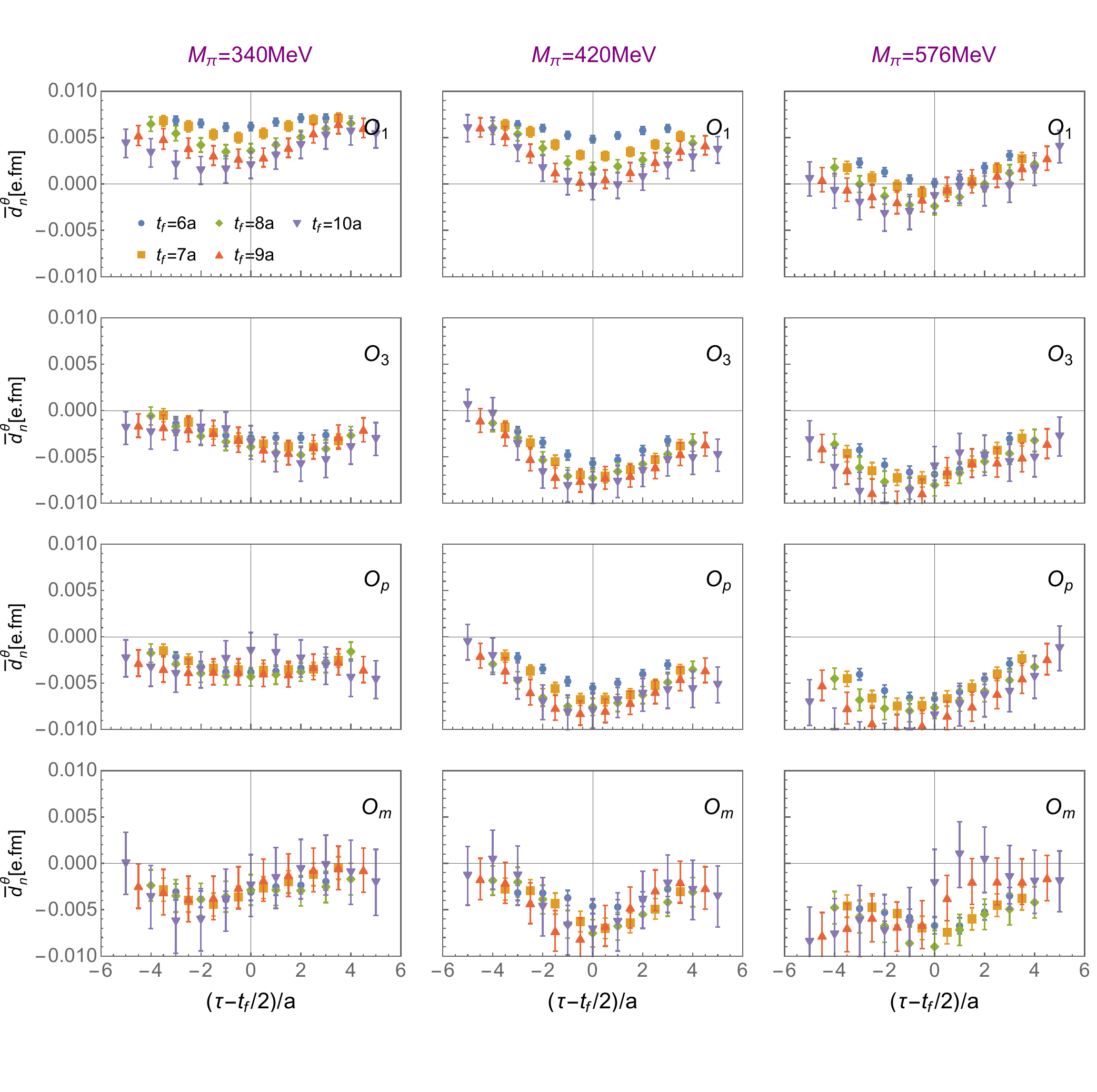}
\vspace{-12mm}
\caption{The EDM results obtained using definition $\bar{d}_n^\theta(t_f,\tau)$ from Eq.~(\ref{eqn:edm_bar}) on
  ensembles 24I-005(left), 24I-010(middle) and 24I-020(right) with electric field strength $|n_z|=2$.
  The topological charge density is constructed from gluon field at gradient flow time $t_{gf}=4a^2$.
  \label{fig:Ori_g4xy_res_mass005Ez2_igf4}
}
\end{figure}

Next, in Figure~\ref{fig:Ori_g4xy_res_mass005Ez2_igf4} we show the results of $\bar{d}^\theta_n(t_f,\tau)$
computed with the same method and parameters.
Comparing with the $\tilde{d}^\theta_n$ results, we observe that the discrepancies between different nucleon
interpolating operators are significantly reduced, indicating that $\bar{d}^\theta_n$ contains less excited state
contamination.
The results obtained using the chiral-covariant operators $O_{3,p,m}$ are close to each other, 
whereas the results from the non-covariant operator $O_1$ have the opposite sign.
It also shows large excited-state contamination, with a sign change appearing as the source-sink separation increases at
the heavier pion mass.

%------------------------------------------------------------------------------

%------------------------------------------------------------------------------

As discussed in Sec~\ref{sec:EDMext}, the nEDM should be extracted using GEVP vectors defined in
Eq.~(\ref{eqn:edm_bgem_GEVP}), which minimizes the excited state contamination by isolating the ground state nucleon in
the presence of the background electric field.
The estimator for the EDM based on GEVP method is 
\begin{equation}
\label{eqn:edm_bgem_GEVP2}
d_n^\theta(t_f,\tau)
  = \frac{1}{\mcE_z} 
    \frac {v^\dag_{L,0}C^{\psi\psi^\dag,\uparrow}_{3pt,\mcE}(t_f,\tau)v_{R,0}}       
          {v^\dag_{L,0}C^{\psi\psi^\dag,\uparrow}_{2pt,\mcE}(t_f)v_{R,0}}
  \xRightarrow{t_f,\tau,(t_f-\tau)\rightarrow\infty}d_n/\bar{\theta}.
\end{equation}

\begin{figure}[ht!]
\centering
\includegraphics[width=.96\textwidth]{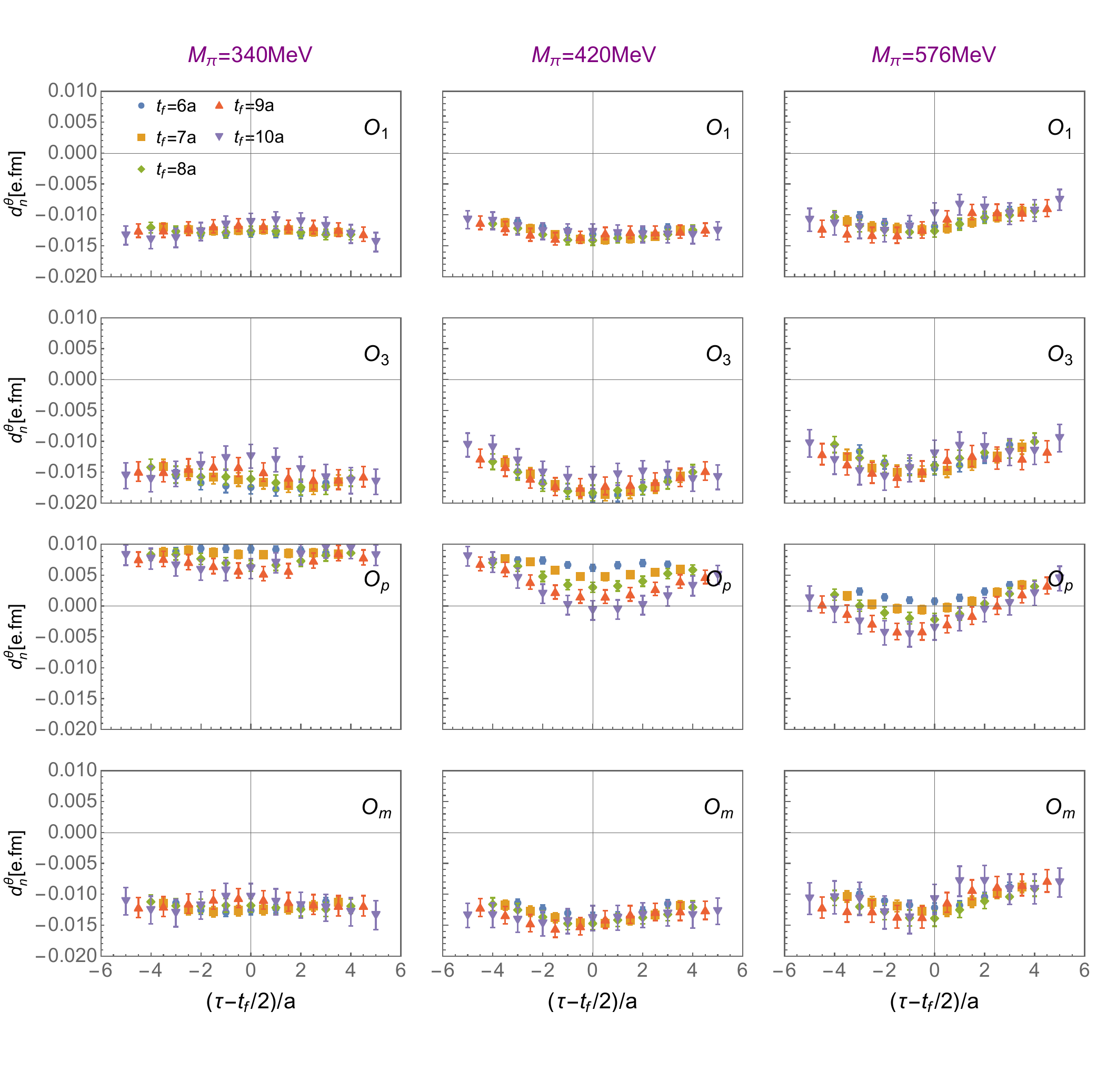}
\vspace{-12mm}
\caption{The EDM results obtained using GEVP method~(\ref{eqn:edm_bgem_GEVP2}) on ensembles
  24I-005(left), 24I-010(middle), and 24I-020(right) with electric field strength $|n_z|=2$. 
  The topological charge density is constructed from gluon field at gradient flow time $t_{gf}=4a^2$.
  \label{fig:GEVP_g4xy_res_mass005Ez2_igf4}
}
\end{figure}

The $d_n^\theta(t_f,\tau)$ results obtained using a $2\times2$ GEVP applied to positive and negative projections of
different interpolating operators are shown in Fig.~\ref{fig:GEVP_g4xy_res_mass005Ez2_igf4}.
These nEDM results are consistent for the GEVP based on operators $O_1$, $O_m$, and $O_3$ but disagree with those for
operator $O_p$.
The likely explanation is that operator $O_p$ couples to different excited states compared to operators $O_{1,3,m}$.
Indeed, as shown in the previous section, the effective mass of the first excited state $\mcN_1$ extracted using the  
$2\times2$ GEVP  with operators $(1\pm\gamma_4)O_p$ is higher than that from other nucleon operators.
This suggests that the contribution of the transition matrix element between the ground state and the first excited
state may be not reliably removed in the GEVP analysis of the correlators constructed with $O_p$.
Although nEDM results from all the operators would likely converge in the large-$t_f$ limit, we conclude that in
practice the operator $O_p$ is less suitable for nEDM calculation even though it is covariant under a chiral rotation~\cite{Ema:2024vfn}. 

It is also peculiar that the results extracted using different
methods in Eqs.~(\ref{eqn:edm_tilde},\ref{eqn:edm_bar}\ref{eqn:edm_bgem_GEVP2}) differ significantly.
In particular, the results obtained from $\tilde{d}_n^\theta$ and $d_n^\theta$ yield results with different sign for all interpolating operators at smaller pion mass. For the operator ${O_{1,3,m}}$, as the pion mass increases, $\tilde{d}_n^\theta$ gradually approaches the negative region and becomes closer to the results for $d_n^\theta$, indicating that the excited state contamination decreases more rapidly at heavier pion mass.
Notably, for the most widely used interpolating operator $O_1$, which is not covariant under chiral rotations, the results $\tilde{d}_n^\theta$, $\bar{d}_n^\theta$ and $d_n^\theta$ differ significantly, which indicates that GEVP method is necessary to reliably obtain the ground-state nEDM.

%------------------------------------------------------------------------------

\begin{figure}[ht!]
\centering
\includegraphics[width=.96\textwidth]{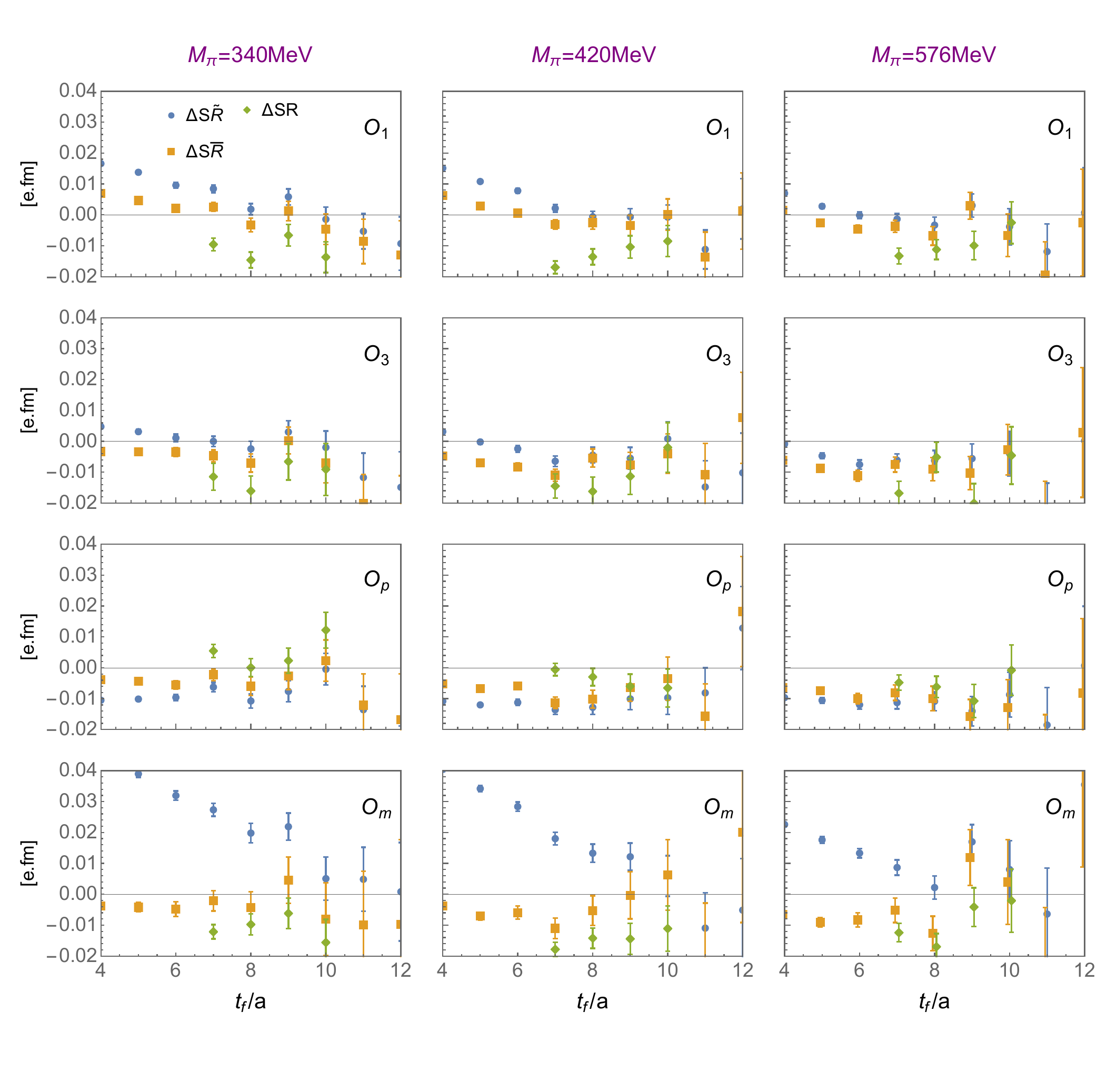}
\vspace{-12mm}
\caption{The comparison of differential ratios $\Delta S\tilde{R}(t_f)$, $\Delta S\bar{R}(t_f)$ and $\Delta SR(t_f)$.}
\label{fig:DeltaR}
\end{figure}

To compare results obtained from different nEDM estimators, one can also sum the ratios of three- and two-point
correlation functions defined in Eqs.~(\ref{eqn:edm_tilde},\ref{eqn:edm_bar},\ref{eqn:edm_bgem_GEVP2}) over the
timeslices of the topological charge between source and sink,
\begin{align}
S\tilde{R}(t_f)&=\frac{1}{\mcE_z} 
  \sum_{\tau=1}^{t_f-1}\frac{\Tr \big[T^+_{S_z} \la N(t_f) \, q(\tau) \, \bar N(0)\ra_{\mcE_z}\big]}
                            {\Tr \big[T^+       \la N(t_f) \, \bar N(0)\ra_{\mcE_z}\big]}\,,
\\
S\bar{R}(t_f)&=\frac{1}{\mcE_z} 
  \sum_{\tau=1}^{t_f-1}\frac{\Tr \big[\gamma_4\Sigma_z \la N(t_f) \, q(\tau) \, \bar N(0)\ra_{\mcE_z}\big]}
                            {\Tr \big[T^+       \la N(t_f) \, \bar N(0)\ra_{\mcE_z}\big]}\,,
\\
SR(t_f)&=\frac{1}{\mcE_z} 
  \sum_{\tau=1}^{t_f-1}\frac{ v^\dag_{L,0}C^{\psi\psi^\dag,\uparrow}_{3pt,\mcE}(t_f,\tau)v_{R,0}}       
                            { v^\dag_{L,0}C^{\psi\psi^\dag,\uparrow}_{2pt,\mcE}(t_f)v_{R,0}}\,,
\end{align}
and define the differential ratios, for example,
\begin{equation}
\Delta SR(t_f) = SR(t_f) - SR(t_f-1)\,.
\end{equation}
When the source-sink separation is sufficiently large, the differential ratios should be saturated with  the ground-state nEDM. 
The results for these differential ratios are shown in Fig~\ref{fig:DeltaR}.
The blue data points are obtained using $\Delta S\tilde{R}(t_f)$, which exhibits strong excited state contamination
similar to that observed for $\tilde{d}_n^\theta$.
A sign change around $t_f=10a$ is observed for the results obtained using $O_1$.
The orange data points correspond to the results $\Delta S\bar{R}(t_f)$, which is constructed using $\bar{d}_n^\theta$.
For operator $O_1$, the results for $\Delta S\bar{R}(t_f)$ change sign earlier than those for $\Delta S\tilde{R}(t_f)$,
and they are negative at small source-sink separations for other operators.
As the pion mass increases, the sign change in both $\Delta S\tilde{R}(t_f)$ and $\Delta S\bar{R}(t_f)$ occurs at
smaller values of $t_f$. 
The green data points represent the differential ratios $\Delta SR(t_f)$ constructed using the GEVP results
for $t_f\ge7a$ because the reference time is fixed at $t_0=5a$.
One can see that these results show good convergence for a range of $t_f$ except for those obtained using
$O_p$, and they approach the ground-state matrix element faster than the other two definitions.
The results from $O_p$ are positive at the lightest pion mass but become negative at the heavier pion mass.
Overall, the differences among the results of $\Delta S\tilde{R}(t_f)$, $\Delta S\bar{R}(t_f)$ and $\Delta SR(t_f)$
become smaller as the pion mass increases.

%------------------------------------------------------------------------------

\begin{figure}[ht!]
\centering
\includegraphics[width=.32\textwidth]{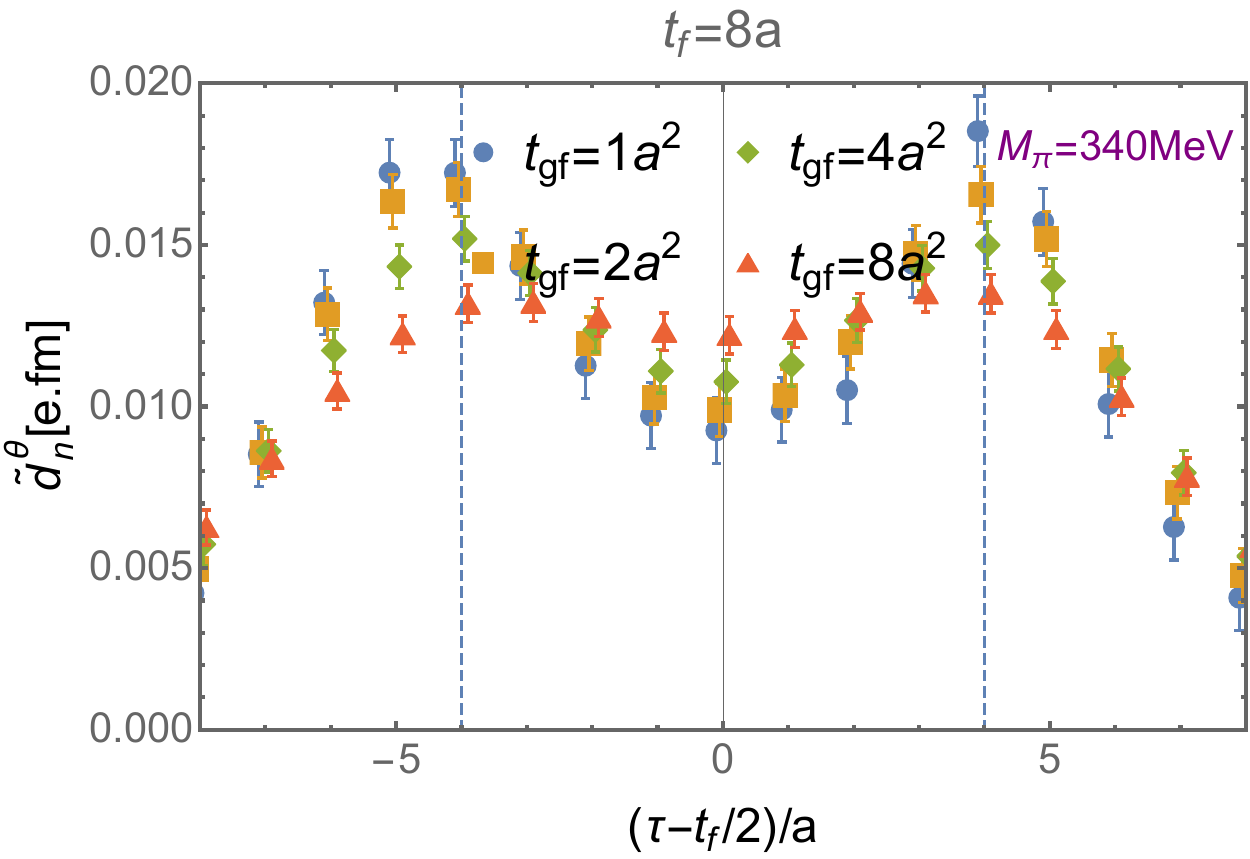}
\includegraphics[width=.32\textwidth]{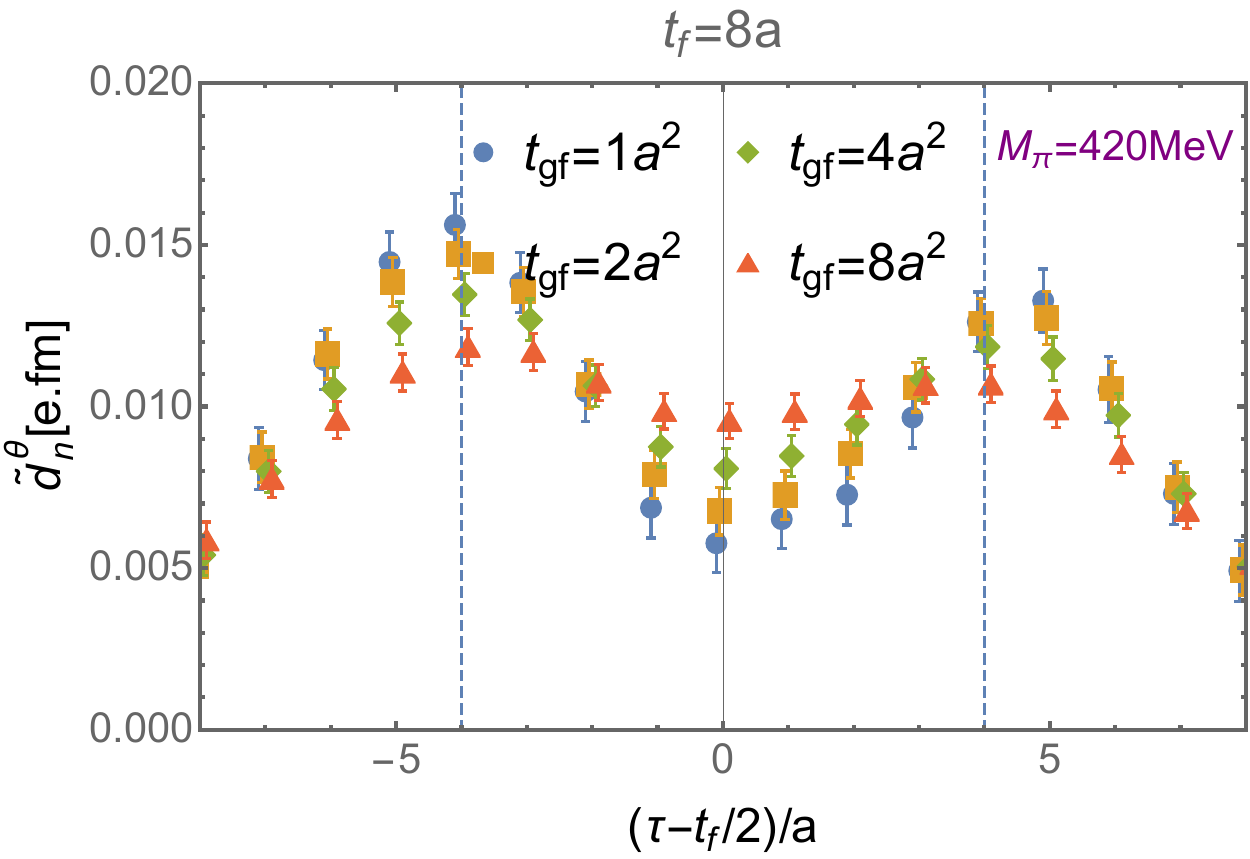}
\includegraphics[width=.32\textwidth]{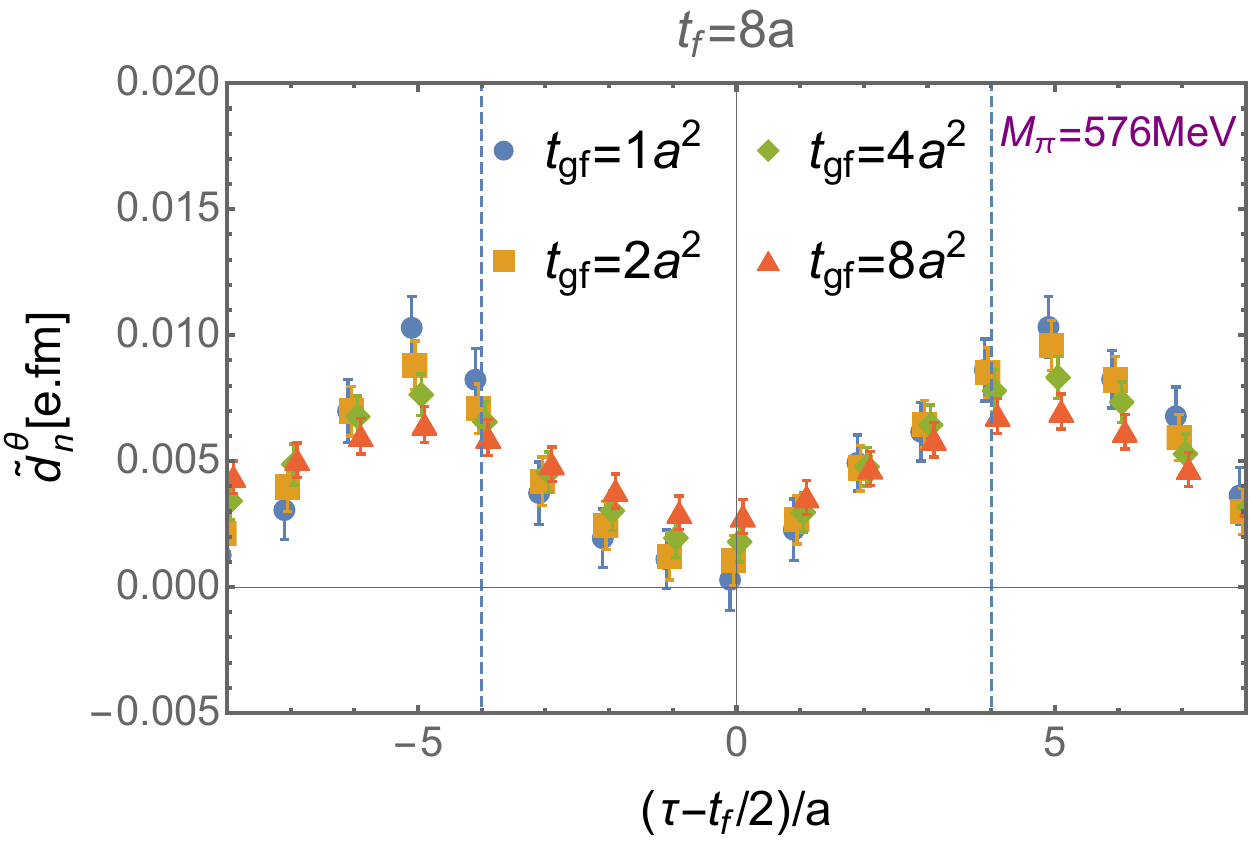}
\\
\includegraphics[width=.32\textwidth]{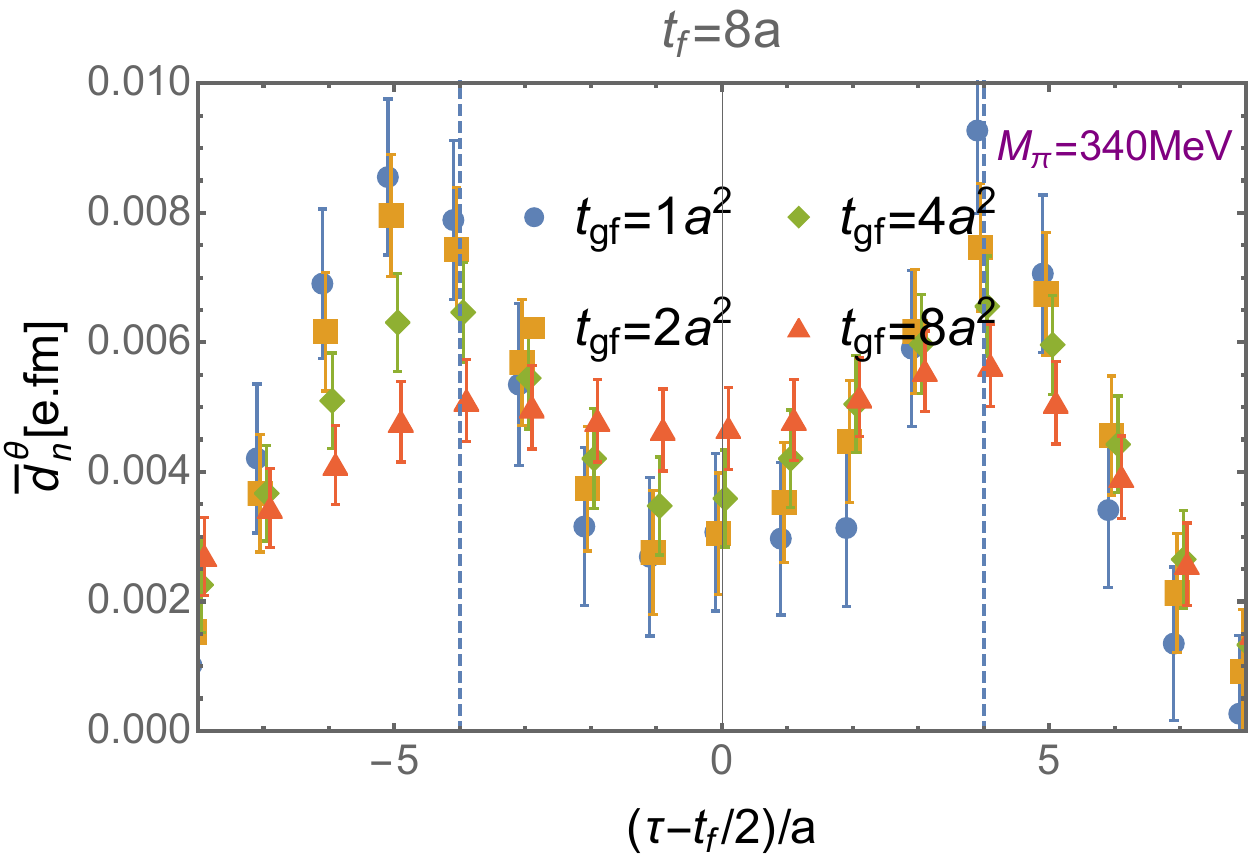}
\includegraphics[width=.32\textwidth]{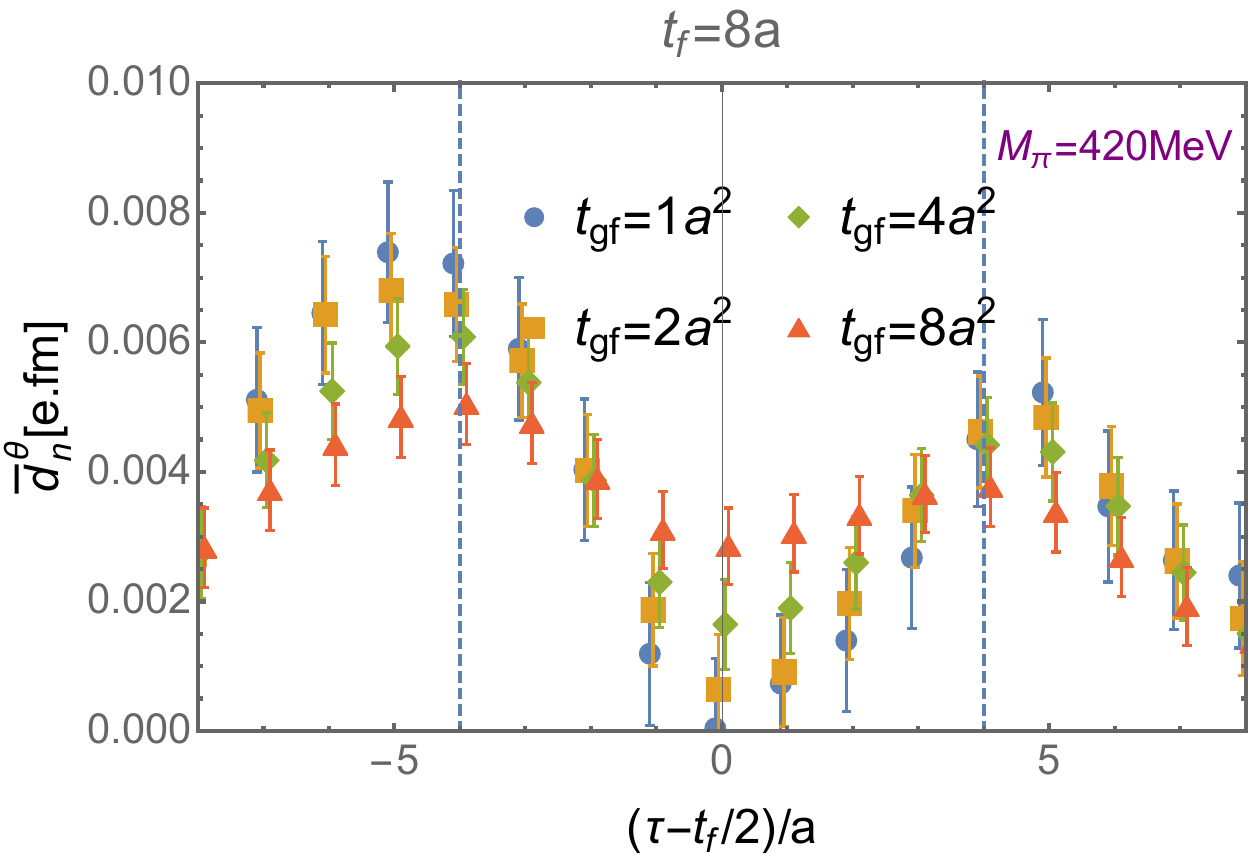}
\includegraphics[width=.32\textwidth]{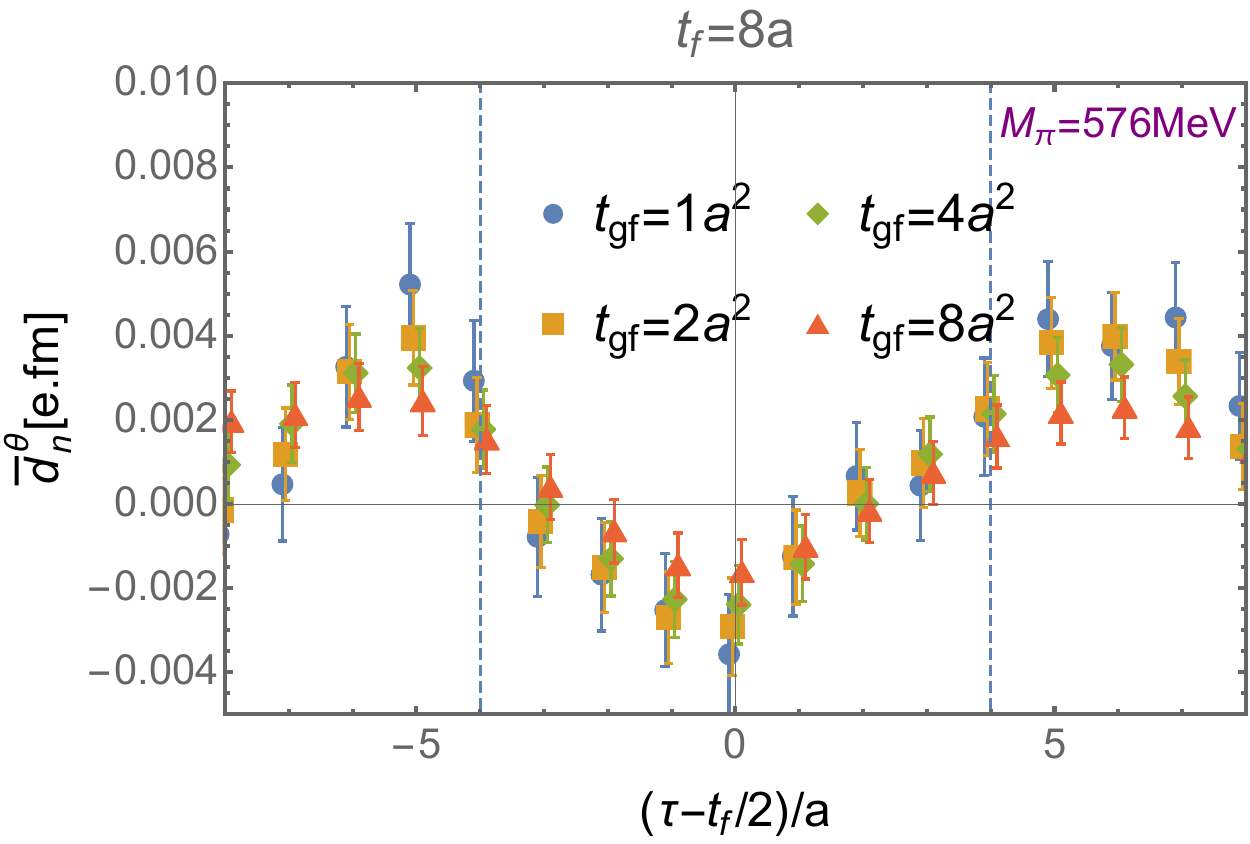}
\\
\includegraphics[width=.32\textwidth]{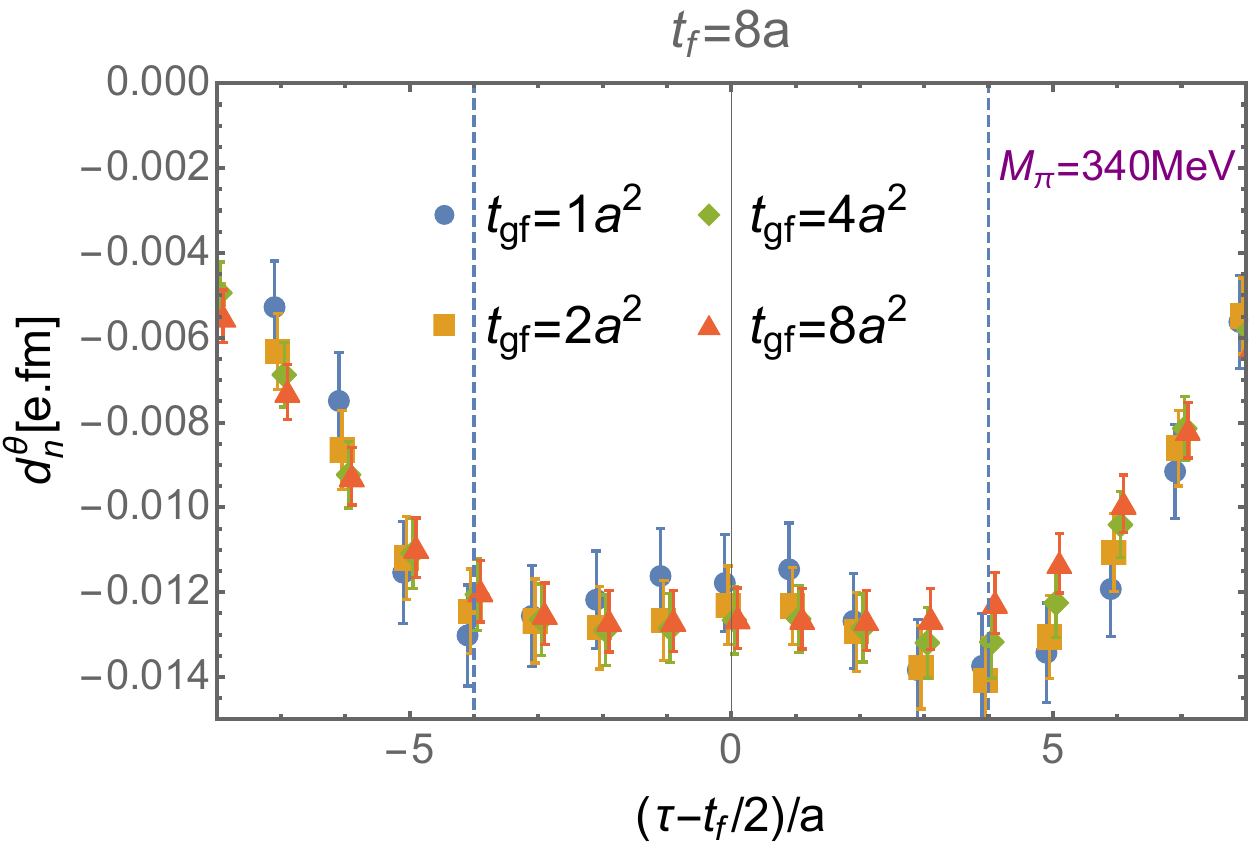}
\includegraphics[width=.32\textwidth]{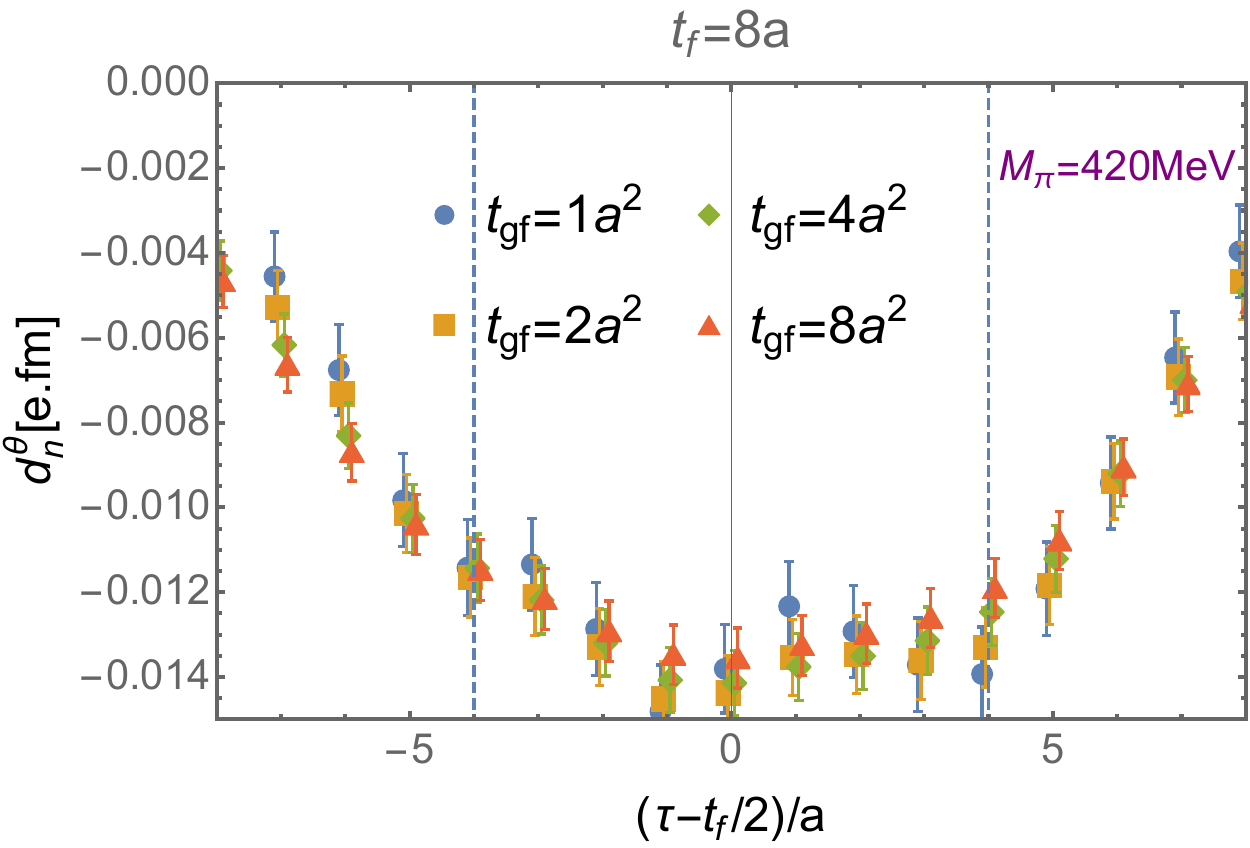}
\includegraphics[width=.32\textwidth]{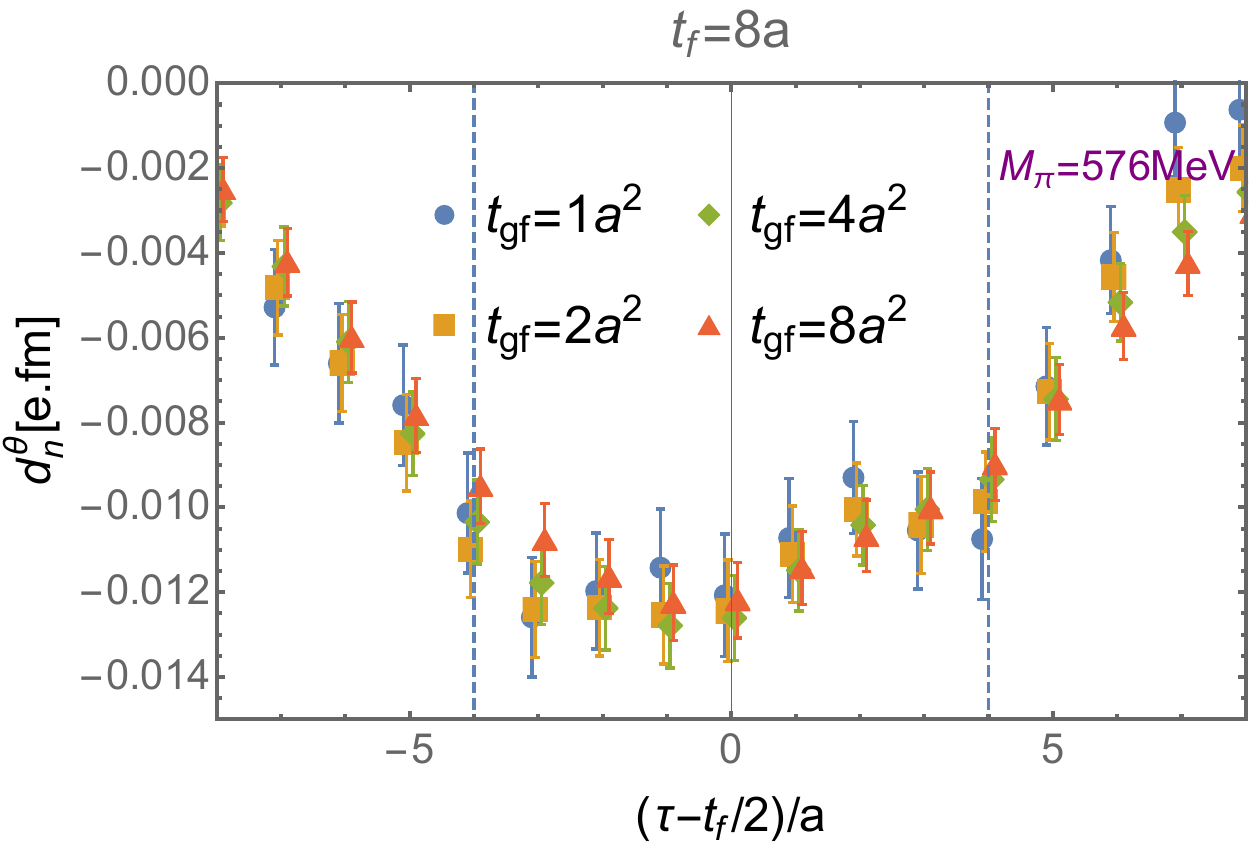}
\caption{The gradient flow-time dependence of the results obtained using  $\tilde{d}_n^\theta$, $\bar{d}_n^\theta$, 
  and GEVP results $d_n^\theta$ on ensembles 24I-005(left), 24I-010(middle) and 24I-020(right). 
  These results are computed with nucleon interpolating operator $O_1$, source-sink separation $t_f=8a$,
  and electric field strength is $|n_z|=2$.
  \label{fig:gf_dep}
}
\end{figure}

The numerical results presented above are obtained using gluon topological charge evaluated at gradient flow time at
$t_{gf}=4a^2$.
In our previous proceeding~\cite{He:2023gwp}, we calculated the nEDM using the definition $\tilde{d}_n$ given in
Eq.~(\ref{eqn:edm_tilde}), and found that the resulting nEDM exhibits a strong dependence on the gradient flow time due
to the diffusion of the gluon fields under the gradient flow.
As we have shown in Fig.~\ref{fig:qcoreeeigendep}, the correlation function of the gluon topological charge operator between
different time slices has a visible dependence on the gradient flow.
One may therefore question whether the nEDM extracted using local topological operator is sensitive to the gradient flow
time. However,as discussed in Sec.~\ref{sec:edm_point}, the nEDM can in principle be extracted from the nucleon two point correlation function in the presence of the electric field and the global topological charge. Therefore, that ground state nEDM is expected to be insensitive to the gradient flow time when the flow time is sufficiently large for the
global topological charge to remain invariant under further flow evolution. Furthermore, gradient flow suppresses the UV
fluctuations of the gauge field while keeping the underlying topological (instanton) invariant. Consequently, the extracted ground state nEDM should be independent of the gradient flow time up to the residual discretization and excited state effects.
In Fig.~\ref{fig:gf_dep}, we compare the results for $\tilde{d}_n^\theta$, $\bar{d}_n^\theta$ and GEVP results $d_n^\theta$.
These results are obtained using operator $O_1$, with the source-sink separation fixed at $t_f=8a$.
The results for $\tilde{d}_n^\theta$ and $\bar{d}_n^\theta$ exhibit a strong dependence on the gradient flow time.
In contrast, the results obtained using GEVP definition shows little sensitivity to the flow time once $t_{gf}\ge 2a^2$. 
%------------------------------------------------------------------------------

%%%%%%%%%%%%%%%%%%%%%%%%%%%%%%%%%%%%%%%%%%%%%%%%%%%%%%%%%%%%%%%%%%%%%%%%%%%%%%%%
\subsection{Dependence of the nEDM on the definition of topological charge density}
\begin{figure}[ht!]
\centering
\includegraphics[width=.48\textwidth]{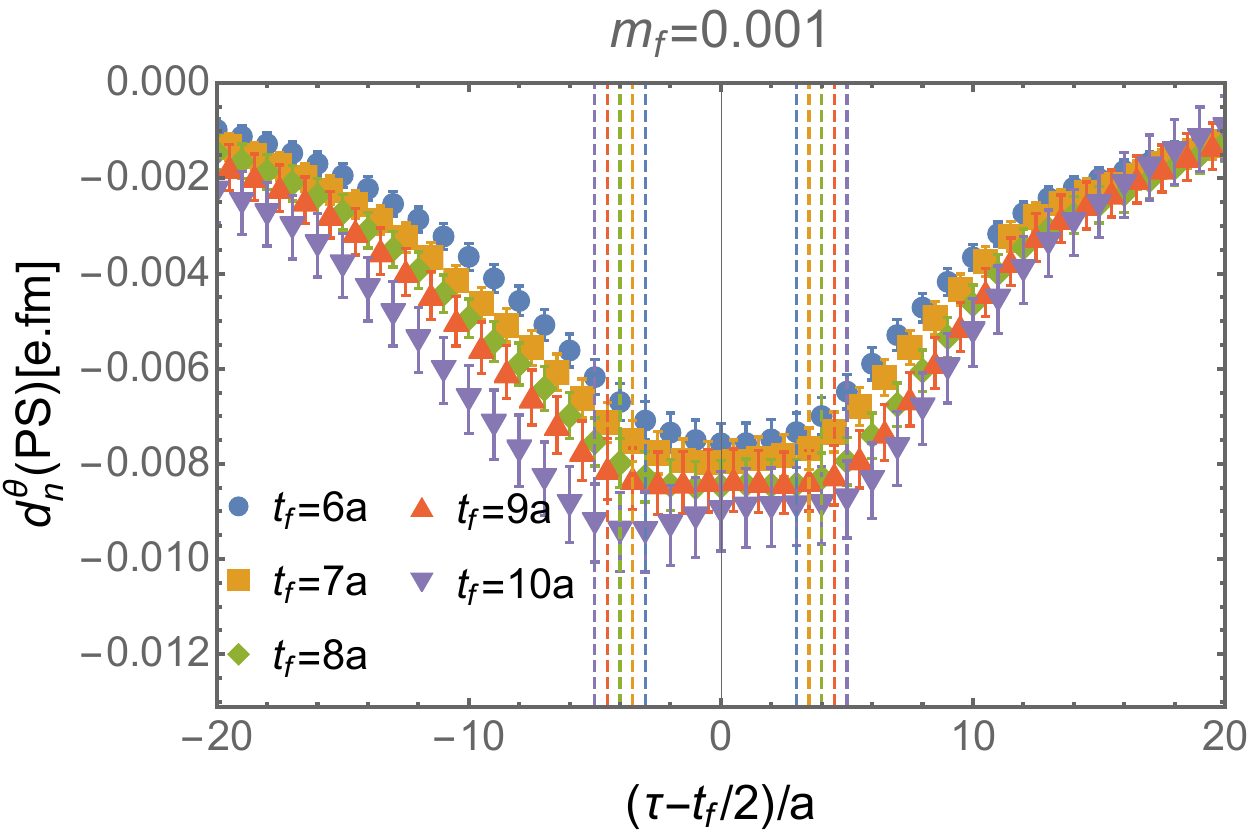}
\includegraphics[width=.48\textwidth]{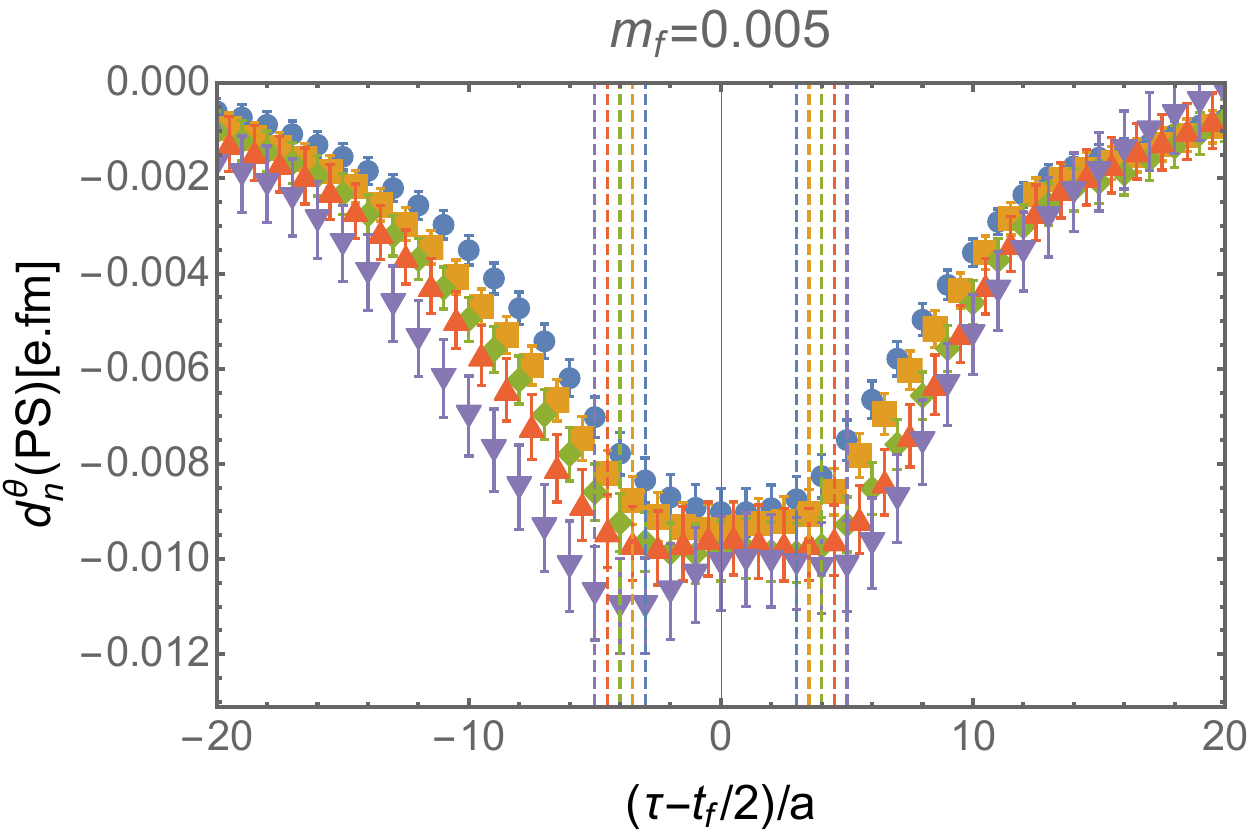}
\\
\includegraphics[width=.48\textwidth]{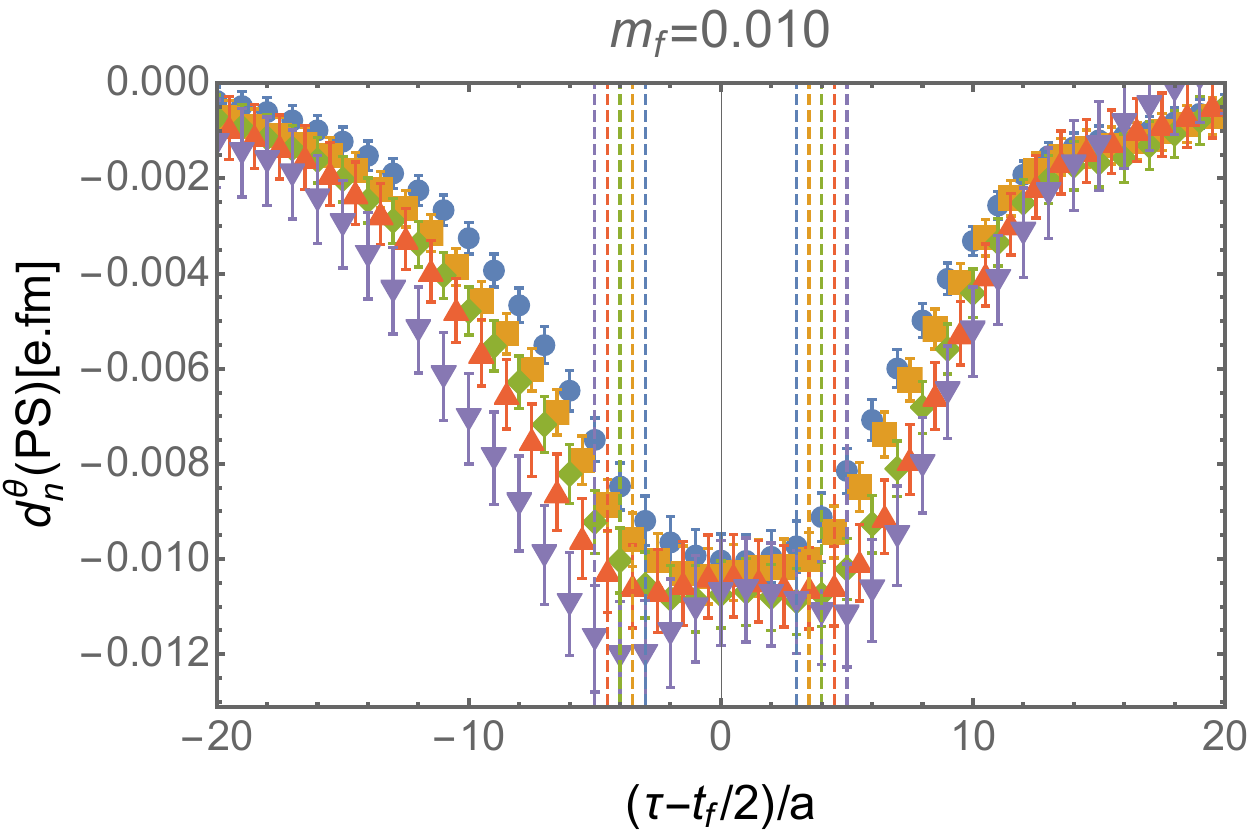}
\includegraphics[width=.48\textwidth]{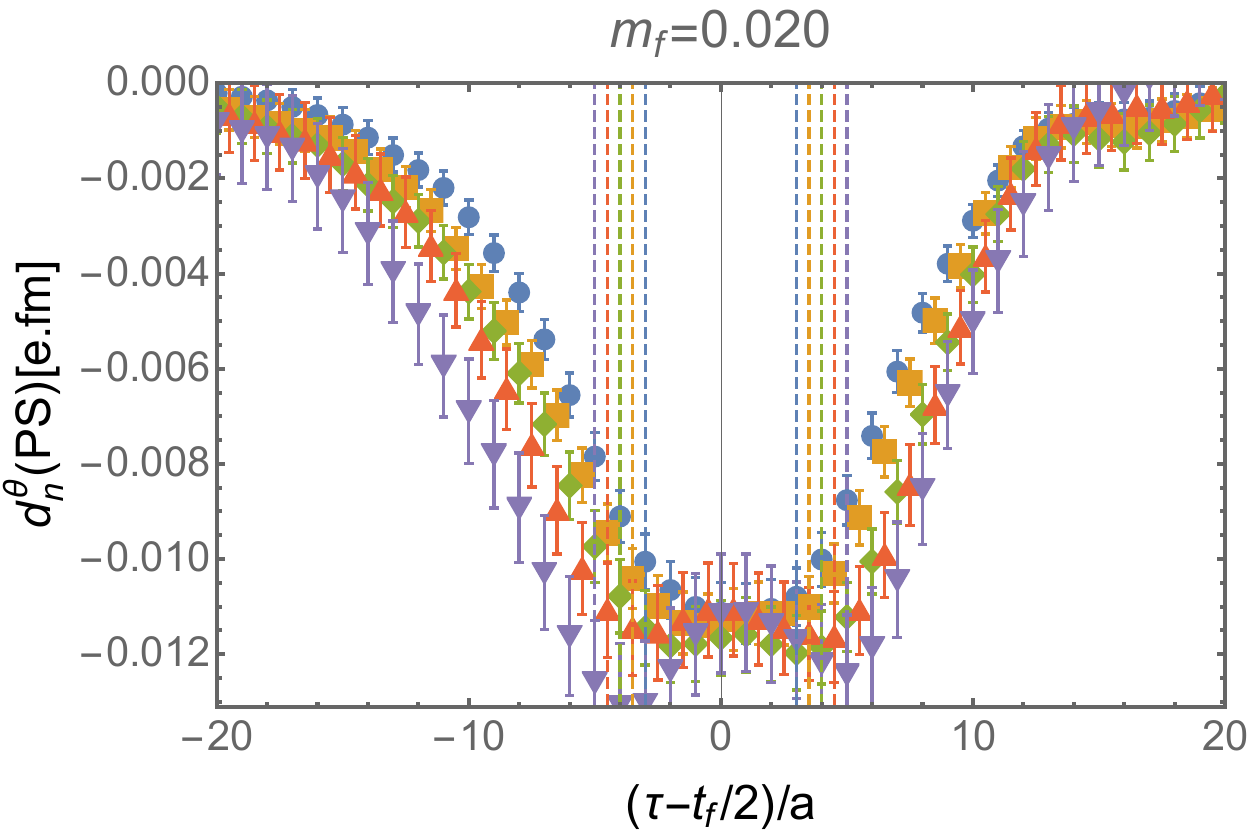}
\caption{GEVP results for the nEDM ($d_n^\theta$) obtained with topological charge computed using the pseudoscalar quark density (ensemble 24I-005). The
  loop quark mass is varied from 0.001 (top-left) to 0.020 (bottom-right). 
  Results are shown for nucleon interpolating operator $O_1$ and the electric field strength is $|n_z|=2$.
  \label{fig:EDM_PS}
}
\end{figure}

In Fig.~\ref{fig:global_g} and Fig.~\ref{fig:global_q}, we have shown that the nEDM results obtained using the global
gluon topological charge and pseudoscalar quark loop are consistent but noisy.
However, the correlation function of local topological charge defined using these two constructions are quite different
as shown in Fig.~\ref{fig:qcoreeeigendep}.
Particularly, the effective masses of pseudoscalar quark density correlators with different quark masses shown in
Fig.~\ref{fig:tp_eff} indicate that they are dominated by pion states.
In Fig.~\ref{fig:EDM_PS}, we present the results $d_n^\theta$ obtained using the GEVP method with the interpolating
operator $O_1$ and the topological charge constructed from the pseudoscalar quark density with varying loop quark mass.
The dashed lines represent the positions of the source and sink nucleon operators.
The nEDM values estimated within the source-sink region become larger as the loop quark mass increases.
These data also show that the results obtained at the smaller loop quark mass exhibit larger
excited state contamination, which may suggest contamination from the $N\pi$ states.
Moreover, when the pseudoscalar quark loop is inserted outside the source-sink separation region, the correlators
corresponding to heavier loop quark mass decay more rapidly than those obtained with the lighter quark masses.
This behavior shows the finite volume effect becomes more pronounced at the lighter loop quark mass.

\begin{figure}[ht!]
\centering
\includegraphics[width=.48\textwidth]{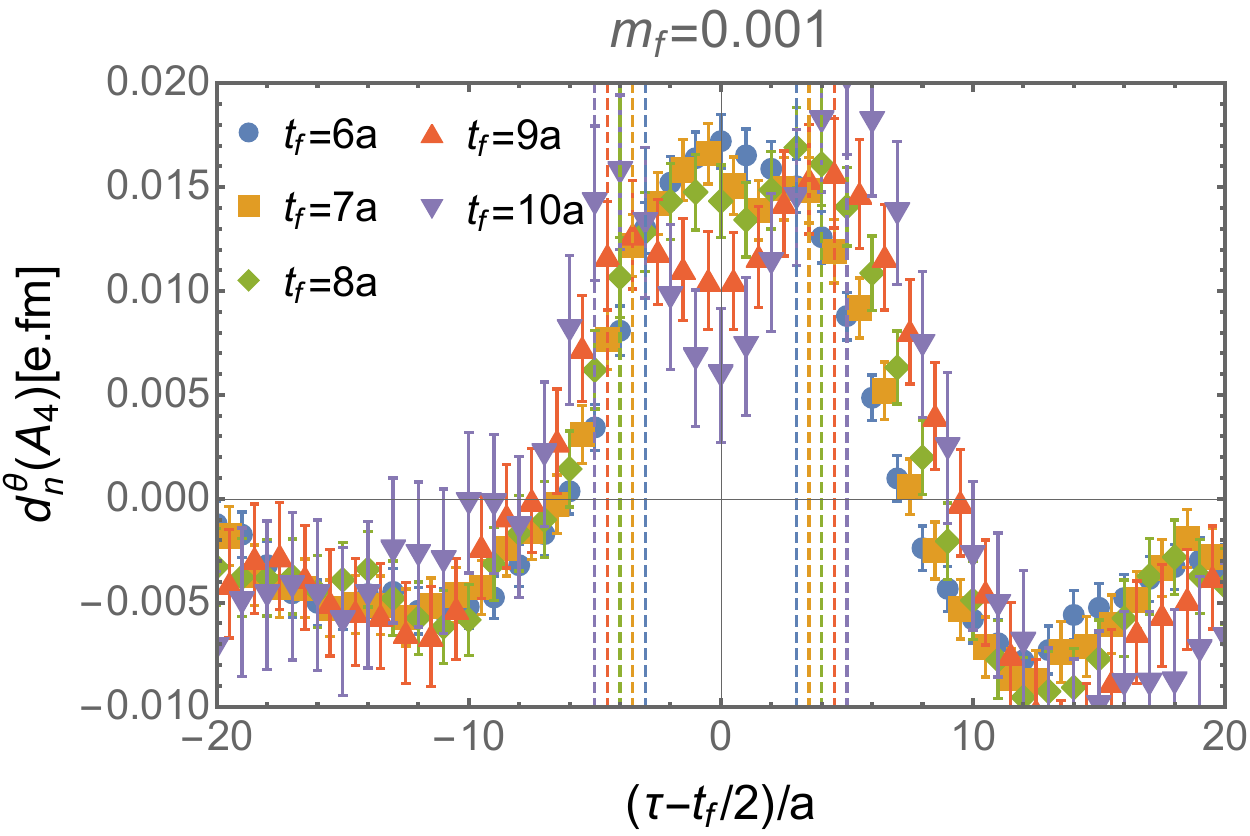}
\includegraphics[width=.48\textwidth]{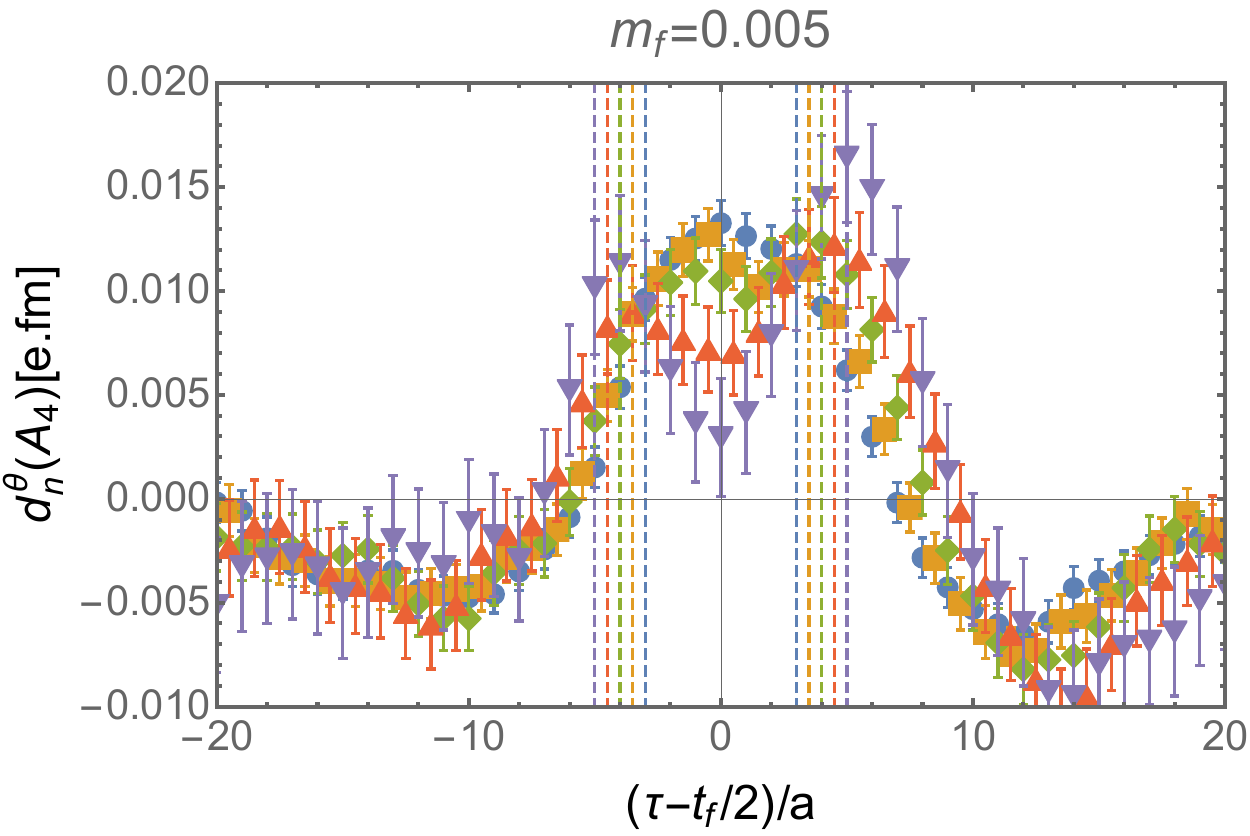}
\\
\includegraphics[width=.48\textwidth]{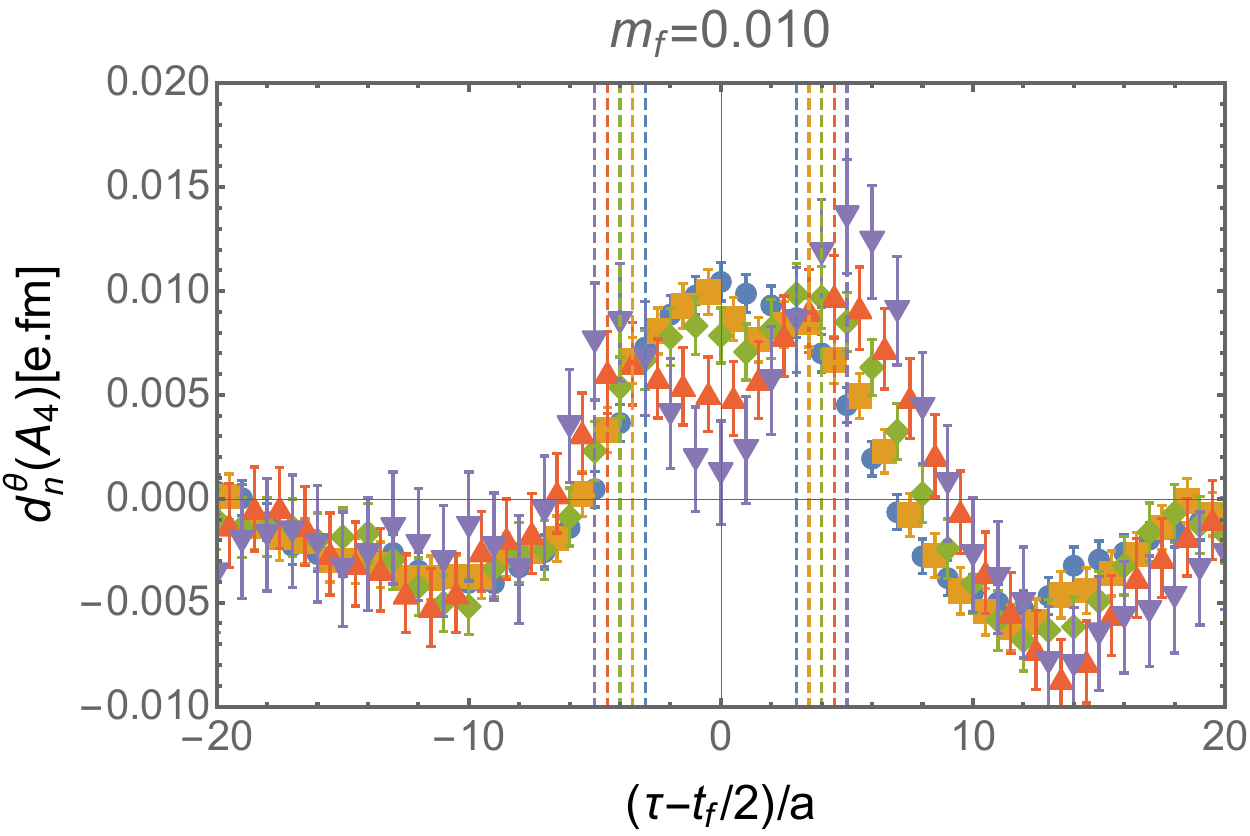}
\includegraphics[width=.48\textwidth]{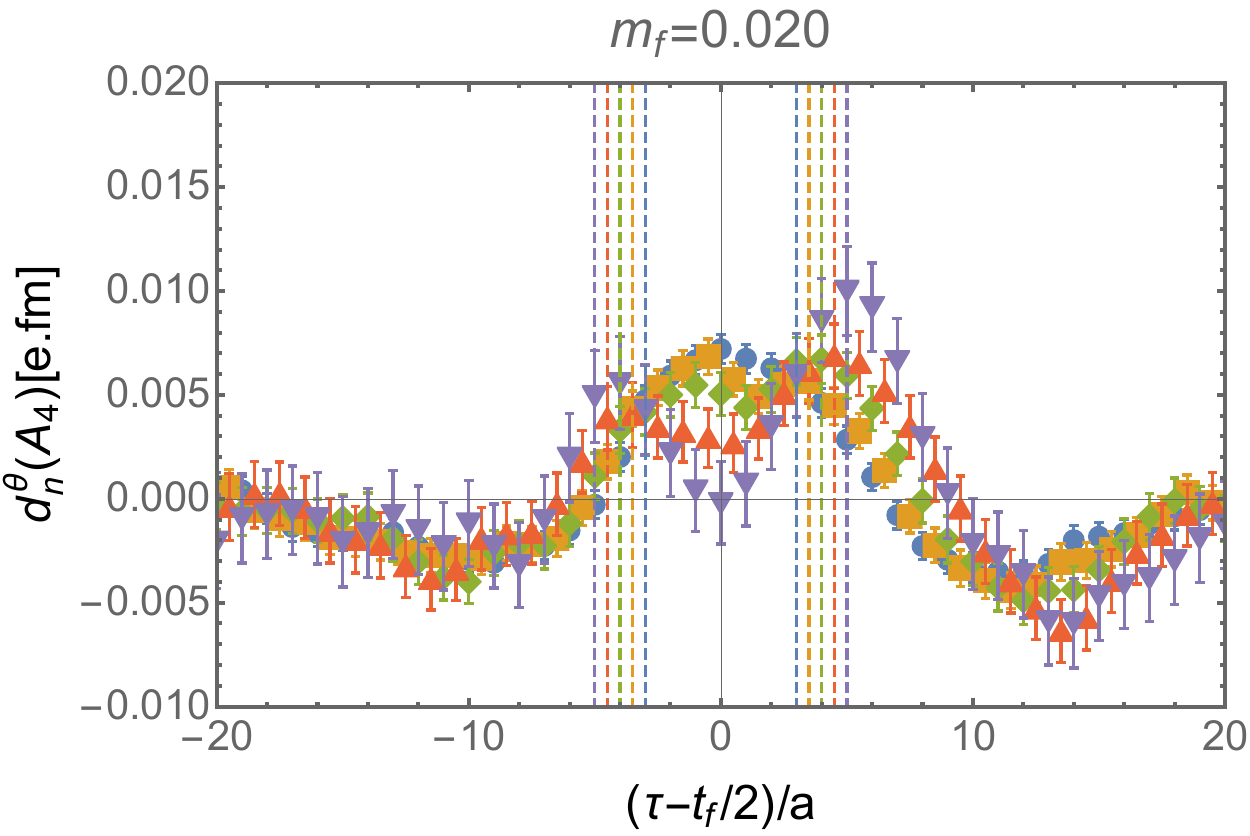}
\caption{Similar to Fig.~\ref{fig:EDM_PS} but with $\partial_\tau A_4(\tau)$~(\ref{eq:EDMA4}) instead of the
  pseudoscalar quark density.
  These results are obtained with nucleon interpolating operator $O_1$ and the electric field strength is $|n_z|=2$.
  \label{fig:EDM_A4}
}
\end{figure}

In Fig.~\ref{fig:EDM_A4}, we plot the contribution from the temporal component of the axial vector quark loop to the nEDM, which is defined as
\begin{equation}
\label{eq:EDMA4}
d_n^\theta(A_4) 
  = \frac{1}{\mcE_z} 
    \frac{ v^\dag_{L,0}C^{\psi\psi^\dag,\uparrow}_{3pt,\mcE}[A_4](t_f,\tau)v_{R,0}}
         {v^\dag_{L,0}C^{\psi\psi^\dag,\uparrow}_{2pt,\mcE}(t_f)v_{R,0}}\,,
\end{equation}  
where the three-point correlation function $C^{\psi\psi^\dag,\uparrow}_{3pt,\mcE}[A_4](t_f,\tau)$ is obtained by
replacing $q_{top}(\tau)$ in Eq.~(\ref{eq:3ptoriQtop}) with $\partial_\tau A_4(\tau)$. The prefactor $-Z_A/2$ appearing in Eq.~(\ref{eq:ABJ}) has not been included in the results presented here.

The data points within the source-sink region do not exhibit plateaus and trend toward zero as the source-sink separation increases.
This behavior is consistent with our expectation that this quantity does not contribute to the ground-state matrix
elements and instead arises purely as excited state contamination.
As the loop quark mass increases, its magnitude also decreases, and the results within source-sink region approach zero
more rapidly compared to the lighter loop quark-mass case. Compared to the results from pseudoscalar quark loops, the results from the axial vector quark loops exhibit a different behavior when operator is inserted outside the source-sink separation region. Instead of decaying directly to zero, the signal first becomes negative and gradually approaches zero, resulting in a pronounced valley-like region.

\begin{figure}[ht!]
\centering
\includegraphics[width=.48\textwidth]{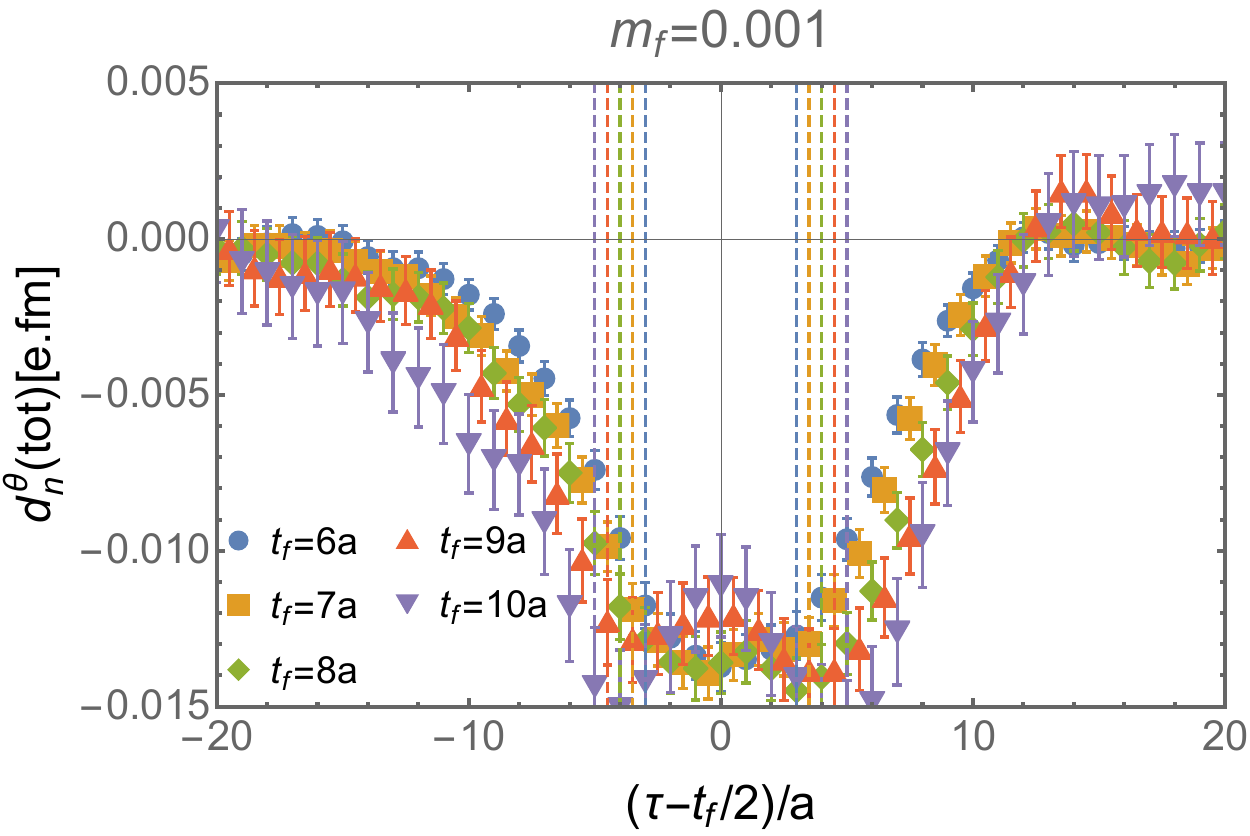}
\includegraphics[width=.48\textwidth]{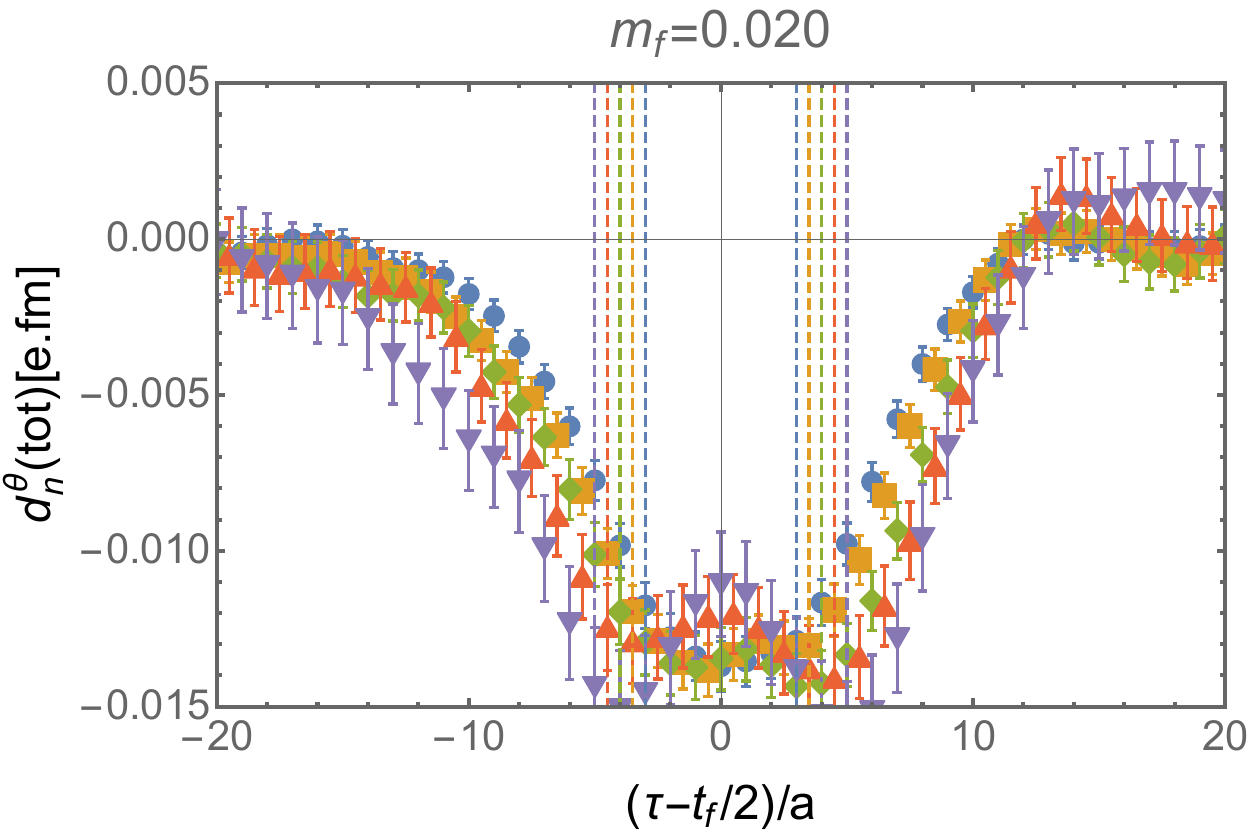}
\\
\includegraphics[width=.48\textwidth]{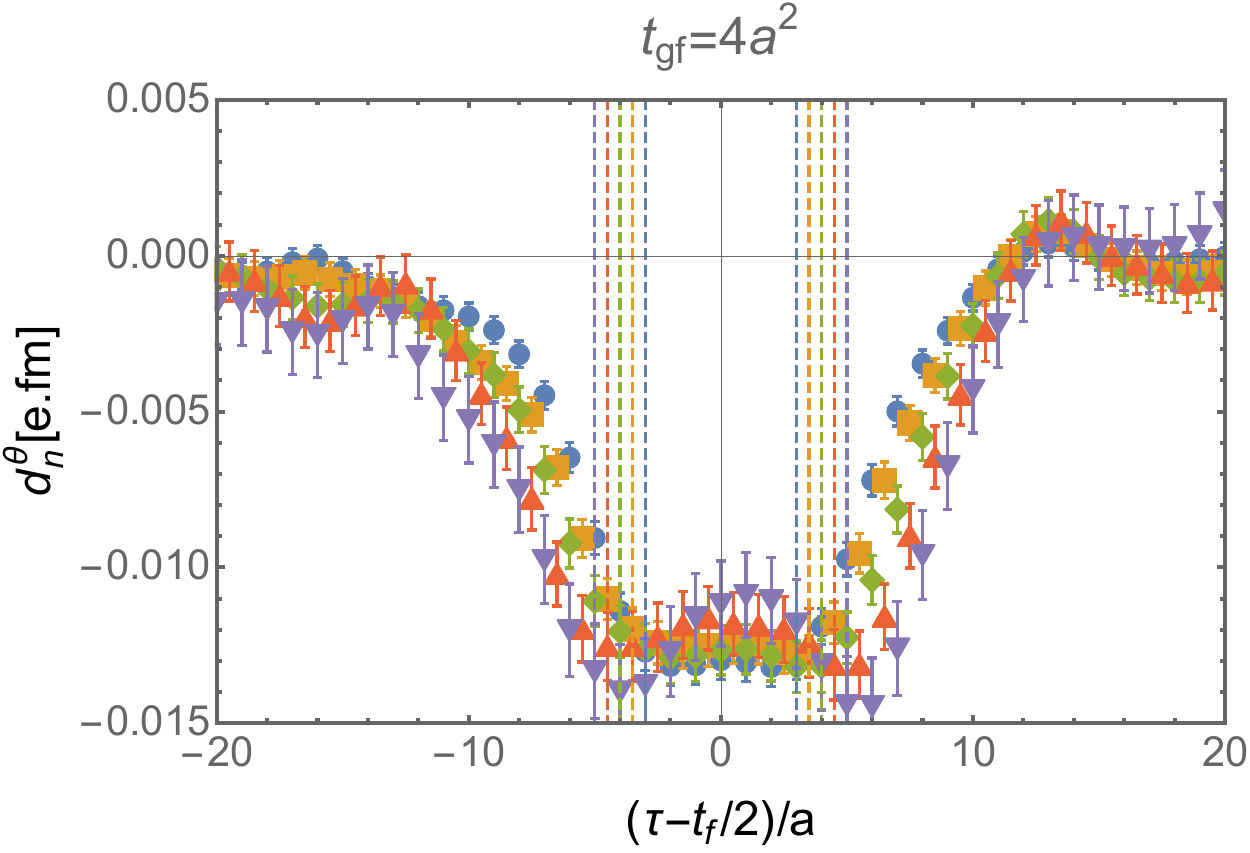}
\caption{Similar to Fig.~\ref{fig:EDM_PS} but using pseudoscalar quark density with the
  axial-vector current correction and varying loop quark masses $m_f=0.001$ and $m_f=0.02$ (top)
  compared to nEDM computed using the gluon-field  topological charge density at flow time $t_{gf}=4a^2$ (bottom).
  These results are obtained with nucleon interpolating operator $O_1$ and the electric field strength is $|n_z|=2$.
  \label{fig:ABJ}
}
\end{figure}

The validity of the ABJ anomaly equation in Eq.~(\ref{eq:ABJ}) does not depend on whether it is evaluated between the hadron state.
We check this in Fig.~\ref{fig:ABJ}, where we combine the correction term $d_n^\theta(A_4)$ with the nEDM values obtained using
the pseudoscalar-density topological charge.
Remarkably, after incorporating this correction the total results using different loop quark masses become consistent
with each other and are in good agreement with those obtained using gluon topological charge.
This indicates the ABJ anomaly equation is satisfied in our case of external states being
nucleon eigenstates deformed by the background electric field. 

Furthermore, when the quark loop operators are inserted outside the source-sink separation region, substantial cancellations occurs between the results from the pseudoscalar and axial vector quark loops. As a result, the total signal decays more rapidly towards zero, similar to the behavior observed for the gluon topological charge operator. This suggests that finite volume effects associated with long-range tails are not a significant source of systematic uncertainty once the contributions from the two quark loops are combined.

%------------------------------------------------------------------------------

%------------------------------------------------------------------------------

%------------------------------------------------------------------------------

%%%%%%%%%%%%%%%%%%%%%%%%%%%%%%%%%%%%%%%%%%%%%%%%%%%%%%%%%%%%%%%%%%%%%%%%%%%%%%%%
\subsection{Neutron EDM from multi-operator GEVP}
In this section, we present the numerical results of the nEDM obtained with the GEVP applied to multiple nucleon operators.
As shown in the previous section, the nEDM computed with pseudoscalar quark density may be subject to large $N\pi$
contamination and finite volume effects.
Even though adding the axial-current correction brings them into agreement with nEDM results using gluonic $q_{top}$,
from now on we focus exclusively on the latter and adopt them for our final results. Nevertheless, we verify that the ABJ anomaly Eq.~(\ref{eq:ABJ}) remains valid when the correlators are constructed using the multi-operator GEVP analysis described below.

Since we have four different nucleon interpolating operators, they can be combined for a multi-operator GEVP analysis.
The corresponding matrix for the two point and three point correlation function are defined as
\begin{equation}
C_{2,ij}(t_f)=\la O_i(t_f) O_j^\dag(0)\ra_{\mcE_z},~~~~C_{3,ij}(t_f,\tau)=\la O_i(t_f) q_{top}(\tau)O_j^\dag(0)\ra_{\mcE_z}
\end{equation}
For each element $O_iO_j^\dag$, the correlation function forms a $2\times2$ matrix spanned by $N$ and $N^*$
correlators, similar to Eq.~(\ref{eq:2ptppbar}).
As before, the corresponding eigenstates are obtained by solving the non-Hermitian GEVP,
\begin{equation}
\label{eq:multiGEVP}
C_{2}(t)v_{R,n}=\lambda_n C_{2}(t_0)v_{R,n}\,,\quad
C_{2}^\dag(t_f)v_{L,n}=\lambda_n C_{2}^\dag(t_0)v_{L,n} \,.
\end{equation}
Here, we have assumed that the external electric field is sufficiently weak so that the eigenvalues remain real. 
The correlation function matrix $C_{2, ij}$ satisfies the relation $C_2^\dag=U C_2U$, where $U$ is a diagonal matrix
constructed from $\sigma_z$ and has the same dimension as $C_{2}$, $i.e.$$,  $~$U=\text{diag}(\sigma_z,\sigma_z,\dots)$.
Consequently, the left- and right-eigenvectors $v_{L,i}$ and $v_{R,i}$ are related as $v_{L,i}=Uv_{R,i}$ for real eigenvalue.
Once the ground state eigenvectors $v_{R,0}$ and $v_{L,0}$ are obtained, the nEDM from multi-operator GEVP can be
computed using the same definition~(\ref{eqn:edm_bgem_GEVP2}) as before.

%------------------------------------------------------------------------------
\begin{figure}[ht!]
\centering
\includegraphics[width=.32\textwidth]{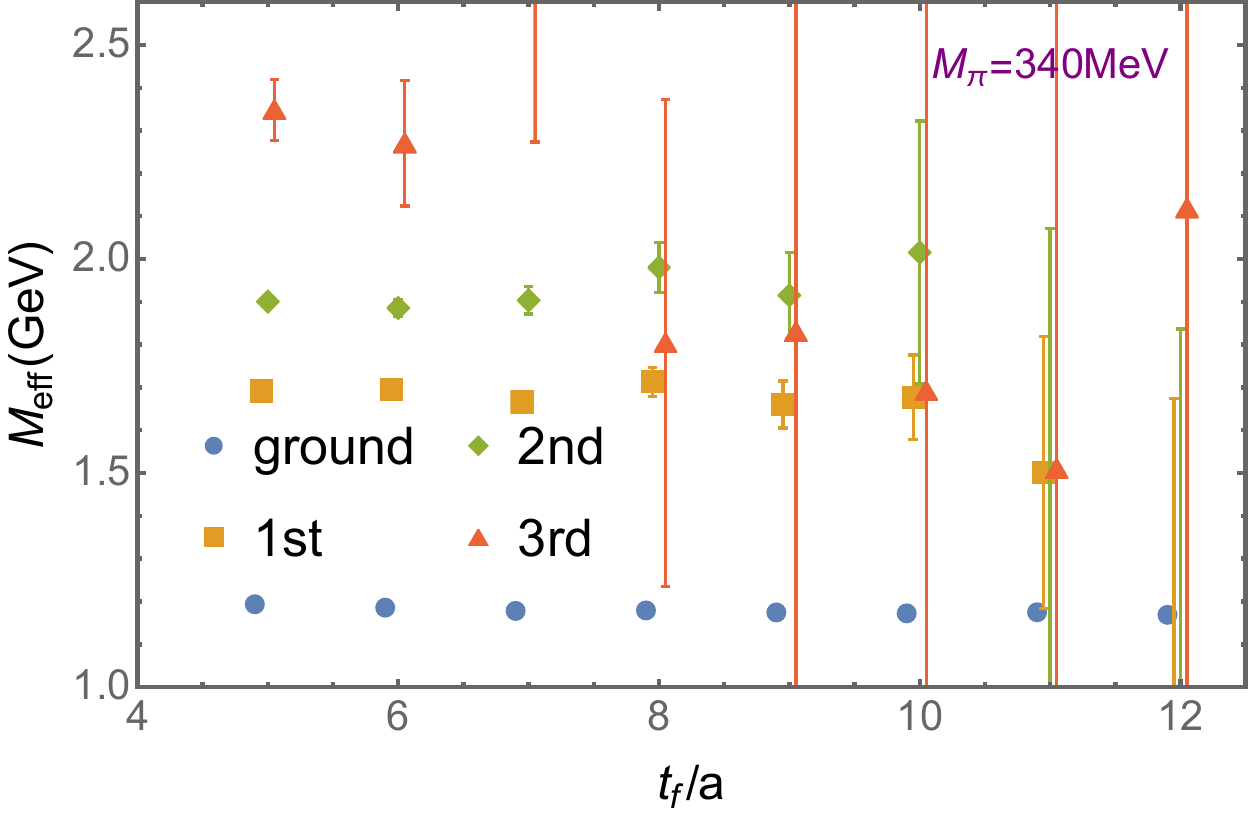}
\includegraphics[width=.32\textwidth]{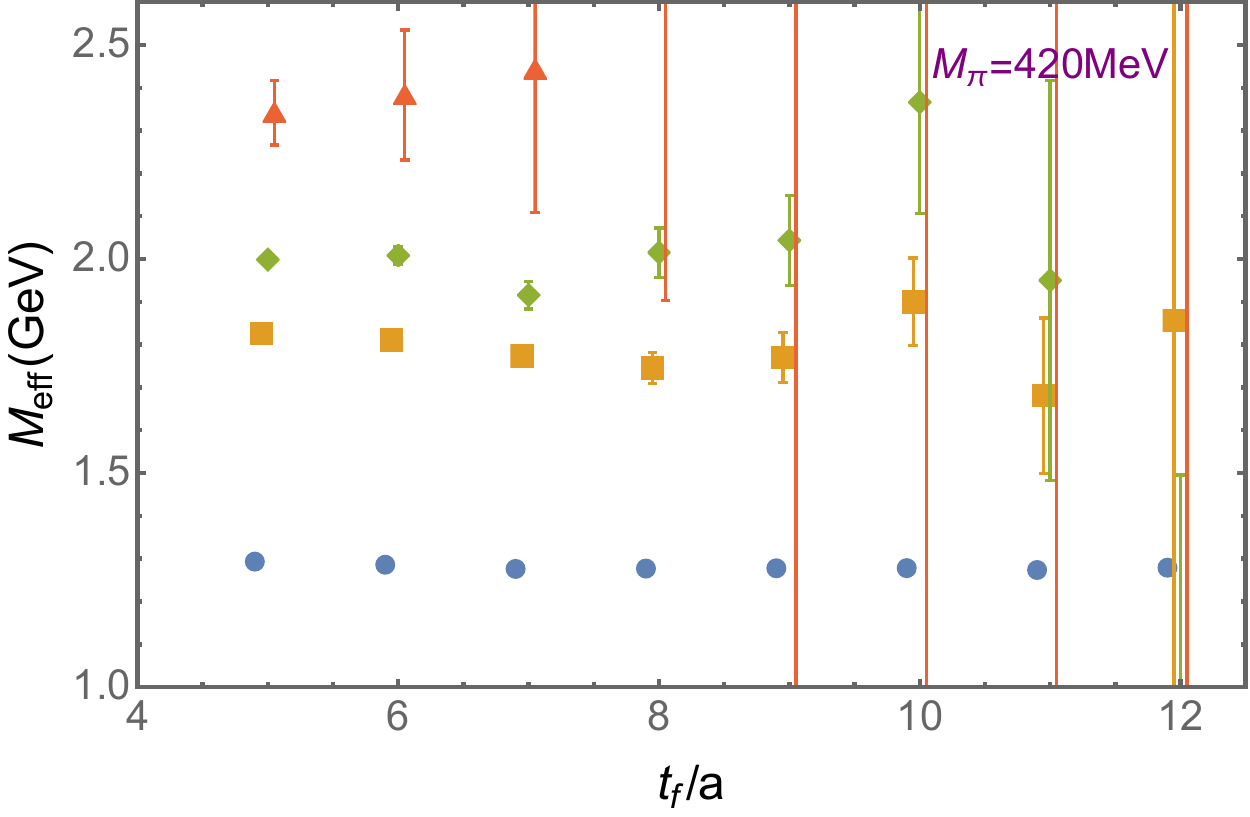}
\includegraphics[width=.32\textwidth]{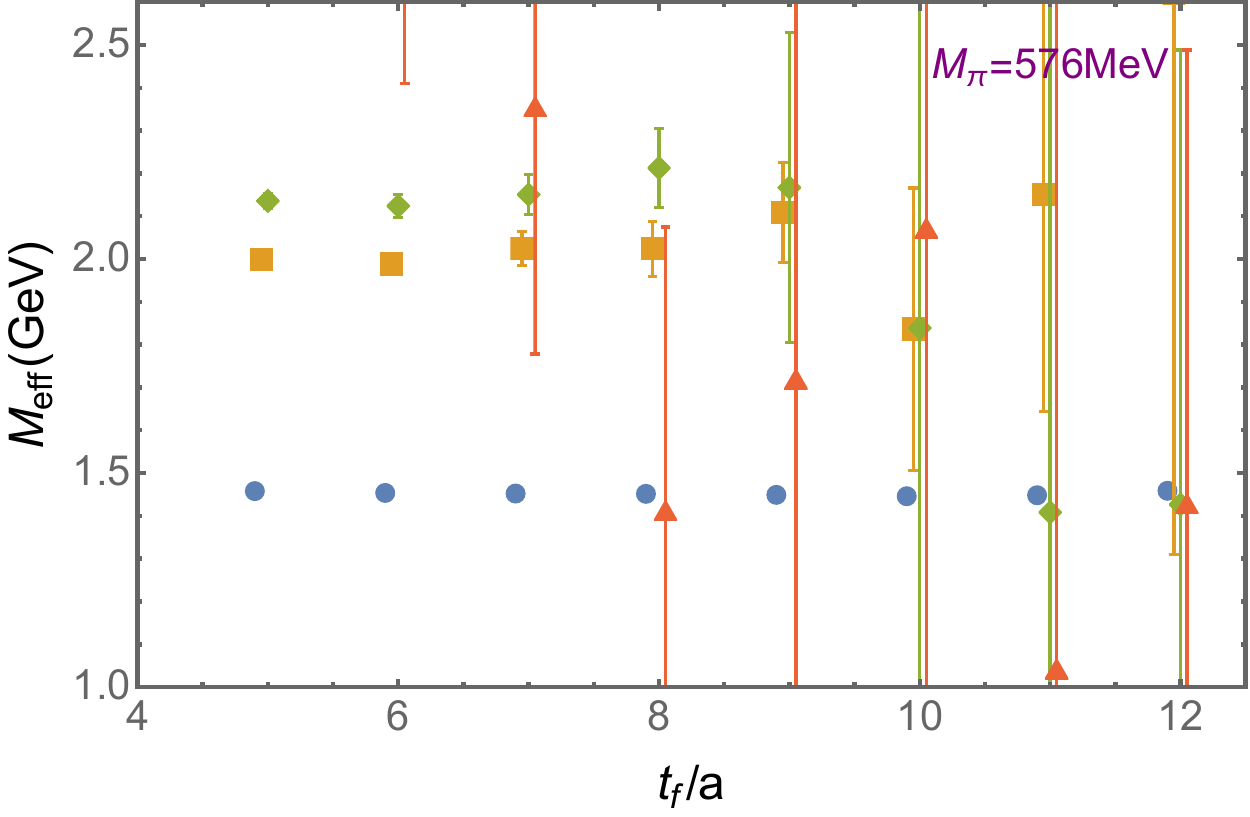}
\caption{
  Comparison of the effective masses of the first four eigenstates from GEVP
  using nucleon operators $\{O_1,O_2\}$ and computed with electric field strength $|n_z|=2$
  on the three ensembles.}
\label{fig:effmass_O1O2GEVP}
\end{figure}

In Figure~\ref{fig:effmass_O1O2GEVP} we present the effective masses of the first four nucleon eigenstates in the presence of electric field calculated using the generalized eigenvalues from~(\ref{eq:multiGEVP})
and constructed from nucleon operators $O_1$ and $O_2$.
The mass gap between the second and the third states is small, and it is evident that the second state is absent when
only the operator $O_p$ is used in the GEVP, as shown in Fig.~\ref{fig:GEVP_res_mass005Ez2}.

\begin{figure}[ht!]
\centering
\includegraphics[width=.96\textwidth]{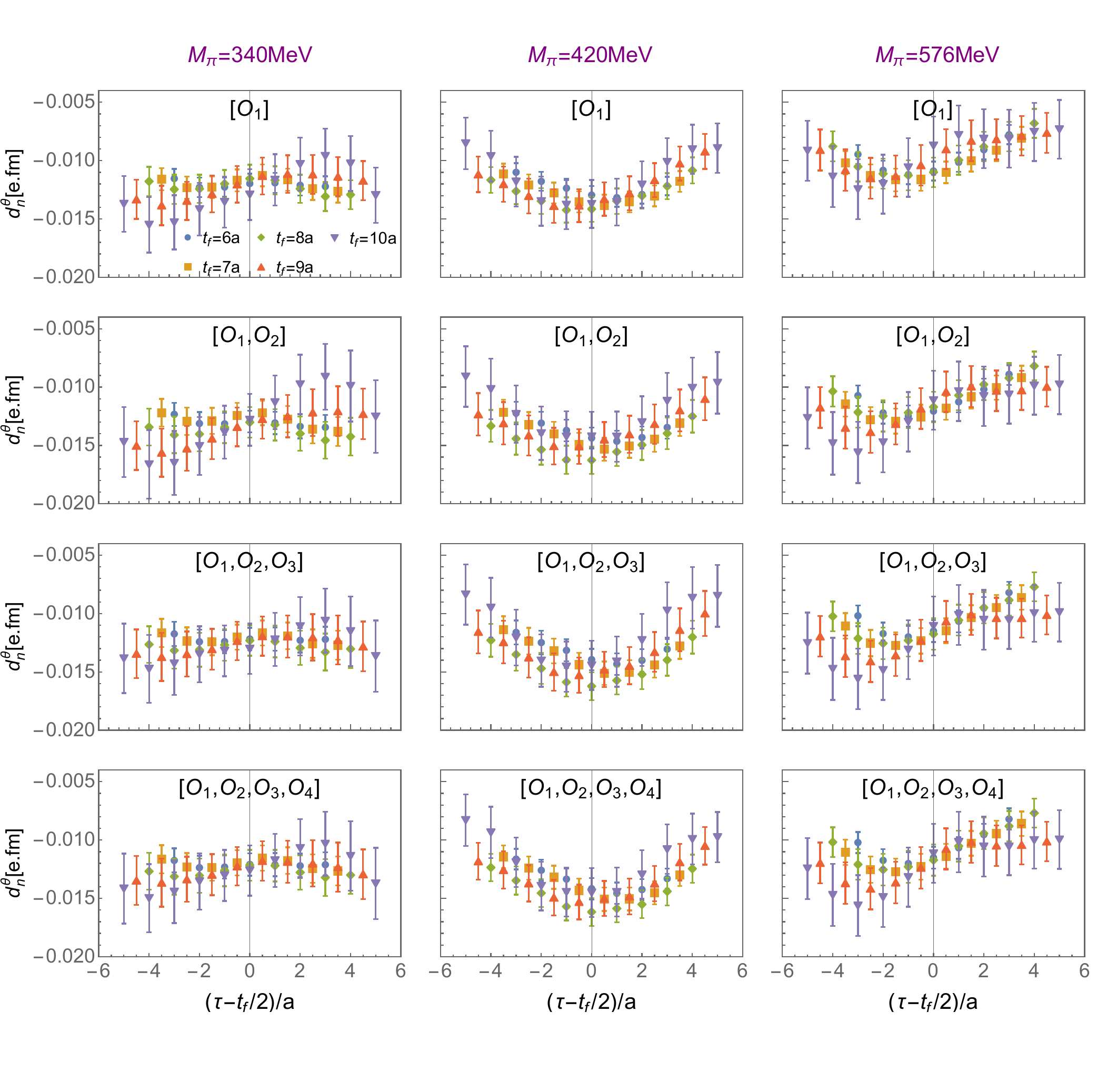}
\vspace{-12mm}
\caption{The EDM results obtained using multiple operators GEVP on ensemble 24I-005(left panel), 24I-010(middle panel)
and 24I-020(right panel) with $|n_z|=1$ electric field. From the top to the bottom panel, the panels display
the nEDM results obtained using operator basis $\{O_1\}$, $\{O_1,O_2\}$,  $\{O_1,O_2,O_3\}$ and $\{O_1,O_2,O_3,O_4\}$,
respectively.}
\label{fig:multiGEVP_res_mass005Ez1_igf4}
\end{figure}

The numerical results for $d_n^\theta$ obtained using multi-operator GEVP and $|n_z|=1$ electric field are shown in
Fig.~\ref{fig:multiGEVP_res_mass005Ez1_igf4}.
From the top to the bottom, the panels display the nEDM results obtained from the GEVP method using operators $\{O_1\}$,
$\{O_1,O_2\}$, $\{O_1,O_2,O_3\}$ and $\{O_1,O_2,O_3,O_4\}$, respectively, and the GEVP reference time $t_0=5a$. The results obtained with $t_0=4a$, $5a$, and $6a$ are found to be consistent within statistical uncertainties.
The topological charge density is constructed from the gluon field at gradient flow time $t_{gf}=4a^2$. As shown in Fig.~\ref{fig:gf_dep}, the nEDM results obtained using the GEVP definition are consistent for all $t_{gf}\ge2a^2$, we therefore adopt the results at $t_{gf}=4a^2$ as our final results and find that the systematic uncertainty associated with the flow time is negligible~\cite{EDM_preparation}.
In this case, the results obtained from the GEVP using $O_1$ alone already exhibit relatively small excited state
contamination.
The GEVP method with additional nucleon operators exhibits mild reduction of excited state contamination compared to the
single-operator case, particularly when the operator $O_2$ is included in the variational basis.
This improvement indicates that the operator $O_2$ has small overlap with the nucleon ground state and relatively large
overlap with the negative parity state $N^*$, therefore including $O_2$ in the variational basis improves the GEVP
projection on the ground state.
When $O_2$ is included, the resulting nEDM value increases slightly in magnitude compared to using only $O_1$.
When using the largest basis $\{O_1,O_2,O_3,O_4\}$ in the GEVP, the magnitude of the nEDM decreases when the pion mass is
decreased from $m_\pi=420$ MeV to $340$ MeV, thus exhibiting behavior consistent with low-energy (chiral) theory.

\begin{figure}[ht!]
\centering
\includegraphics[width=.96\textwidth]{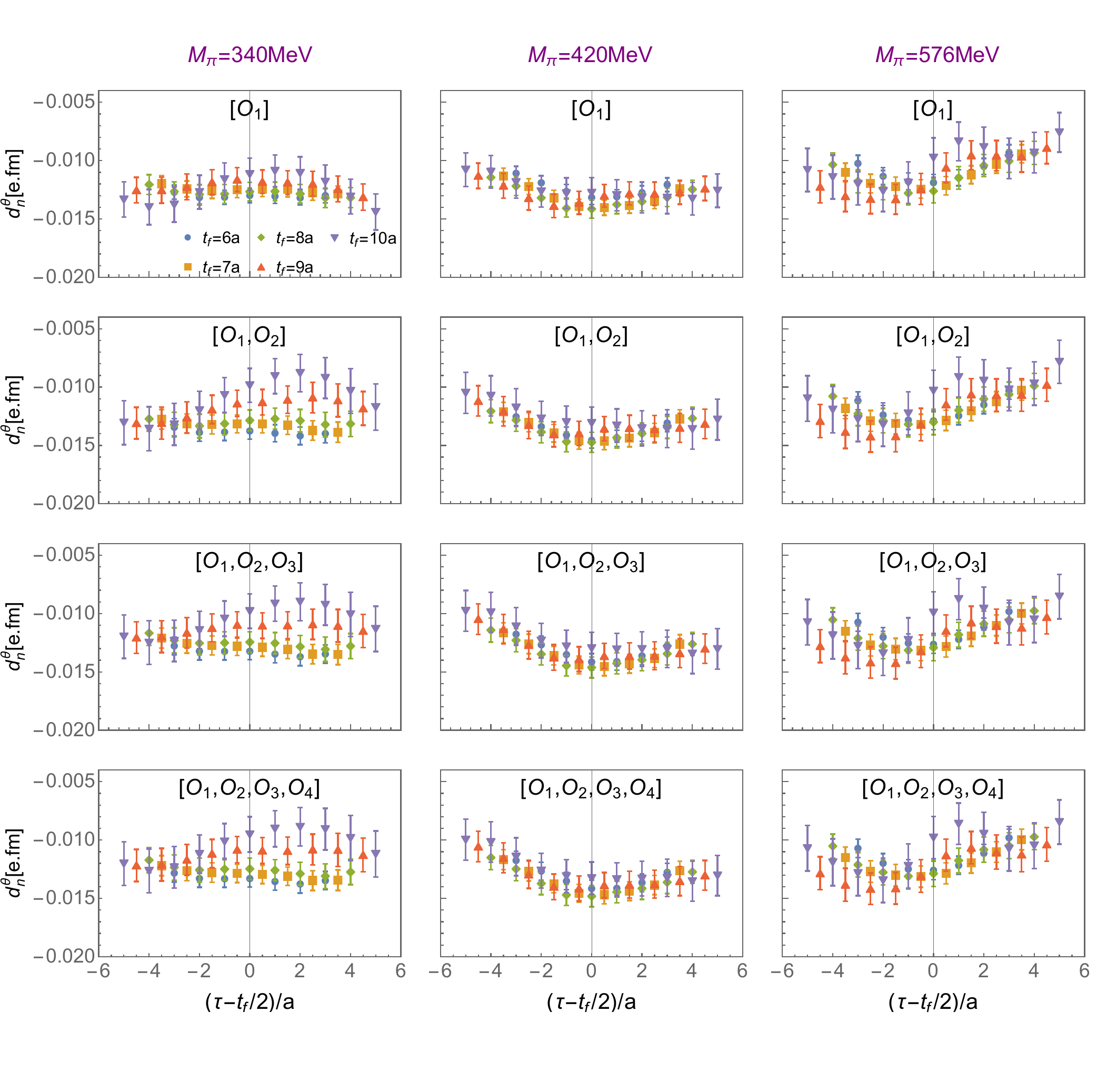}
\vspace{-12mm}
\caption{Similar to Fig.~\ref{fig:multiGEVP_res_mass005Ez1_igf4} but with $|n_z|=2$ electric field.}
\label{fig:multiGEVP_res_mass005Ez2_igf4}
\end{figure}

The corresponding results computed with $|n_z|=2$ electric field are shown in 
Fig.~\ref{fig:multiGEVP_res_mass005Ez2_igf4}.
Compared to the $|n_z|$=1 results, they have smaller statistical uncertainties.
Again, the results obtained using $O_1$ alone show small excited-state contamination.
However, when we include additional nucleon operators, the results show stronger excited state contamination at
$m_\pi=340$ MeV.
This behavior could be due to statistical fluctuations, or it may indicate that higher-order effects $\mcE_z$
become more significant at $|n_z|=2$ electric field compared to $|n_z|=1$.
In contrast, for the ensembles with $m_\pi=420$ and 576 MeV, the results remain largely unchanged as the nucleon basis
in GEVP is extended, indicating reduced excited state contamination in this case.

\begin{figure}[ht!]
\centering
\includegraphics[width=.32\textwidth]{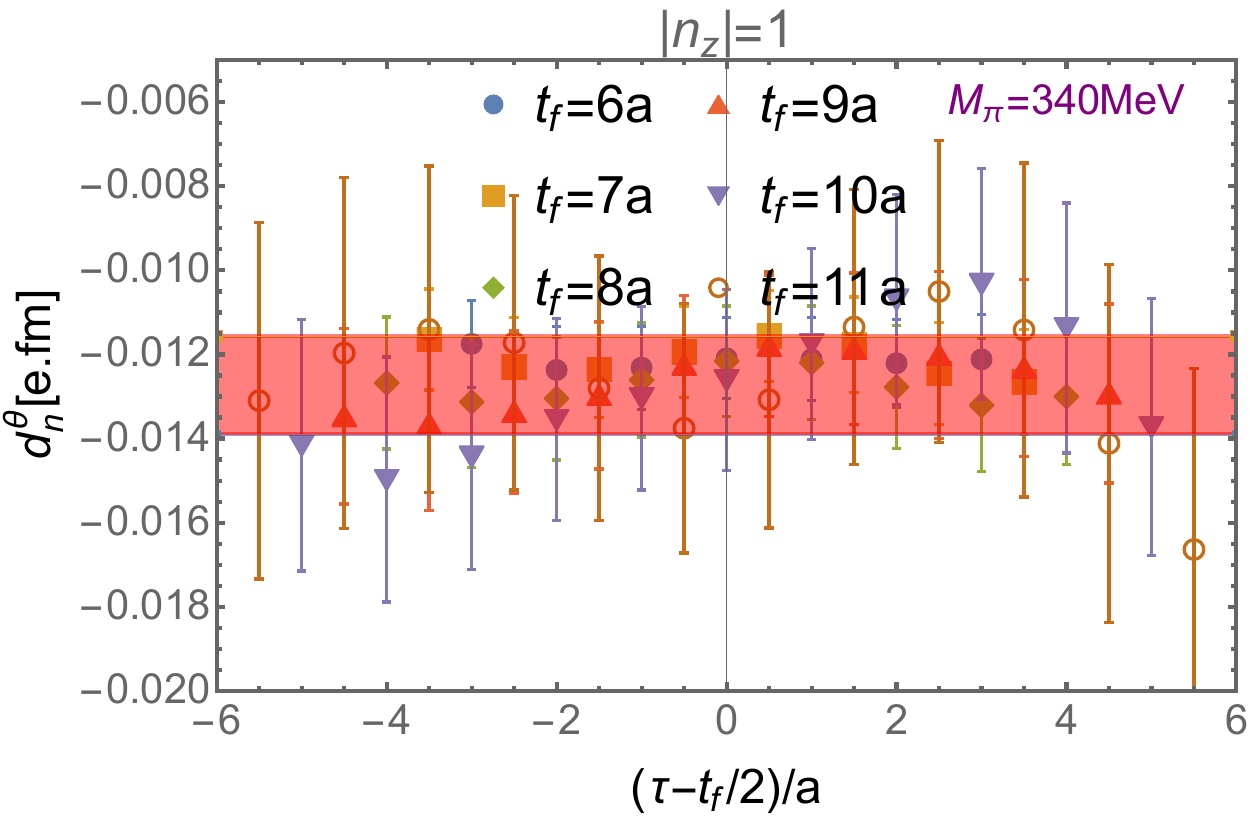}
\includegraphics[width=.32\textwidth]{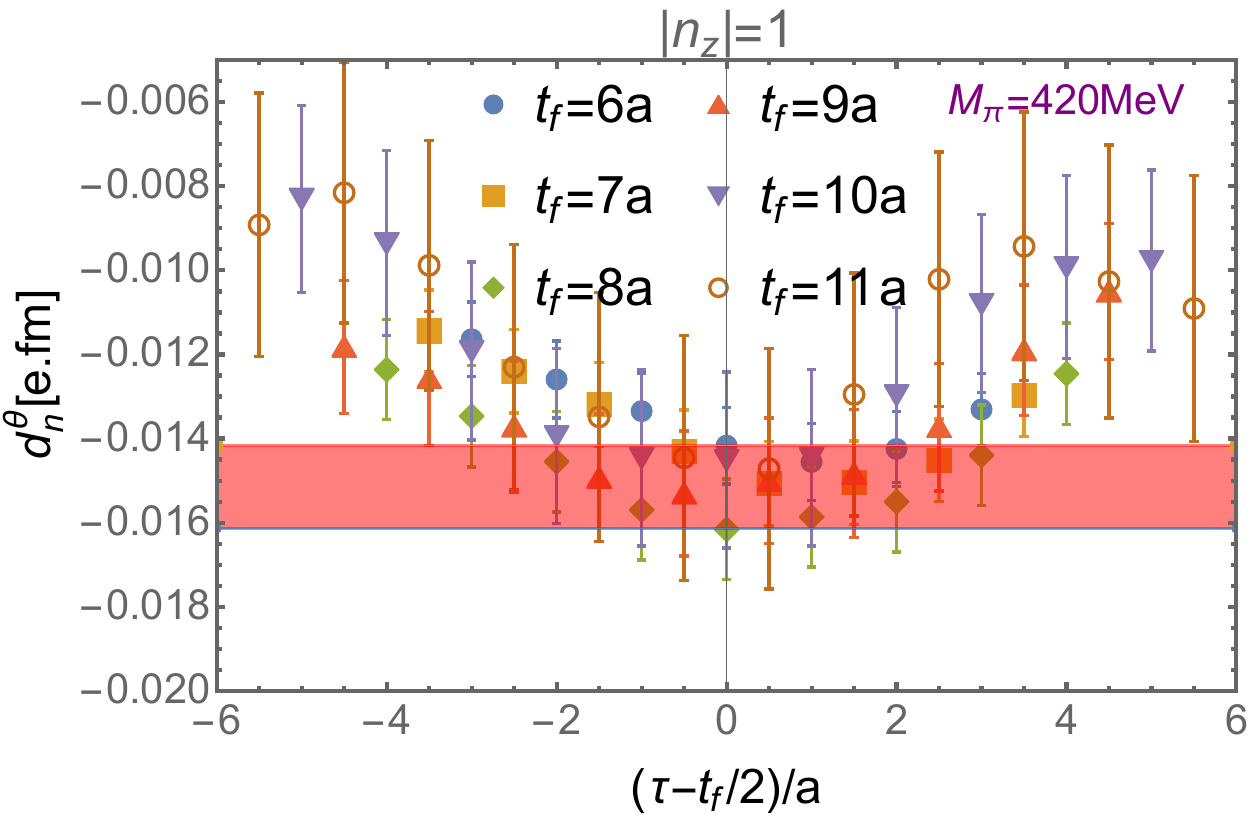}
\includegraphics[width=.32\textwidth]{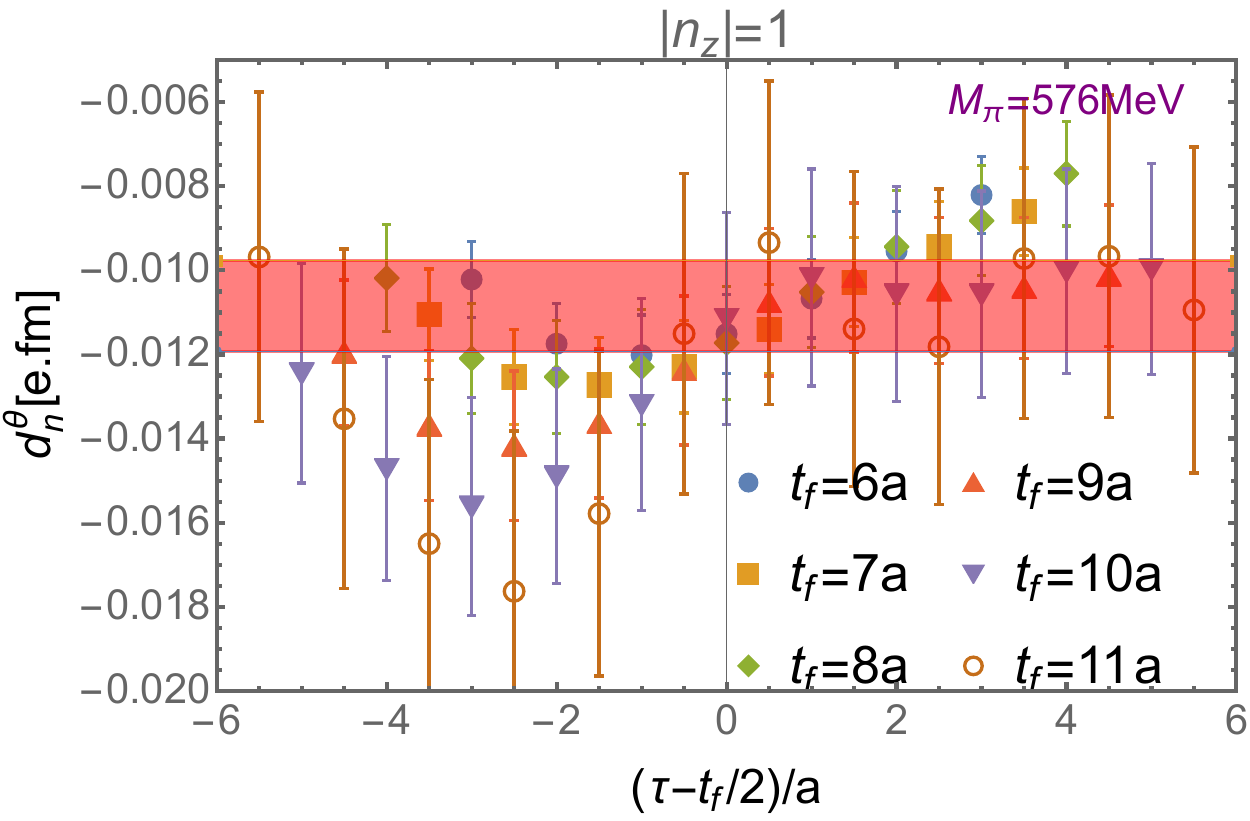}
\\
\includegraphics[width=.32\textwidth]{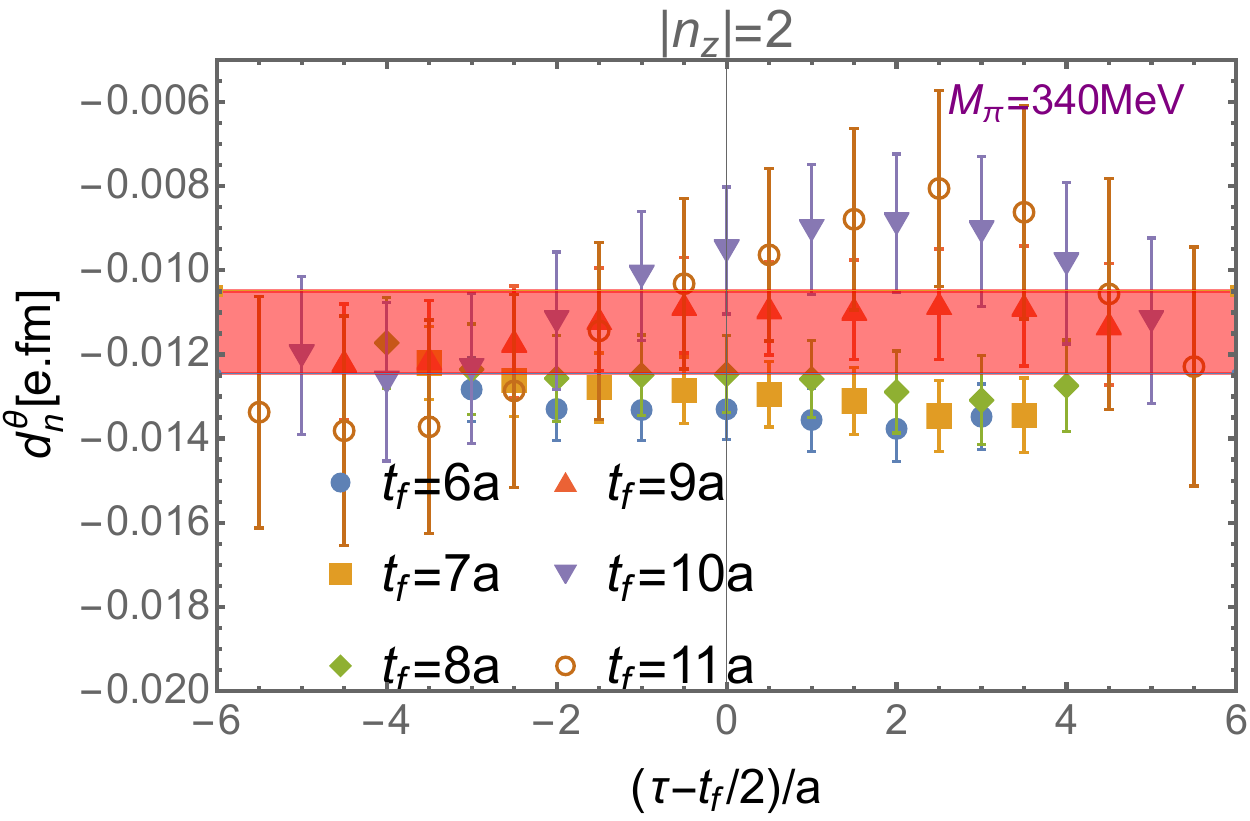}
\includegraphics[width=.32\textwidth]{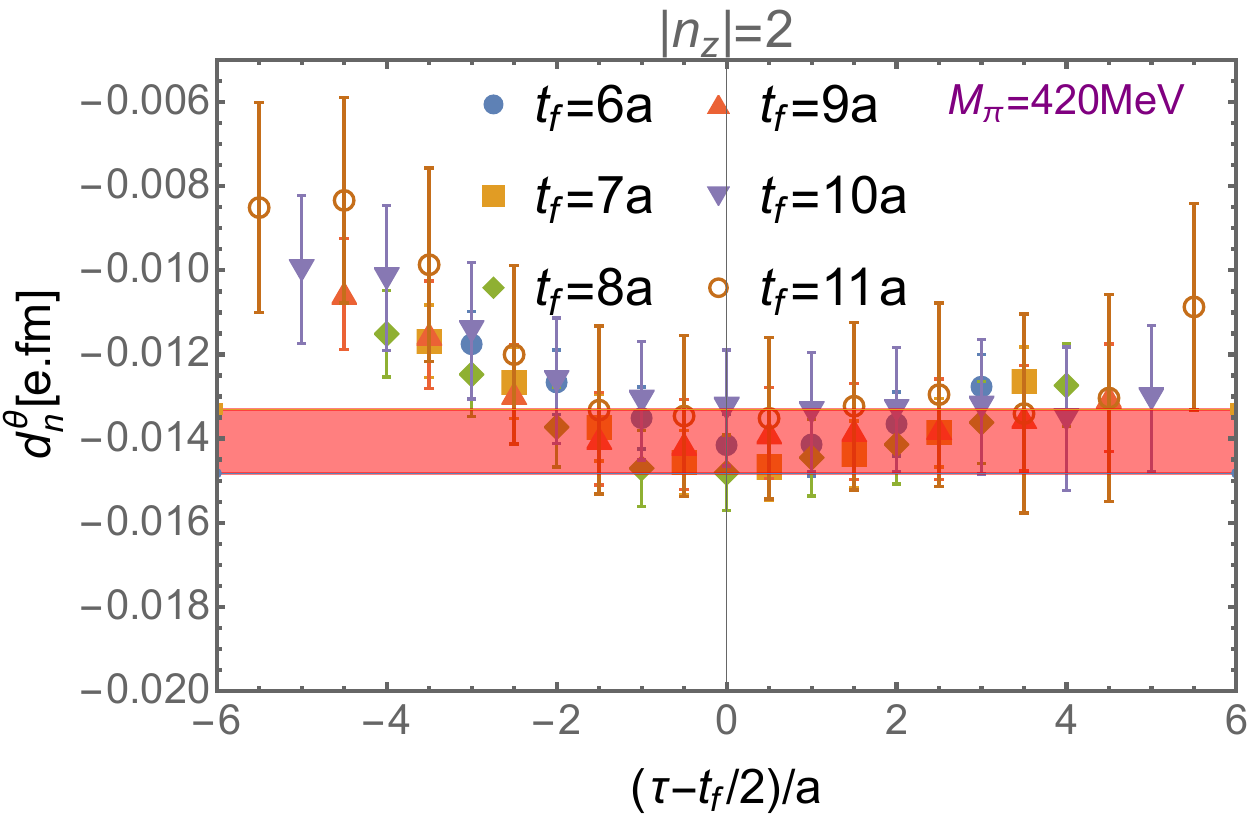}
\includegraphics[width=.32\textwidth]{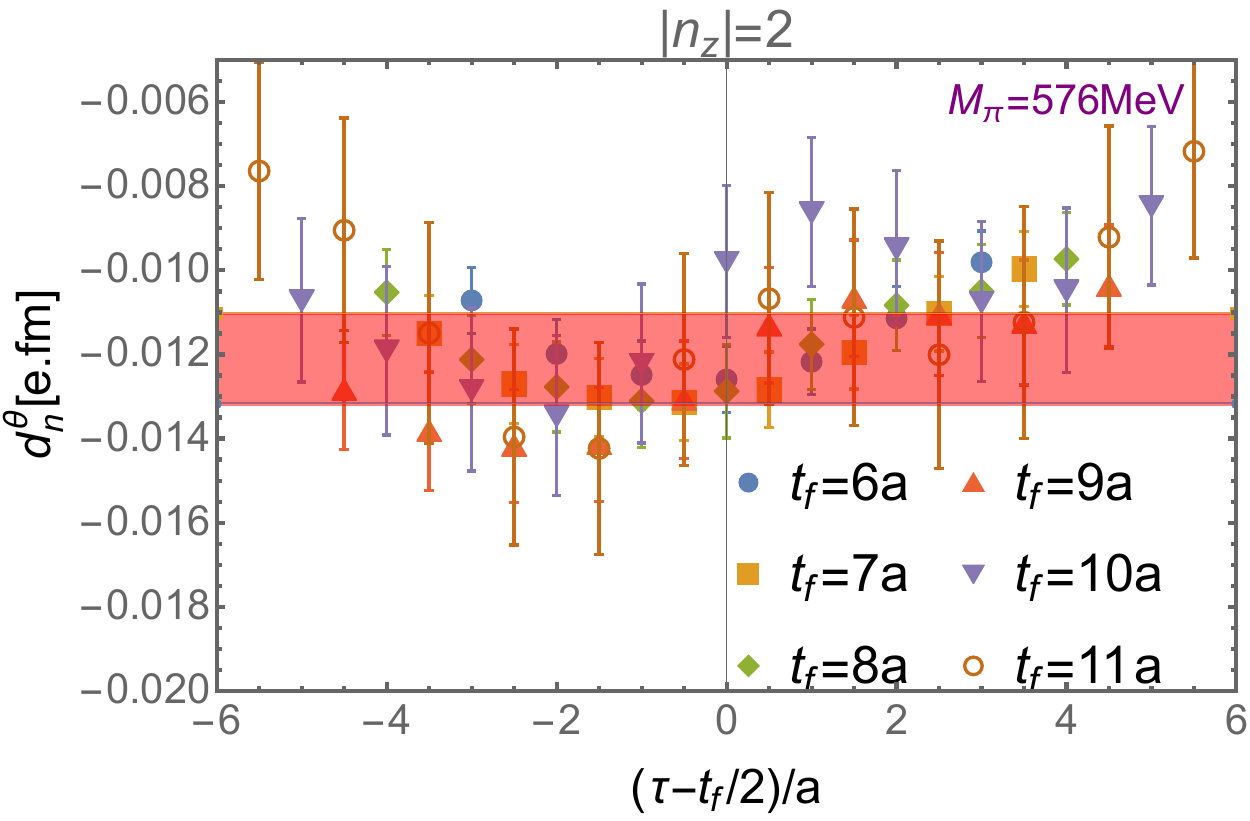}
\caption{The EDM results obtained using GEVP with operators $\{O_1,O_2,O_3,O_4\}$ on ensemble 24I-005(left),
  24I-010(middle), and 24I-020(right) with $|n_z|=1,2$ electric field. 
  The red bands represent the ground-state nEDM results obtained using a constant fit to the data points with $t_f=(8,9,10)$ in most cases, except for the results with $|n_z|=2$ on ensemble 24-005 and 24I-020, where the fit region is chosen to as $t_f=(9,10,11)$ to ensure a good $\chi^2/d.o.f$. For each $t_f$, we select the three central points for the fit.}
\label{fig:multiGEVP_res_igf4}
\end{figure}

To obtain the final results, we perform a constant fit to the results with $t_f=(8,9,10)a$. 
For each $t_f$, we select the three central points and carry out a correlated constant fit. 
To estimate the systematic uncertainty of the fit, we perform a similar fit to the data with $t_f=(9,10,11)a$ and take
the difference between the two fits as their systematic uncertainty. 
The fit results are shown as red bands in Fig.~\ref{fig:multiGEVP_res_igf4}.
In most cases, the fits yield good $\chi^2/d.o.f$, typically smaller than 1.
The only exception is in the case with $|n_z|=2$ electric field and $m_\pi=340$ MeV, where the results with different
$t_f$ show larger variation: the fit to $t_f=(8,9,10)a$ data gives $\chi^2/d.o.f~\sim 1.5$, while the fit to
$t_f=(9,10,11)a$ data yields $\chi^2/d.o.f~\sim 1$. 
A similar behavior is observed for $|n_z|=2$ and with pion mass of $m_\pi=576$ MeV. 
In these two cases, we adopt the latter fit as the central value and take the difference between the two fits as the
systematic uncertainty.
The final results of nEDM on these three ensembles are summarized in Table~\ref{tab:nedm_res}.

%------------------------------------------------------------------------------

%------------------------------------------------------------------------------

%------------------------------------------------------------------------------

%------------------------------------------------------------------------------
\begin{table*}[htbp]
\centering
\caption{The final nEDM results on the ensembles with different pion masses.
  \label{tab:nedm_res}}
\begin{tabular}{|c|c|c|c|} 
\hline\hline
\multirow{2}{*}{Ensemble} & \multirow{2}{*}{$M_\pi$(MeV)}  &  \multicolumn{2}{c|}{
$-d_n^\theta[e.\text{fm}]$($G\tilde{G}$)}
\\
\cline{3-4} & &  $|n_z|=1$  & $|n_z|=2$ \\
\hline
24I-020 &  576  &   0.0109(11)(11) &  0.0122(11)(05)    \\
24I-010 &  420  &  0.0152(10)(05)  &  0.0141(08)(03)    \\
24I-005 &  340 &  0.0128(12)(04)  &   0.0115(10)(12)    \\
\hline\hline
\end{tabular}
\end{table*}

\begin{figure}[ht!]
\centering
\includegraphics[width=.6\textwidth]{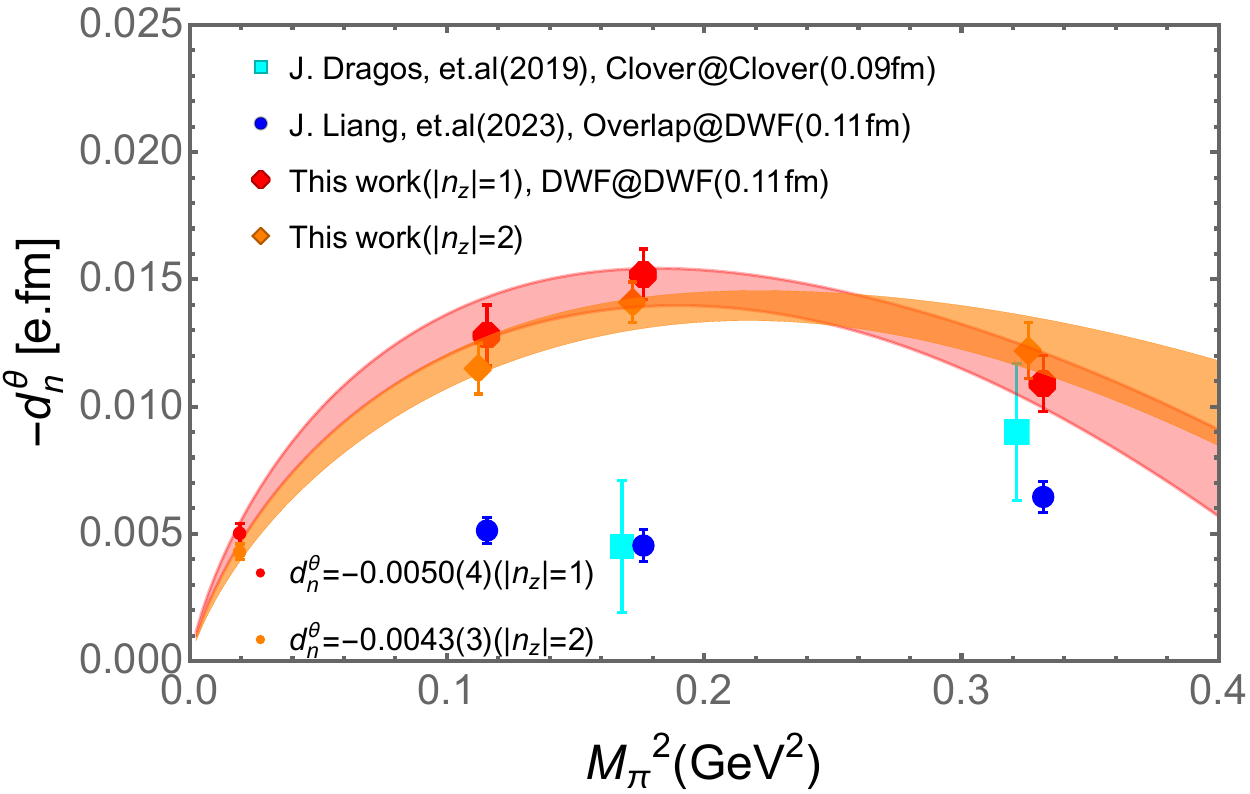}
\caption{The comparison of our nEDM result to other previous works.
  The cyan data point is from ~\cite{Dragos:2019oxn}, which is obtained using clover fermion for both valence quark and
  sea quark, with a lattice spacing of $a\approx0.09$ fm.
  The blue data points are from ~\cite{Liang:2023jfj}, in which the authors use the same ensembles as in this work 
  but with overlap-action valence  quarks. 
  The red and orange bands represent the chiral extrapolation results for two different electric field strengths
  $|n_z|=1,2$.
  \label{fig:EDM_compa}
}
\end{figure}

Comparison of nEDM results obtained in this work with those from other groups is shown in Fig.~\ref{fig:EDM_compa}.
The cyan data point is from Ref.~\cite{Dragos:2019oxn}, where the authors use the clover fermion action for both
valence quarks and sea quarks, with a lattice spacing of approximately 0.09 fm.
The blue points are from Ref. \cite{Liang:2023jfj}, which employs the same gauge ensembles as our calculation but use
the overlap fermion action for the valence quarks.
Our results are shown with the filled red circles for the $|n_z|=1$ electric field and orange diamonds for $|n_z|=2$.
We combine the results with the different pion masses for a chiral extrapolation using the following Ansatz~\cite{Crewther:1979pi, OConnell:2005mfp},
\begin{equation}
d_n^\theta=c_0 m_\pi^2+c_1m_\pi^2\log\left(\frac{m_\pi^2}{m^{2}_{N,phy}}\right)\,.
\end{equation}

The red and orange bands represent the chiral extrapolation results for the two electric field strengths.
At the physical point, we obtain $d_n^\theta=-0.0050(4)(4)$ for $|n_z|=1$ and $d_n^\theta=-0.0043(3)(3)$ for $|n_z|=2$, where the first uncertainty is statistical and the second one is the systematic uncertainty associated with the choice of constant fit range used to extract the ground state nEDM. We take the result with $|n_z|=1$ as our final result since it has smaller higher-order corrections in $\mcE_z$.
The difference between the results with $|n_z|=1,2$ is taken as an additional systematic uncertainty. Adding this with the systematic uncertainty arising from the fit region in quadrature, our final result
at the physical point is 
\begin{equation}
d_n^\theta/\bar\theta=-0.0050(4)^\text{stat}(8)^\text{sys} \,.
\end{equation}

Remaining systematic uncertainties including discretization error, finite volume effects and autocorrelation of topological charge, are beyond the scope of the present study.
Both aforementioned groups extract the nEDM using the form factor method, whereas we employ the background field
method. Unlike in the form factor method, there is no systematic uncertainty due to the extrapolation of the electric dipole form factor to the forward limit.
In addition, although the authors of Refs.~\cite{Dragos:2019oxn,Liang:2023jfj} use the global topological charge, the summation region of the topological charge density is
truncated to reduce the gauge field noise.
The choice of summation region can introduce additional systematic uncertainties.
In contrast, we use the local topological charge operator and employ the Feynman-Hellmann theorem, which does not
require introducing a cut-off in the summation of the topological charge over the lattice time extent.
Furthermore, we find that the commonly used nucleon interpolating operator $O_1$ is subject to significant excited state
contamination in the nEDM calculation, at least in our method.
We have observed that the extracted values can even change sign if excited states are not treated properly.
This is reminiscent of nEDM results that change sign depending on whether the $N\pi$ state is included or not in the fit Ansatz as observed in Ref.~\cite{Bhattacharya:2021lol}.
In our calculation, the dominant contamination arises from the mixing between the ground state and first excited state in the presence of electric field, and the corresponding mass gap is comparable to the mass difference between the nucleon $N$ and its negative-parity partner state $N^*$.
We control this effect by employing the GEVP method spanning positive and negative parity-projected nucleon operators to
reduce the excited state contamination, and observe consistency between the results obtained with different subsets of
nucleon operators.
We speculate that this may be the crucial advantage over the traditional approach that typically employs single positive parity-projected nucleon operator 
$\psi_N=(1+\gamma_4)\mcO$ and thus may struggle to remove such contamination using only conventional multi-state fits.
This subtle excited-state contamination may also contribute to the difference between our results and those of
other groups.

The summary of recent lattice results for the contribution of QCD $\Theta$-term to nEDM is presented in
Table~\ref{tab:nedm_summ}. 
The calculations in Refs.~\cite{Alexandrou:2020mds} and~\cite{Bhattacharya:2021lol} employ the direct extraction of
the nEDM using the form factor method with global topological charge at the physical point, which results in relatively
large statistical uncertainties.
As we have shown in Figs.~\ref{fig:global_g},\ref{fig:global_q}, it is very challenging to obtain a good nEDM signal when
global topological charge is used in the nEDM calculation.
Instead, our method based on the local topological charge provides an alternative strategy for computing the nEDM
directly in the limit of forward kinematics, which should lead to comparatively improved statistical precision. Despite this advantage, the pion masses used in the present study are relatively heavy; extending this calculation to lighter pion masses will therefore be crucial to robustly control the chiral extrapolation.

%------------------------------------------------------------------------------

%------------------------------------------------------------------------------
\begin{table*}[htbp]
\centering
\caption{Summary of lattice QCD results for the QCD $\Theta$ term-induced neutron EDM.
  \label{tab:nedm_summ}}
\begin{tabular}{lc|rl} 
\hline\hline
&&&  $[10^{-3}\,e\cdot\mathrm{fm}]$ \\
\hline
Dragos et al(2019) & \cite{Dragos:2019oxn}
  & $d_n/\bar\theta=$ & $-1.5(7)$    \\
Alexandrou et al(2020) & \cite{Alexandrou:2020mds}  
  & $|d_n/\bar\theta|=$ & $\phantom{-}0.9(2.4)$  \\
Bhattacharya et al (2021) & \cite{Bhattacharya:2021lol} 
  & $d_n/\bar\theta=$ & $-28(18)(54)$  \\
Liang et al (2023) & \cite{Liang:2023jfj} 
  & $d_n/\bar\theta=$ & $-1.48 (0.14) (0.31)$\\
This work & 
  & $d_n/\bar\theta=$ &  $-5.0(0.4)(0.8)$ \\
\hline\hline
\end{tabular}
\end{table*}

%%%%%%%%%%%%%%%%%%%%%%%%%%%%%%%%%%%%%%%%%%%%%%%%%%%%%%%%%%%%%%%%%%%%%%%%%%%%%%%%
%%%%%%%%%%%%%%%%%%%%%%%%%%%%%%%%%%%%%%%%%%%%%%%%%%%%%%%%%%%%%%%%%%%%%%%%%%%%%%%%
\section{Summary and Conclusions}\label{sec:conc}
In this work, we have calculated the $\Theta$-induced neutron EDM using the background field method in the framework of  lattice QCD.
One of the advantages of this method is that only nucleon two-point correlation functions 
are needed, and three-point correlation functions of the vector current are not required.
Another advantage is that the nEDM can be directly extracted as the nucleon forward matrix element at rest rather than 
as the electric dipole form factor $F_3(Q^2)$ must be extrapolated to the forward limit, 
as is done in the conventional form factor method.
The lattice ensembles used in this calculation have lattice spacing $a\sim0.11$fm and three pion masses of 340MeV,
420MeV and 576MeV.
We have studied only one lattice spacing value, and have no way to assess the discretization error.
However, we believe they must be moderate due to automatic improvement of the chirally-symmetric Domain-Wall action that
we have used. 

In the background field method, the nEDM is related to the energy shift of the neutron in the presence of the background
electric field.
However, we found that the nEDM extracted using global topological charge suffers from large statistical uncertainty.
To improve the signal, we propose a new method to determine the nEDM from the matrix element of the local topological charge
within the nucleon ground state in the presence of electric field.
This method is based on the Feynman-Hellmann theorem, which relates the energy shift induced by an external source to
the matrix element of the corresponding local operator.
The  nucleon ground state in the presence of background field is obtained by solving the generalized eigenvalue
problem from the correlators between states of opposite parity. 

To validate this method, we compare the nEDM obtained using different nucleon interpolating operators.
We find that the results are consistent across different choices of interpolating operators, regardless of whether the
operators are covariant under chiral transformation.
However, significant differences can arise if the external state is not projected to the correct eigenstate, due to
excited state contamination. 
We also compare the results using topological charge defined from the gluon field with gradient flow and the pseudoscalar
quark density and find that the latter may suffer from large $N\pi$ state contamination.
However, as the loop quark mass increases, the excited state contamination is reduced and the results start to agree with
those obtained using the gluon topological charge.
Furthermore, we verify that the ABJ anomaly relation is satisfied by combining contributions from the axial-vector
divergence and the pseudoscalar quark density.

We compare our results with those reported in Ref~\cite{Dragos:2019oxn,Liang:2023jfj}, which are obtained using form
factors method at pion masses similar to those used in this calculation.
In these calculations, the summation region of the topological charge density is restricted in order to improve the
signal.
However, this procedure might introduce the additional systematic uncertainties associated with the shape and the size
of the cut-off imposed on the summation region.
The statistical precision of $\theta$-induced nEDM is typically very poor, which makes it difficult to determine whether
the effects of the CP-violating interaction have been truly saturated and the cut-offs do not introduce bias.
The direct calculations at the physical point reported in Ref~\cite{Alexandrou:2020mds,Bhattacharya:2021lol} 
and performed with the global topological charge without cut-off do not show statistically-significant nEDM signal.

The most important distinction of our method is that by employing the GEVP we can remove excited state contamination caused
by mixing with the nucleon parity partner and its excited states.
We find that this procedure applied to the most widely used interpolating operator $O_1$ changes the result dramatically
so that the sign is changed.
Notably, the sign of resulting nEDM values in Ref.~\cite{Bhattacharya:2021lol} also change sign  depending on whether
the $N\pi$ states are included in the fit.
In our analysis we cannot discriminate $N\pi$ states from nucleon-like excited states, although we find that the energy gaps do not depend on the pion mass.
Still, our variational studies with multiple nucleon operators indicate that substantial contamination may be present in the EDM calculations and must be carefully treated.

In the future, we plan to investigate the result at lighter pion mass to improve the chiral extrapolation to the physical point and to explore finer lattices to examine discretization effects, and larger volume to assess the finite volume effects.
Further, this method can be easily applied to other $\CPV$ operators, and might be especially beneficial for the nEDM
induced by Weinberg's three-gluon operator and 4-quark $\CPV$ operators. 

%%%%%%%%%%%%%%%%%%%%%%%%%%%%%%%%%%%%%%%%%%%%%%%%%%%%%%%%%%%%%%%%%%%%%%%%%%%%%%
\section{Acknowledgements}
We thank Tanmoy Bhattacharya, Kaori Fuyuto, Rajan Gupta, Keh-Fei Liu, Jian Liang, Emanuele Mereghetti, Andrea Shindler for fruitful discussions. We thank Michael Abramczyk for valuable discussions and contributions during the initial stages of this project. FH and SS are supported by the National Science Foundation under award PHY-2412963. FH is also supported by the U.S. Department of Energy, Office of Science, Office of Nuclear Physics, under Grant No. DE-SC0013065. TB is supported by DOE grant DE-SC0010339. LJ is supported by DOE grant DE-SC0021147, DE-SC0010339, and DE-SC0026314. TI is supported by US DOE Contract DESC0012704(BNL) and the Scientific Discovery through Advanced Computing (SciDAC) program LAB 22-2580, and also Laboratory Directed Research and Development (LDRD No. 23 - 051) of BNL and RIKEN-BNL Research Center. (Any opinions,
findings, and conclusions or recommendations expressed in this material are those of the author(s) and do not
necessarily reflect the views of the National Science Foundation.)
HO is supported by the JSPS KAKENHI (Nos. 21K03554, 22H00138).

The research reported in this work made use of computing and long-term storage
facilities of the USQCD Collaboration, which are funded by the Office of Science
of the U.S. Department of Energy.
The calculations were performed with the Grid~\cite{Boyle:2015tjk}, Hadron~\cite{antonin_portelli_2023_8023716} and Qlua~\cite{Pochinsky:QLua} software suites.

%\clearpage
\appendix
%%%%%%%%%%%%%%%%%%%%%%%%%%%%%%%%%%%%%%%%%%%%%%%%%%%%%%%%%%%%%%%%%%%%%%%%%%%%%%%%
%%%%%%%%%%%%%%%%%%%%%%%%%%%%%%%%%%%%%%%%%%%%%%%%%%%%%%%%%%%%%%%%%%%%%%%%%%%%%%%%
\section{Conventions in Minkowski and Euclidean space}\label{sec:appenx_EM}
Using the convention in \cite{Abramczyk:2017oxr,Bhattacharya:2021lol}, the Euclidean gamma matrices are related to their Minkowski counterparts through
\begin{equation}
\gamma^4_E=\gamma^0_M\,,
\quad\gamma^i_E=-i\gamma^i_M\,,
\quad\gamma^5_E=-\gamma^5_M\,,
\quad\sigma^{ij}_E=-\sigma^{ij}_M\,,
\quad\sigma^{4i}_E=-i\sigma^{0i}_M
\end{equation}
with $\gamma^5_E=\gamma^1_E\gamma^2_E\gamma^3_E\gamma^4_E$ and $\gamma^5_M=i\gamma^0_M\gamma^1_M\gamma^2_M\gamma^3_M$.
The momentum conventions in Minkowski and Euclidean space are related by
\begin{equation}
q_E^4=iq_M^0,
\quad q_E^i=q_M^i,
\quad x_E^4=ix^0_M,
\quad x_E^i=x_M^i,
\quad q_E\cdot q_E=-q_M\cdot q_M\,.
\end{equation}
The transformations of derivatives and vector potential between Minkowski and Euclidean space are given by
\begin{equation}
\partial^4_E=-i\partial^0_M,
\quad \partial^i_E=-\partial^i_M,
\quad A^4_E=-iA^0_M,
\quad A^i_E=-A^i_M,
\end{equation}
then the covariant derivative transforms $D^4_E=-iD^0_M$ and $D^i_E=-D^i_M$.
The gauge field strength tensor transforms as
\begin{equation}
G^{4i}_E= iG^{0i}_M,
\quad G^{ij}_E=G^{ij}_M.
\end{equation}
The color electromagnetic field transforms as
\begin{equation}
\mathcal{E}^i=-G^{4i}_E=-iG^{0i}_M=i\epsilon^i,
\quad\mathcal{H}^i=\frac{1}{2}\epsilon_{ijk}G^{jk}_E=\frac{1}{2}\epsilon_{ijk}G^{jk}_M=-H^i.
\end{equation}
The fermionic Lagrangian density transforms according to
\begin{equation}
\bar{\psi}(i\slashed{D}-m)_M\psi=-\bar{\psi}(\slashed{D}+m)_E\psi.
\end{equation}
Similarly for the gauge field,
\begin{equation}
(G_{a\mu\nu}G^{a\mu\nu})_M=-2(G^{a0i}_MG^{a0i}_M-G^{aij}_MG^{aij}_M)=2(G^{a0i}_EG^{a0i}_E+G^{aij}_EG^{aij}_E)=(G_{a\mu\nu}G^{a\mu\nu})_E.
\end{equation}
Therefore, the full action transforms as
\begin{equation}
\begin{aligned}
iS_M  &= i\int d^4 x_M \left[\bar{\psi}(i\slashed{D}-m)\psi-\frac{1}{4}G_{a\mu\nu}G^{a\mu\nu}(x)\right]
\\    &= -\int d^4 x_E \left(\bar{\psi}(\slashed{D}_E+m)\psi+\frac{1}{4}\left[G_{a\mu\nu}G^{a\mu\nu}(x)\right]_E\right)=-S_E.
\end{aligned}
\end{equation}
The Levi–Civita symbol in Minkowski and Euclidean space is defined by
\begin{equation}
\epsilon_E^{1234}=\epsilon_M^{0123}=1,
\quad \epsilon_{1234,E}=-\epsilon_{0123,M}=1.
\end{equation}
Accordingly, the global topological charge is given by
\begin{equation}
\begin{aligned}
Q_{top,M} &= \int d^4x_M\frac{g^2}{32\pi^2}\epsilon_M^{\mu\nu\alpha\beta}Tr[G_{\mu\nu}G_{\alpha\beta}]_M
\\&= -\int d^4x_E\frac{g^2}{32\pi^2}\epsilon_E^{\mu\nu\alpha\beta}Tr[G_{\mu\nu}G_{\alpha\beta}]_E,
  = -Q_{top,E}.
\end{aligned}
\end{equation}
The topological charge on a single time slice transforms as
\begin{equation}
\label{eq:tolog}
\begin{aligned}
q_{top,M} &= \int d^3x_M\frac{g^2}{32\pi^2}\epsilon_M^{\mu\nu\alpha\beta}Tr[G_{\mu\nu}G_{\alpha\beta}]_M
\\&= -i\int d^3x_E\frac{g^2}{32\pi^2}\epsilon_E^{\mu\nu\alpha\beta}Tr[G_{\mu\nu}G_{\alpha\beta}]_E,
   = -iq_{top,E}.
\end{aligned}
\end{equation}

Finally, the QCD action including the $\theta$ term transforms as
\begin{equation}
e^{iS_M+i\bar{\theta} Q_{top,M}}=e^{-S_E-i\bar{\theta} Q_{top,E}}.
\end{equation}
The tensor interaction in Euclidean space can be related to the one defined in Minkowski space, 
\begin{equation}
\begin{aligned}
(\sigma^{4\nu}q_\nu)_E &= \sigma^{4j}_Eq_E^j=-i\sigma^{0j}_Mq_M^j=(i\sigma^{0\nu}q_\nu)_M,\\
(\sigma^{i\nu}q_\nu)_E &= \sigma^{i4}_Eq_E^4+\sigma^{ij}_Eq_E^j=\sigma^{i0}_Mq_M^0-\sigma^{ij}_Mq_M^j=(\sigma^{i\nu}q_\nu)_M.
\end{aligned}
\end{equation}
The CP violating matrix element of the vector current in Minkowski is expressed as
\begin{equation}
\la p',\lambda'|\bar{q}\gamma^\mu q|p,\lambda\ra_\CPV 
  = \bar{u}(p',\lambda')\left(F_1\gamma^\mu+(F_2+i\gamma_5F_3)\frac{i\sigma^{\mu\nu}q_\nu}{2m}\right) u(p,\lambda).
\end{equation}
If we transform to  Euclidean space, we have 
\begin{equation}
\begin{aligned}
\la p',\lambda'|\bar{q}\gamma^4_E q|p,\lambda\ra_\CPV 
  &= \la p',\lambda'|\bar{q}\gamma^0_M q|p,\lambda\ra_\CPV
\\&= \bar{u}(p',\lambda')\left(F_1\gamma^0_M+(F_2+i\gamma^5_MF_3)\frac{(i\sigma^{0\nu}q_\nu)_M}{2m}\right) u(p,\lambda)  
\\&= \bar{u}(p',\lambda')\left(F_1\gamma^4_E+(F_2-i\gamma^5_EF_3)\frac{(\sigma^{4\nu}q_\nu)_E}{2m}\right) u(p,\lambda) 
\\
\la p',\lambda'|\bar{q}\gamma^i_E q|p,\lambda\ra_\CPV 
  &= -i\la p',\lambda'|\bar{q}\gamma^i_M q|p,\lambda\ra_\CPV
\\&= -i\bar{u}(p',\lambda')\left(F_1\gamma^i_M+(F_2+i\gamma^5_MF_3)\frac{(i\sigma^{i\nu}q_\nu)_M}{2m}\right) u(p,\lambda)
\\&= \bar{u}(p',\lambda')\left(F_1\gamma^i_E+(F_2-i\gamma^5_EF_3)\frac{(\sigma^{i\nu}q_\nu)_E}{2m}\right) u(p,\lambda).
\end{aligned}
\end{equation}
So the Euclidean CP violating matrix element of vector current should be expressed as
\begin{equation}
\la p',\lambda'|\bar{q}\gamma_E^\mu q|p,\lambda\ra_\CPV 
  = \bar{u}(p',\lambda')\left(F_1\gamma^\mu+(F_2-i\gamma_5F_3)\frac{\sigma^{\mu\nu}q_\nu}{2m}\right)_E u(p,\lambda),
\end{equation}
and the effective Euclidean Dirac operator is expressed as
\begin{equation}
(\slashed{D}+m)\psi(x)=(\gamma_\mu\partial^\mu+iJ^\mu {\cal A}_\mu(x)+m)\psi(x),
\end{equation}
where $J^\mu$ is the EM current. The plane-wave fields $\psi_p(x)$ and $\mcA_{q,\mu}(x)$ depend on 
the Euclidean 4-momenta $p$, $q$ as
\begin{equation}
\begin{gathered}
\label{eqn:momenta_euc}
\psi_p(x)\sim e^{i p x}\,,
\,\, \partial_\mu\psi_p(x) \leftrightarrow i p_\mu u_p\,;
\quad
\mcA_{q,\mu}(x) \sim e^{i(p^\prime-p) x} = e^{i q x}\,, 
\,\, \partial_\nu \mcA_{q,\mu}(x) \leftrightarrow i q_\nu \mcA_\mu\,.
\end{gathered}
\end{equation}

The Dirac operator becomes 
\begin{equation}
\lp[ i \slashed{p} + m_N 
  -\big(\frac12 F_{\mu\nu}\sigma^{\mu\nu}\big)\frac{F_2(0) - iF_3(0)\gamma_5}{2m_N} \rp] \,.
\end{equation}

The ABJ anomaly relation in Minkowski space is
\begin{equation}
Z_A\partial_\mu(\bar{\psi}\gamma^\mu\gamma^5\psi)
=2im\bar{\psi}\gamma^5\psi-\frac{g^2}{16\pi^2}\epsilon^{\mu\nu\alpha\beta}Tr[G_{\mu\nu}G_{\alpha\beta}].
\end{equation}

Using the above relations, we can obtain the ABJ relation in Euclidean space,
\begin{equation}
Z_A\partial_E^\mu(\bar{\psi}\gamma_E^\mu\gamma_E^5\psi)
  = 2m\bar{\psi}\gamma_E^5\psi-\frac{g^2}{16\pi^2}\epsilon_E^{\mu\nu\alpha\beta}Tr[G_{\mu\nu}G_{\alpha\beta}]_E.
\end{equation}
Since the gluon topological charge density is defined as $\rho_Q=\frac{g^2}{32\pi^2}\epsilon_E^{\mu\nu\alpha\beta}Tr[G_{\mu\nu}G_{\alpha\beta}]_E$, we have 
\begin{equation}
\rho_Q=m\bar{\psi}\gamma_E^5\psi-Z_A\frac{\partial_E^\mu(\bar{\psi}\gamma_E^\mu\gamma_E^5\psi)}{2}.
\end{equation}
Alternatively, if we use the pseudoscalar quark loop to define the topological charge, we arrive at
\begin{equation}
Q_{top,M} = \int d^4x_Mim\bar{\psi}\gamma^5\psi=-\int d^4x_Em\bar{\psi}\gamma_E^5\psi = -Q_{top,E}.
\end{equation}

%%%%%%%%%%%%%%%%%%%%%%%%%%%%%%%%%%%%%%%%%%%%%%%%%%%%%%%%%%%%%%%%%%%%%%%%%%%%%%%%
%%%%%%%%%%%%%%%%%%%%%%%%%%%%%%%%%%%%%%%%%%%%%%%%%%%%%%%%%%%%%%%%%%%%%%%%%%%%%%%%
\section{Gamma matrices in the Dirac basis}\label{sec:appenx_diracbas}
The gamma matrices used in Sec~\ref{sec:edm_mag} are defined in the Dirac basis,
\begin{equation}
\label{eqn:gammamatr_dirac}
\begin{gathered}
\gamma_1=\left(\begin{array}{cc} 0 & -i\sigma_x \\ i\sigma_x & 0\end{array}\right)\,,
\quad\gamma_2=\left(\begin{array}{cc} 0 & -i\sigma_y \\ i\sigma_y & 0\end{array}\right)\,,
\quad\gamma_3=\left(\begin{array}{cc} 0 & -i\sigma_z \\ i\sigma_z & 0\end{array}\right)\,,\\
\gamma_4=\left(\begin{array}{cc} \mathbb{I} & 0 \\ 0 & -\mathbb{I}\end{array}\right)\,,
\quad\gamma_5=\gamma_1\gamma_2\gamma_3\gamma_4=\left(\begin{array}{cc} 0 & -\mathbb{I} \\ -\mathbb{I} & 0\end{array}\right)\,.
\end{gathered}
\end{equation}

%%%%%%%%%%%%%%%%%%%%%%%%%%%%%%%%%%%%%%%%%%%%%%%%%%%%%%%%%%%%%%%%%%%%%%%%%%%%%%
%%%%%%%%%%%%%%%%%%%%%%%%%%%%%%%%%%%%%%%%%%%%%%%%%%%%%%%%%%%%%%%%%%%%%%%%%%%%%%
\section{Anti-Hermitian perturbation theory
  \label{sec:app_antiherm}}

Since the electric field on a Euclidean lattice is imaginary, it mixes positive-parity ($N$) and negative-parity ($N^*$) states via an anti-Hermitian interaction.
It is instructive to understand the consequences of this fact  in a simple example.
Consider a two-level system with energies $\vep_{0,1} = \bar\vep \mp \frac12\Delta\vep$.
We drop the average energy $\bar\vep$ without loss of generality.
The goal is to understand how to solve the Euclidean-time evolution of the system,
$\pd\psi t = -\mcH\psi$, with the Hamiltonian
\begin{equation}
\label{eqn:nonherm_example}
\mcH + V
= \lp(\begin{array}{cc} -\frac12 \Delta\vep & \\ & +\frac12\Delta\vep\end{array}\rp)
+ \lp(\begin{array}{cc} & \alpha \\ -\alpha & \end{array}\rp)
= -\frac12\Delta\vep\sigma_3 + i \alpha\sigma_2,
\end{equation}
where $\Delta\vep$ and $\alpha$ are real positive.
To solve the above equation 
one must exponentiate the non-Hermitian
Hamiltonian $\mcH$ that has eigenvalues
\begin{equation}
\lambda_{0,1} = \mp \sqrt{\lp(\frac12\Delta\vep\rp)^2 - \alpha^2}
\end{equation}
which are real if the anti-Hermitian perturbation is small, $\alpha<\frac12\Delta\vep$,
but the corresponding eigenvectors are not orthogonal.
Consider a general matrix $X = b \sigma_3 + i a \sigma_2$, where $a,b$ are real.
Since $X^2 = b^2 - a^2$, its exponent is
\begin{equation}
\begin{aligned}
e^{b\sigma_3 + ia\sigma_2} 
&= \sum_{n=0}^\infty \Big[\frac{(b^2 - a^2)^n}{(2n)!} \Big]
  + \frac{b\sigma_3 + ia\sigma_2}{\sqrt{b^2-a^2}} 
  \sum_{n=0}^\infty \Big[\frac{(b^2 - a^2)^{n+\frac12}}{(2n+1)!}\Big] \\
&= \cosh\sqrt{b^2 - a^2} + \frac{b\sigma_3 + ia\sigma_2}{\sqrt{b^2-a^2}}\sinh\sqrt{b^2-a^2} \\
&= \mcP_+ e^{\sqrt{b^2 - a^2}} + \mcP_- e^{-\sqrt{b^2 - a^2}} \,.
\end{aligned}
\end{equation}
where the projectors are not Hermitian for $a\neq 0$,
\begin{equation}
\mcP_\pm = \frac12\lp(1\pm \frac{b\sigma_3 + ia\sigma_2}{\sqrt{b^2-a^2}}\rp) = \mcP_\pm^2\,,
\quad \det\mcP_\pm = 0\,.
\end{equation}
The exponentials determine the evolution of the new eigenstates.
The energy gap becomes smaller for larger $\alpha$
 unlike in the usual Hermitian theory, where the energy levels are repelled.
Note that the above holds only for anti-Hermitian perturbations smaller than half of the unperturbed energy gap,
{\it i.e.} $a<b$ or $\alpha<\frac12\Delta\vep$.
If $a>b$ ($\alpha>\frac12\Delta\vep$) then both exponentials become complex phases. The argument does not directly apply to the case of color electric field in QCD. The transfer matrix of the Wilson gauge action is Hermitian and non-negative~\cite{Luscher:1976ms}, of which prerequisite that our setup with the constant background electric field breaks.

In a more general case, one can calculate the energy eigenvalues and eigenstates 
as a perturbation in $V$ order-by-order,
\begin{equation}
\label{eqn:gen_epair_eqn_right}
\lp(\mcH + V\rp)\lp(\psi_\lambda + a^{(1)}_{\mu\lambda} \psi_\mu + \ldots\rp) 
  = \lp( E_\lambda + E^{(1)}_\lambda + E^{(2)}_\lambda \ldots\rp)
    \lp( \psi_\lambda + a^{(1)}_{\mu\lambda}\psi_\mu + \ldots\rp)\,,\
\end{equation}
where the unperturbed  eigenvectors $\psi_\lambda$ are orthogonal, and the corresponding energy levels
$E_\lambda$ are real for the Hermitian Hamiltonian $\mcH$.
The perturbation $V$ is an arbitrary operator (it may be anti-Hermitian), for which we require only
that its diagonal elements $\la \lambda|V|\lambda\ra=0$ so that the energy levels $E_\lambda$ do not get
(potentially complex) corrections at the first order.
This fits our application in which the original, unperturbed theory is parity-conserving and has
non-degenerate definite-parity eigenstates, while the anti-Hermitian perturbation operator is
parity-odd and its diagonal elements are zero in the respective eigenvector basis.
We will further assume that all matrix elements of $V$ vanish between degenerate energy states. In our case, for example, the matrix elements of the dipole moment $\mcD_z$ vanish between the degenerate spin-up/down states of the same unperturbed nucleon-like state with definite parity. 

The $O(V)$ and $O(V^2)$ terms of Eq.~(\ref{eqn:gen_epair_eqn_right}) yield\footnote{
  Following the standard method, we set undefined $a^{(1)}_{\lambda\lambda}=0$.}
\begin{equation}\label{eq:vemf}
\begin{gathered}
E^{(1)}_\lambda = 0\,,\quad
a^{(1)}_{\mu\lambda}|_{\mu\ne\lambda} = \frac{V_{\mu\lambda}}{E_\lambda - E_\mu}\,,\\
E^{(2)}_\lambda = \sum_{\mu\ne\lambda} \frac{ V_{\lambda\mu} V_{\mu\lambda} } {E_\lambda - E_\mu} \,.
\end{gathered}
\end{equation}
The above expression yields, to order $O(V)$, the right eigenvector 
\begin{equation}
\psi^{(0+1)R}_\lambda = \psi_\lambda 
  + \sum_{\mu\ne\lambda} \frac{V_{\mu\lambda}}{E_\lambda - E_\mu}\psi_\mu \,.
\end{equation}
An equation similar to Eq.~(\ref{eqn:gen_epair_eqn_right}) can be solved for the left
eigenvector to give
\begin{equation}
\psi^{(0+1)L\dag}_\lambda = \psi^\dag_\lambda 
  + \frac{V_{\lambda\mu}}{E_\lambda - E_\mu}\psi^\dag_\mu,
\end{equation}
corresponding to the same eigenvalue $E_\lambda + O(V^2)$.
Note that the left- and right- eigenvectors are bi-orthogonal,
\begin{equation}
\psi^{(0+1)L\dag}_{\lambda'} \psi^{(0+1)R}_\lambda 
  = \delta_{\lambda\,\lambda'} + O(V^2)\,,
\end{equation}
while the eigenvectors are orthogonal themselves only if the perturbation is Hermitian, {\it e.g.},
\begin{equation}
\psi^{(0+1)R\dag}_{\lambda'} \psi^{(0+1)R}_\lambda 
  = \delta_{\lambda\,\lambda'} 
  + 
  \frac{(V - V^\dag)_{\lambda'\lambda}}{E_\lambda - E_{\lambda'}}\big|_{\lambda\ne\lambda'}
  + O(V^2)\,,
\end{equation}
In addition, the eigenvalues will receive corrections that will ``attract'' (reduce
the gaps between) the energy levels,
\begin{equation}
E^{(2)}_\lambda = \sum_{\mu\ne\lambda} \frac{ V_{\lambda\mu} V_{\mu\lambda} } {E_\lambda - E_\mu}
  \stackrel{V^\dag=-V}\longrightarrow 
    \sum_{\mu\ne\lambda} \frac{ \lp(-|V_{\lambda\mu}|^2\rp)}{E_\lambda - E_\mu}\,,
\end{equation}
unlike in the case of Hermitian perturbation, which generally increases the gaps between energy levels. 

So far we focus only on the electric-field perturbation and do not explicitly introduce the topological charge. As discussed in the main text, the expression of nEDM is derived using Feynman-Hellman theorem. If one instead wishes to derive it using perturbation theorem, it is necessary to introduce two perturbations, the background electric field term, $i\mcE_z \mcD_z$, and topological charge term $i\bar{\theta}q_{top}$. The corresponding perturbation series can be expanded as
\begin{equation}
\label{eqn:EDM_pert}
\lp(\mcH + i\mcE_zD_z+i\bar{\theta}q_{top}\rp)\lp(\psi_\lambda + a^{(1)}_{\mu \lambda} \psi_\mu + b^{(1)}_{\mu\lambda} \psi_\mu\ldots\rp) 
  = \lp( E_\lambda + E^{(1)}_{\lambda}(\mcE_z) + E^{(1)}_{\lambda}(\bar{\theta}) + E^{(2)}_{\lambda}(\mcE_z,\bar{\theta}) \ldots\rp)
    \lp(\psi_\lambda + a^{(1)}_{\mu \lambda} \psi_\mu + b^{(1)}_{\mu\lambda} \psi_\mu\ldots\rp)
\end{equation}
where $E^{(1)}_{\lambda}(\mcE_z)$ and $E^{(1)}_{\lambda}(\bar{\theta})$ denote the energy shifts at $O(\mcE_z)$ and $O(\bar{\theta})$, respectively. Both vanish according to Eq.~(\ref{eq:vemf}). The coefficients $a^{(1)}_{\mu \lambda}$ and $b^{(1)}_{\mu \lambda}$ are given by
\begin{equation}
a^{(1)}_{\mu\lambda}|_{\mu\ne\lambda} = \frac{\la \mu|i\mcE_zD_z|\lambda\ra}{E_\lambda - E_\mu},~~~b^{(1)}_{\mu\lambda}|_{\mu\ne\lambda} = \frac{\la \mu|i\bar{\theta}q_{top}|\lambda\ra}{E_\lambda - E_\mu}
\end{equation}
$E^{(2)}_{\lambda}(\mcE_z,\bar{\theta})$ is the perturbative energy shift at  order $O(\bar\theta \mcE_z)$, which is given by
\begin{equation}
E^{(2)}_{\lambda}(\mcE_z,\bar{\theta})=\sum_{\mu\neq\lambda}\la \lambda|i\bar{\theta}q_{top}|\mu\ra a^{(1)}_{\mu\lambda}+\sum_{\mu\neq\lambda}\la \lambda|i\mcE_z\mcD_z|\mu\ra b^{(1)}_{\mu\lambda}
\end{equation}
For the ground state, one can take $|\lambda\ra=|N\ra$ and $|\mu\ra=|N^*\ra$. According to Eq.~(\ref{eq:eneshift}), the nucleon energy shift can be related to the nEDM through $E^{(2)}_{\lambda}(\mcE_z,\bar{\theta})= i\bar{\theta}\mcE_z (d_n/\bar{\theta})$, one can readily verify that the nEDM $d_n$ obtained in this way is consistent with the one given in Eq.~(\ref{eq:nEDMdef}).

The simple system~(\ref{eqn:nonherm_example}) has either real or complex-conjugate pairs of eigenvalues. 
This is a general property of matrices which are similar to their hermitian conjugates, $\mcH^\dag = S\mcH S^{-1}$. If $\lambda$ is an eigenvalue of $\mcH$,  then so is $\lambda^*$ :
\begin{equation}
\det(\mcH-\lambda)=0
\quad\Leftrightarrow\quad
\det(\mcH^\dag - \lambda^*) 
= \det(S\mcH S^{-1} - \lambda^*)
= \det(\mcH - \lambda^*) = 0\,.
\end{equation}
The similarity matrix $S$ relates the right- and left-side eigenvectors, $v_L = Sv_R$.
Such symmetry may be related to $PT$-symmetry~\cite{Bender:2023cem}.

%%%%%%%%%%%%%%%%%%%%%%%%%%%%%%%%%%%%%%%%%%%%%%%%%%%%%%%%%%%%%%%%%%%%%%%%%%%%%%%%
%%%%%%%%%%%%%%%%%%%%%%%%%%%%%%%%%%%%%%%%%%%%%%%%%%%%%%%%%%%%%%%%%%%%%%%%%%%%%%%%
\section{Estimates from nucleon polarizability\label{sec:appenx_polariz}}
In this section, we estimate the effects of the nucleon polarizability on the anomalous magnetic moment
estimator~(\ref{eq:ratiokappa}) and whether the presence of an Euclidean electric field results in complex-valued energy levels.
Estimates presented here are crude and intended to give only order-of-magnitude values.
Considering only a single spin-(1/2) excited state with $\Delta m_N = m_{N^*} - m_N \approx 0.5$ GeV, the second-order expression for the polarizability~(\ref{eqn:nucl_polariz_pert2}) yields an estimate for the matrix element of the electric dipole moment operator,
\begin{equation}
|\la N_*|D_z| N\ra| \approx \sqrt{\frac{2\pi\alpha_E\Delta m_N}{e^2}} \approx 0.46\,\mathrm{fm}\,,
\end{equation}
where we used $\alpha_E\approx1.2(1)\cdot10^{-3}\,\mathrm{fm}^3$ ~\cite{ParticleDataGroup:2024cfk},
and $e \approx \sqrt{4\pi\alpha_{EM}}\approx0.303$.
Assuming that $Z_{N^*}/Z_N\approx1$ and neglecting the vacuum polarizability,
a potential bias to $\kappa^{est}$ can be estimated as
\begin{equation}
|\Delta\kappa^{est}| \approx 2m_N^2\,\frac{Z_{N^*}}{Z_N}\,\frac{|\la N_*|D_z| N\ra|}{\Delta m_N} \approx9.2\,,
\end{equation}
which is much larger than the neutron anomalous magnetic moment $\kappa_n\approx-1.91$.
While this is likely an overestimate (in practice we find $\Delta\kappa/\kappa\approx3$), 
the fact that this bias is overlap-dependent and unavoidable renders
the $\kappa$ estimator~(\ref{eq:ratiokappa}) essentially useless.

Another question is whether the anti-Hermitian contribution of the electric dipole energy shift from an Euclidean electric field results in real-valued energies of nucleon-like states.
This is the case only if such anti-Hermitian contribution to the Hamiltonian is smaller than the mass gaps between the
original positive- and negative-parity nucleon-like states, similar to the regular perturbation theory example discussed in Appendix~\ref{sec:app_antiherm}.
Using the above dipole matrix element estimates and the typical value of the electric field 
$\mcE_0=6\pi/(L_s L_t)\approx0.039\,\mathrm{GeV}^2$ on our lattices,
the dipole energy shift is
\begin{equation}
|\la N^*|\mcE_0 \mcD_z|N\ra|\approx0.089\,\mathrm{GeV}\,.
\end{equation}
Even with the electric field up to $2\mcE_0$ used in our calculations, this matrix element is still well below the mass gap
$\Delta m_N$.
Note that the nucleon polarizability is likely governed by mixing with the $N\pi$ state~\cite{Wang:2023omf}, 
in which case the energy gap in our estimates should be $\Delta m=m_\pi\approx0.34\,\mathrm{GeV}$ in our lightest-pion
ensemble.
This would not substantially change the  crude estimates above.

%%%%%%%%%%%%%%%%%%%%%%%%%%%%%%%%%%%%%%%%%%%%%%%%%%%%%%%%%%%%%%%%%%%%%%%%%%%%%%%%
%%%%%%%%%%%%%%%%%%%%%%%%%%%%%%%%%%%%%%%%%%%%%%%%%%%%%%%%%%%%%%%%%%%%%%%%%%%%%%%%
\section{The effective mass of negative-parity component
\label{sec:appenx_of}}
In Sec.~\ref{sec:NSE}, we showed the effective mass of negative parity operator $\psi_{N^*}$ and found that the effective mass $m_{P^-}(t_f)$ exhibits an upward trend as $t_f$ increases. This behavior originates from the negative sign of the second term in the correlation $\la \psi_{N^*}(t_f)\psi^\dagger_{N^*}(0)\ra$, as shown in Eq.~(\ref{eq:NN2pt}). To investigate this effect more explicitly, we calculate the overlap factor of the negative parity component with eigenstates and use them to reconstruct the correlation function.

The squared overlap factors of optimal operators $O_{\mathcal{N}_n}$ with the eigenstates can be expressed as
\begin{equation}
\la \Omega^L |O^{L}_{\mathcal{N}_n}(0)|\mathcal{N}_n^R\rangle\langle \mathcal{N}_n^L|O^{\dagger R}_{\mathcal{N}_n}(0)|\Omega^R\rangle
  = (v^\dagger_{L,n}  C^{\psi\psi^\dag}_{2pt,\mcE}(t_0)v_{R,n}e^{m_{\mcN_n}t_0})
\end{equation}
Here, no summation over $n$ is needed. $v_{L}$ and $v_{R}$ denote the left and right eigenvectors associated with the $n-$th eigenstate. The optimized operators $O_{\mathcal{N}_n}$ are related to the parity projected operator basis $\psi_i$ through 
\begin{equation}
O^{L}_{\mathcal{N}_n}=(v^\dagger_{L})_{ni}\psi_i,~~~O^{\dagger R}_{\mathcal{N}_n}=\psi^\dagger_i(v_{R})_{in}
\end{equation}

The overlap between interpolating field $\psi_i$ and eigenstates can be obtained from
\begin{equation}
\begin{aligned}
\label{eq:am2}
\la \Omega^L |\psi_i(0)|\mathcal{N}_n^R\rangle \langle \mathcal{N}_n^L |\psi^\dagger_j(0)|\Omega^R\rangle
  &= \sum_{k,k'}(v^\dagger_L)^{-1}_{ik}\la \Omega^L |O^{L}_{\mathcal{N}_k}(0)|\mathcal{N}_n^R\rangle
    \langle \mathcal{N}_n^L|O^{\dagger R}_{\mathcal{N}_{k'}}(0)|\Omega^R\rangle (v_R)^{-1}_{k'j}
\\&= (v^\dagger_L)^{-1}_{in}\la \Omega^L |O^{L}_{\mathcal{N}_n}(0)|\mathcal{N}_n^R\rangle
  \langle \mathcal{N}_n^L|O^{\dagger R}_{\mathcal{N}_{n}}(0)|\Omega^R\rangle (v_R)^{-1}_{nj}
\\&= (v^\dagger_L)^{-1}_{in}(v^\dagger_{L,n}  C^{\psi\psi^\dag}_{2pt,\mcE}(t_0)v_{R,n}e^{m_{\mcN_n}t_0})(v_R)^{-1}_{nj}
\,.
\end{aligned}
\end{equation}
Using the above formula, one can calculate the overlap factors between the parity projected components of interpolating operators and the eigenstates in the presence of the background electric field. The squared overlap factors for the operator $O_1$ are shown in the left panel of Fig.~\ref{fig:overlap}, where $\psi_{1N}$ and $\psi_{1N^*}$ are defined as $\psi_{1N}=\frac14(1+\gamma_4)(1+\Sigma_z)O_1$ and 
$\psi_{1N^*}=\frac14(1-\gamma_4)(1+\Sigma_z)O_1$. One can see that the squared overlap factor of the negative parity operator $\psi_{1N*}$ with the ground state is negative. This term can be written as
\begin{equation}
\label{eq:amneg}
\la \Omega^L |\psi_{1N^*}(0)|\mathcal{N}_0^R\rangle \langle \mathcal{N}_0^L |\psi_{1N^*}(0)|\Omega^R\rangle
  = -\mcE_z^2\left|\sum_i\frac{\la N_i^*|D_z|N_0\ra}{m_{N_0}-m_{N^*_i}}Z_i-\sum_{n} \frac{\la\Omega|\mcD_z|n^-\ra}{  E_{n}} Z^{n^-}_{N_0}\right|^2
\end{equation}
which is exactly the coefficient of second term in the correlator $\la \psi_{N^*}(t_f)\psi^\dagger_{N^*}(0)\ra$ appearing in Eq.~(\ref{eq:NN2pt}). 

Once the overlap factors of operator $\psi_{1N*}$ with the eigenstates have been determined, Eq.~(\ref{eq:NN2pt}) can be
used to reconstruct the two point correlation function of $\psi_{1N^*}$  and the corresponding effective mass. In
Fig.~{\ref{fig:overlap}, we show the blue data points for the effective mass reconstructed by including the ground state
and first excited state using the overlap factors extracted from Eq.~(\ref{eq:am2}). The reconstructed effective mass
(blue data points) is consistent with that obtained directly from the lattice correlator (yellow data point) and
exhibits the same upward tendency. Note that the reconstructed effective mass has smaller uncertainties at large $t_f$,
since the eigenstate masses $m_{\mcN_n}$ are fixed by averaging the data shown in Fig.~\ref{fig:effmass_O1O2GEVP} over
the range $t_f\in[6a,8a]$ when reconstructing the two point correlator.

\begin{figure*}[t]
\centering
\raisebox{8mm}{\includegraphics[width=.48\textwidth]{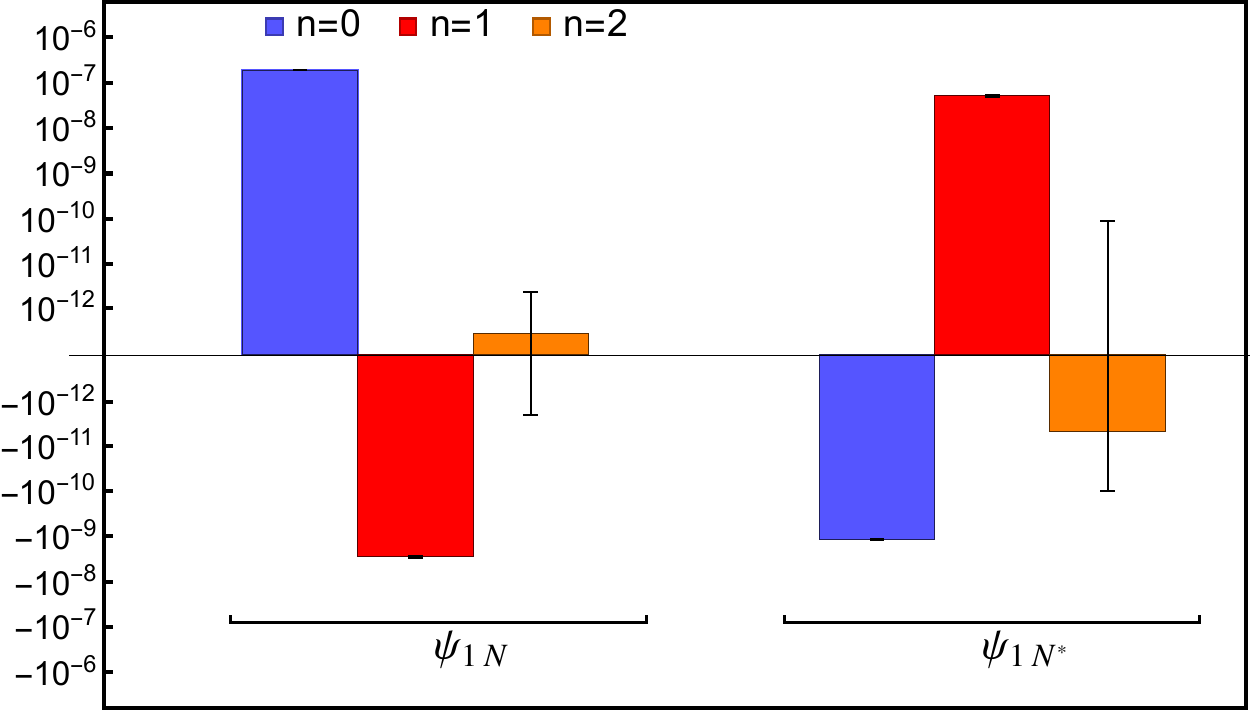}}
\includegraphics[width=.48\textwidth]{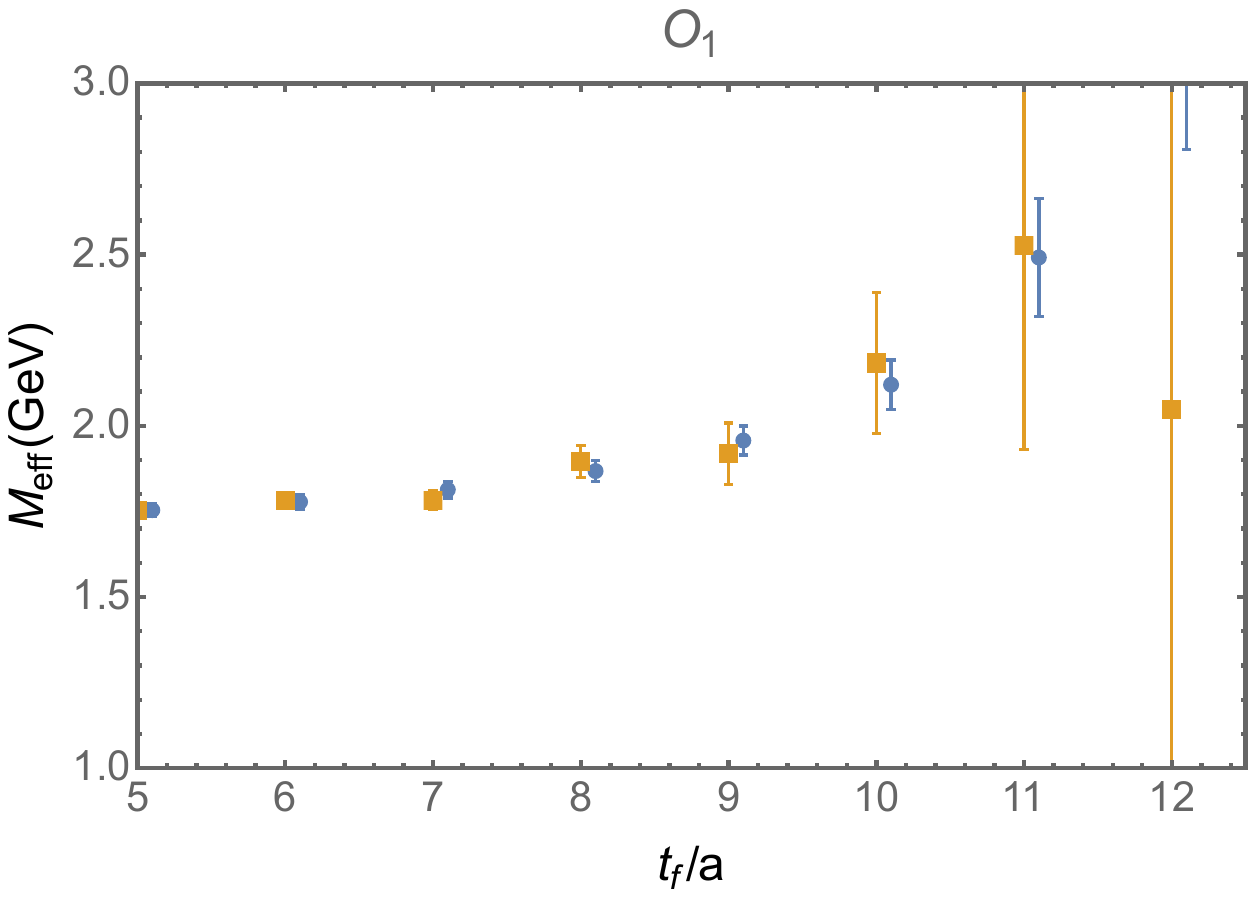}
\caption{The squared overlap factors of the parity components of the interpolating operator $O_1$ with the first three
eigenstates for electric field strength $|n_z|=2$ (left panel) on ensemble 24I-005. The effective mass of negative
parity component of $O_1$ is shown in the right panel.  The blue data points are obtained from the reconstructed
correlator of $\psi_{1N^*}$, while the yellow data points are extracted directly from the original lattice correlator. }
\label{fig:overlap}
\end{figure*}

%%%%%%%%%%%%%%%%%%%%%%%%%%%%%%%%%%%%%%%%%%%%%%%%%%%%%%%%%%%%%%%%%%%%%%%%%%%%%%%%
%%%%%%%%%%%%%%%%%%%%%%%%%%%%%%%%%%%%%%%%%%%%%%%%%%%%%%%%%%%%%%%%%%%%%%%%%%%%%%%%
\section{Transition matrix element between $N$ and $N^*$}\label{sec:appenx_tmax}
The transition matrix elements between the $N$ and $N^*$ are given by~\cite{Devenish:1975jd}
\begin{equation}
\begin{aligned}
\la N^*(p') | J_{EM}^\mu(0) |N(p)\ra
  &= \bar{u}_{N^*}(p')[q^2\gamma^\mu-\slashed{q}q^\mu]u_N(p)G_1(Q^2)
\\&+ \bar{u}_{N^*}\frac{\sigma^{\mu\nu}q_\nu}{2}(m_{N^*}+m_N)u_N(p)G_2(Q^2)
\\  
\la N(p') | J_{EM}^\mu(0) |N^*(p)\ra
  &= \bar{u}_{N}(p')[q^2\gamma^\mu-\slashed{q}q^\mu]u_{N^*}(p)G_1(Q^2)
\\&+ \bar{u}_N\frac{\sigma^{\mu\nu}q_\nu}{2}(m_{N^*}+m_N)u_{N^*}(p)G_2(Q^2)
\end{aligned}
\end{equation}
where $u_{N^*}$ and $u_{N}$ represent the Dirac spinor for the $N^*$ state and $N$ state, respectively. They are related by $u_{N^*}=\gamma_5u_N$ in the rest frame.
The matrix element of electric dipole operator can be expressed as
\begin{equation}
\begin{aligned}
&\la N^*(p') | \mcD_z |N(p)\ra = \int d^3\vec{x}e^{-i(\vec{p}'-\vec{p}).\vec{x}}x_z \la N^*(p') | J_{EM}^4(0) |N(p)\ra
\\&= -i\int d^3\vec{x}\frac{d e^{-i\vec{q}.\vec{x}}}{dx_i}x_z\Big(\bar{u}_{N^*}(p')[\gamma_i q_4-q_i\gamma_4]u_N(p)G_1(Q^2)
-\bar{u}_{N^*}\frac{\sigma^{4i}}{2}(m_{N^*}+m_N)u_N(p)G_2(Q^2)\Big)
\\&= -\Big(\bar{u}_{N^*}(p')[\gamma_z (E_{N^*}-E_N)G_1(Q^2)+i\frac{\sigma^{4z}}{2}(m_{N^*}+m_N)G_2(Q^2)]u_N(p)\Big)(2\pi)^3\delta^3(\vec{p}'-\vec{p}).
\end{aligned}
\end{equation}

Similarly,  
\begin{equation}
\begin{aligned}
\la N(p') | \mcD_z |N^*(p)\ra 
  =-\Big(\bar{u}_{N}(p')[
    \gamma_z (E_N-E_{N^*})G_1(Q^2)
    +i\frac{\sigma^{4z}}{2}(m_{N^*}+m_N)G_2(Q^2)]u_{N^*}(p)\Big)
\, (2\pi)^3\delta^3(\vec{p}'-\vec{p}).
\end{aligned}
\end{equation}

\section{Statistical Improvement from AMA and LMA}\label{sec:appenx_error}
In Fig.~\ref{fig:lma}, we present the nEDM results obtained using the GEVP analysis with the AMA correction alone and with the combination of AMA and LMA, as described in Sec.~\ref{sec:LMA}. We use 200 low modes of the preconditioned Dirac operator to construct the all-to-all nucleon correlators for each ensemble. Once the all-to-all low mode contribution is included, the statistical uncertainty is significantly reduced, especially for the lighter pion mass ensembles, the uncertainty can be reduced by nearly a factor of two.

\begin{figure*}[t]
\centering
\includegraphics[width=.32\textwidth]{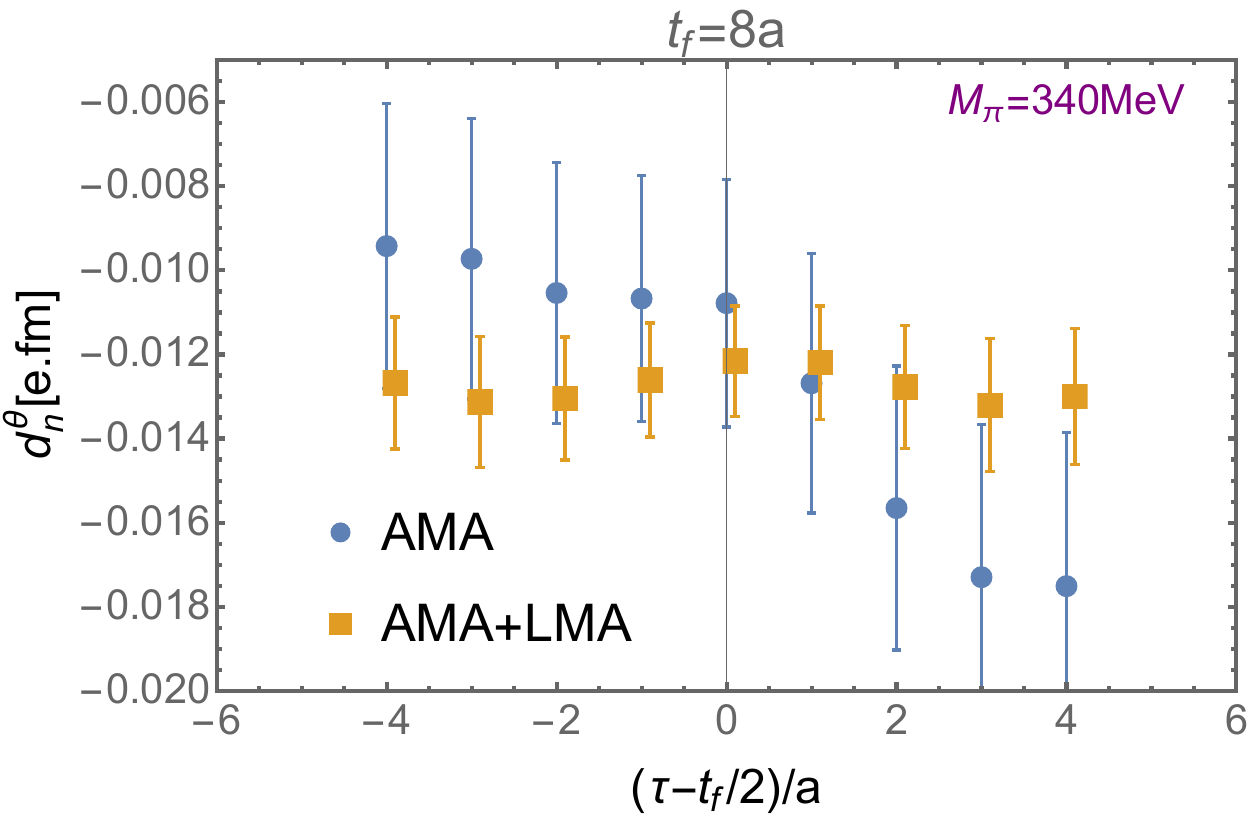}
\includegraphics[width=.32\textwidth]{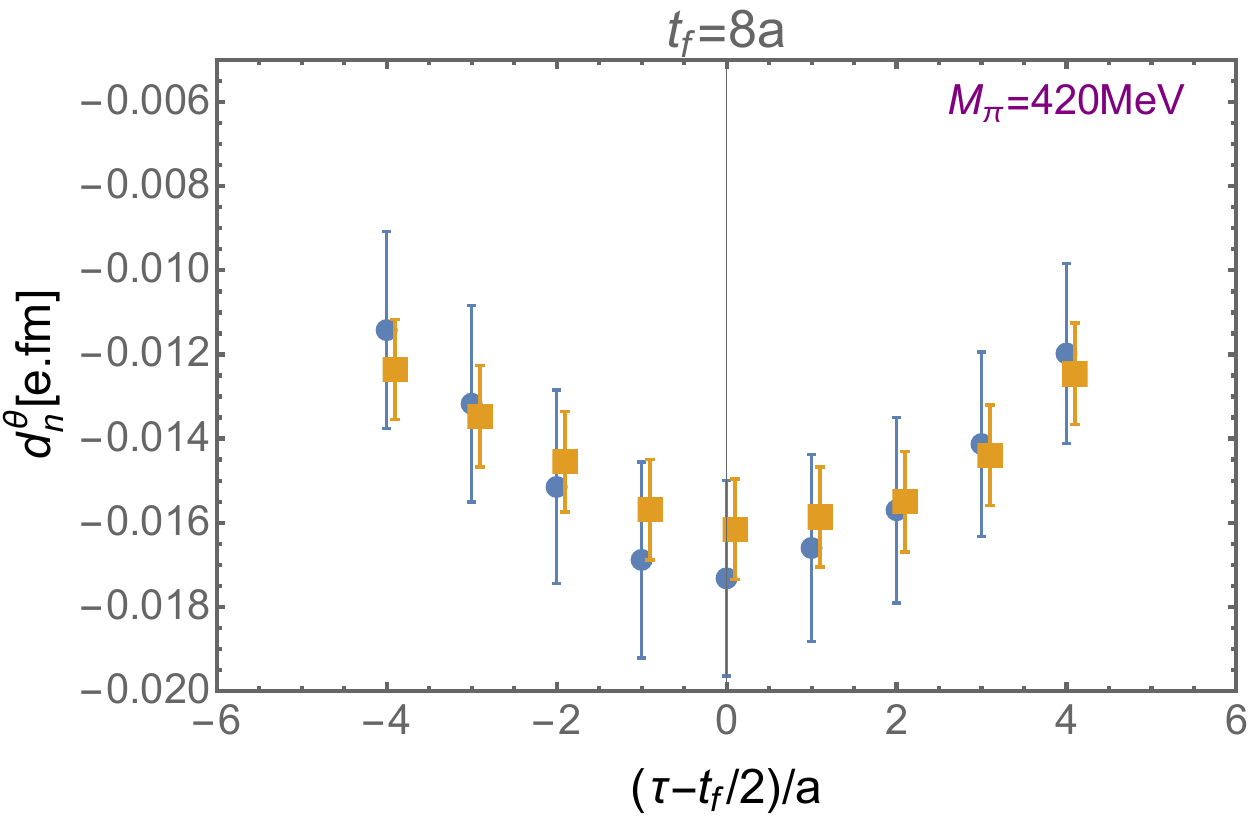}
\includegraphics[width=.32\textwidth]{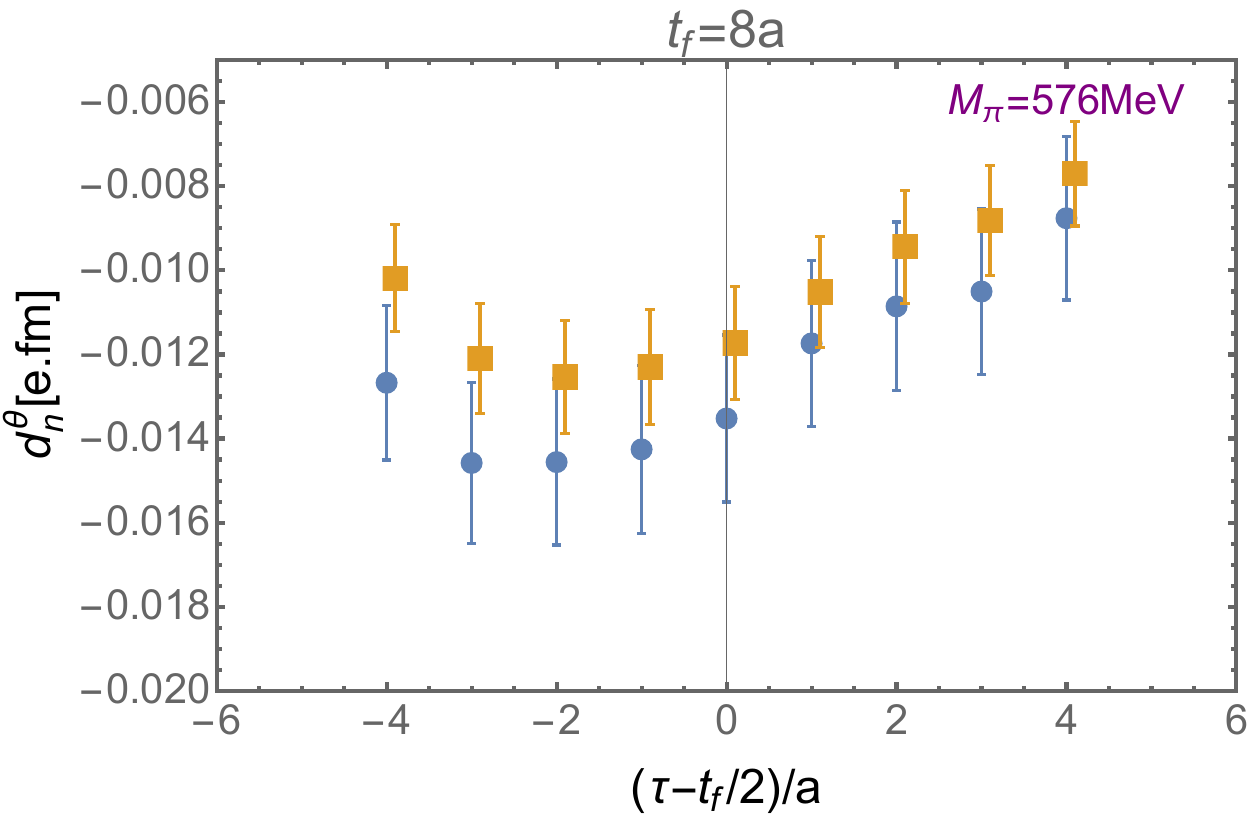}
\caption{Comparison of the nEDM results obtained using the AMA correction alone and the combination of AMA and LMA analysis. The calculations are performed on the ensembles 24I-005(left), 24I-010(middle) and 24I-020(right) with electric field strength $|n_z|=1$. The source sink separation is fixed at $t_f=8a$.}
\label{fig:lma}
\end{figure*}

\bibliography{ref}

\end{document}